\documentclass[reprint, amsmath, amssymb, aps]{revtex4-2}
\usepackage{graphicx}
\usepackage{dcolumn}
\usepackage{bm}
\usepackage{tabstackengine}
\usepackage[dvipsnames]{xcolor}
\usepackage[normalem]{ulem}
\usepackage{amsmath}

\newcommand\notice{\textcolor{black}}

\mathchardef\mhyphen="2D
\begin{document}

\title{Impacts of Point Defects on Shallow Doping in Cubic Boron Arsenide: \newline A First Principles Study}
\author{Shuxiang Zhou, Zilong Hua, and Kaustubh K. Bawane}
\affiliation{Idaho National Laboratory, Idaho Falls, Idaho 83415, USA}

\author{Hao Zhou and Tianli Feng}
\affiliation{Department of Mechanical Engineering, University of Utah, Salt Lake City, Utah 84112, USA}

\begin{abstract}

Cubic boron arsenide (BAs) stands out as a promising material for advanced electronics, thanks to its exceptional thermal conductivity and ambipolar mobility. However, effective control of p- and n-type doping in BAs poses a significant challenge, mostly as a result of the influence of defects. In the present study, we employed density functional theory \notice{(DFT)} to explore the impacts of the common \notice{point defects and impurities} on p-type doping Be$_\text{B}$ and Si$_\text{As}$, and \notice{on} n-type doping Si$_\text{B}$ and Se$_\text{As}$. 
We \notice{found} that the most favorable point defects formed by C, O, and Si are C$_\text{As}$, O$_\text{B}$O$_\text{As}$, Si$_\text{As}$, C$_\text{As}$Si$_\text{B}$, and O$_\text{B}$Si$_\text{As}$, which have formation energies of less than $1.5$ eV.
While only \notice{the} O impurity detrimentally affects the p-type dopings, for n-type dopings, C, O, and Si impurities are all harmful. Interestingly, the antisite defect pair As$_\text{B}$B$_\text{As}$ benefits both p- and n-type doping.
The doping limitation analysis presented in this study can potentially pave the way for strategic development in the area of BAs-based electronics.

\end{abstract}

\maketitle

\section{\label{sec:intro}Introduction}

Over the last decade, cubic boron arsenide (BAs) has attracted an extremely high level of research interest, due to its exceptional thermal conductivity and ambipolar mobility. The thermal conductivity of BAs (i.e., $\sim$1300 W/mK at room temperature, which is only surpassed by diamond for bulk materials) was predicted via theory \cite{broidoInitioStudyUnusual2013,lindsayFirstPrinciplesDeterminationUltrahigh2013,fengFourphononScatteringSignificantly2017a}, then validated by experiments \cite{kangExperimentalObservationHigh2018, tianUnusualHighThermal2018,liHighThermalConductivity2018}. The simultaneously high room-temperature electron and hole mobilities were also predicted by theory \cite{liuSimultaneouslyHighElectron2018} and later confirmed through recent experiments \cite{shinHighAmbipolarMobility2022, yueHighAmbipolarMobility2022}. Beyond its remarkable transport properties, BAs is applicable to existing III–V semiconductor technologies \cite{hartElectronicStructureBAs2000} and has an electronic structure similar to that of Si, but with a wider band gap \cite{wentzcovitchTheoryStructuralElectronic1986}. These attributes position BAs as a promising material for advanced electronics and efficient heat management.

One critical need in high-performance electronic applications relates to the development of methodologies for controlling p- and n-type doping to generate the desired ionizable delocalized (shallow) impurity states \cite{joshiDopingLimitationsCubic2022}. To accomplish this, the behaviors of defects such as intrinsic defects and impurities must be comprehensively understood, as they widely existed in experiments \cite{lv_experimental_2015, kimThermalThermoelectricTransport2016, tianSeededGrowthBoron2018}. Recently, the synthesis of millimeter-sized BAs crystals is achieved via the chemical vapor transport method \cite{xingGaspressureChemicalVapor2018, tianSeededGrowthBoron2018, gamageEffectBoronSources2019, gamageEffectNucleationSites2019}, however, a substantial variance is still observed in the measured thermal properties of BAs \cite{tianUnusualHighThermal2018, lvExperimentalStudyProposed2015, kimThermalThermoelectricTransport2016, xingMultimillimetersizedCubicBoron2018}. Additionally, p-type semiconducting behavior is observed experimentally \cite{gamageEffectBoronSources2019, xingMultimillimetersizedCubicBoron2018, chuCrystalGrowthProperties2003}, indicating the presence of intrinsic defects and/or impurities in BAs samples. For BAs, extensive researches \cite{zhengAntisitePairsSuppress2018, protikInitioStudyEffect2016, favaHowDopantsLimit2021, chenEffectsImpuritiesThermal2021, shinHighAmbipolarMobility2022, tangCompetitionPhononvacancyFourphonon2023} have been performed on the thermal conductivity reduction caused by intrinsic defects and impurities, however, comparably less emphasis has been placed on comprehending how intrinsic defects and impurities affect electronic properties. Notably, first principles calculations \cite{zhengAntisitePairsSuppress2018, chaePointDefectsDopants2018} were utilized to explore the formation of intrinsic defects, identifying the antisite pair As$_\text{B}$B$_\text{As}$ as being the most prevalent intrinsic defect type. As the antisite pair As$_\text{B}$B$_\text{As}$ is usually charge-neutral, it cannot explain the observed p-type semiconducting behavior. Recently, theoretical studies attribute the p-type semiconducting behavior of BAs to impurities such as Si, C, and H \cite{chaePointDefectsDopants2018, lyonsImpurityderivedPtypeConductivity2018}, whereas experiments have validated the presence of Si, C, O, H, Te, and I impurities in BAs \cite{chenEffectsImpuritiesThermal2021}. Therefore, impurities may solely induce p-type behavior in BAs and have the potential to influence both p- and n-type doping limitations. 

Furthermore, the first principles calculations identified specific dopants (e.g., Be$_\text{B}$ and Si$_\text{As}$ for p-type, and Si$_\text{B}$ and Se$_\text{As}$ for n-type) characterized as shallow dopants \cite{chaePointDefectsDopants2018, lyonsImpurityderivedPtypeConductivity2018}. A high p-dopability of BAs is also reported \cite{lyonsImpurityderivedPtypeConductivity2018, bushickBandStructureCarrier2019, mengPressureDependentBehaviorDefectModulated2020, wangNativePointDefects2021}. However, how various point defects influence the behaviors of shallow dopants has never comprehensively been investigated. For the present work, we conducted a thermodynamic analysis demonstrating the interaction between point defects and shallow dopants, utilizing density functional theory (DFT). 



\begin{figure*}[t]
\includegraphics[width=0.91\textwidth]{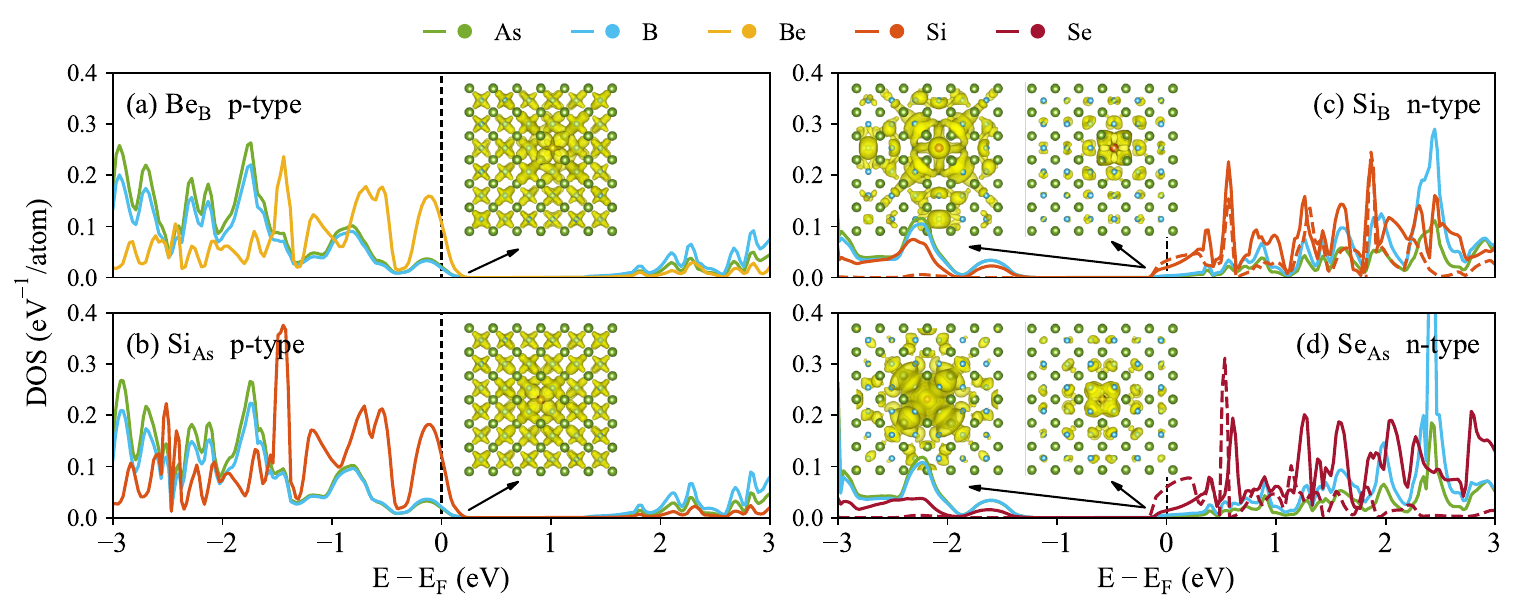}
\caption{\label{fig:dop_dos0} 
Calculated electronic DOS of BAs doped by (a) Be$_\text{B}$, (b) Si$_\text{As}$, (c) Si$_\text{B}$, and (d) Se$_\text{As}$. The solid curves represent the DOS projection on $p$ orbitals of each element, while the dashed curves represent the $s$ orbital projection. The inset figures are the isosurfaces of charge density near the CBM or VBM ($<$0.1 eV), depicted at $4\times10^{-5} \AA^{-3}$. The differently colored atoms represent different species.}
\end{figure*}

\section{\label{sec:compute}Computational Details}

In the present work, the DFT calculations were carried out using the projector augmented-wave method   \cite{blochlProjectorAugmentedwaveMethod1994b, kresseUltrasoftPseudopotentialsProjector1999b}, as implemented in the Vienna ab initio Simulation Package code \cite{kresseInitioMolecularDynamics1993b, kresseEfficientIterativeSchemes1996b}. As the main target of this work is to explore different configurations involving both dopants and point defects, we applied the generalized gradient approximation (GGA) as formulated by Perdew, Burke, and Ernzerhof functional revised for solids (PBEsol) \cite{perdewRestoringDensityGradientExpansion2008a} to reduce the computational cost. Besides that, the total charge of the defect systems was kept neutral, following Ref. \cite{joshiDopingLimitationsCubic2022}. 
For all calculations, spin polarization, a plane-wave cutoff energy of 700 eV, and an energy convergence criterion of $10^{-6}$ eV were applied. For the defect calculations, a $3\times3\times3$ supercell of the conventional cubic cell (216 atoms) and a $2\times2\times2$ $\Gamma$-centered k-point mesh were employed, as used in Ref. \cite{joshiDopingLimitationsCubic2022}. 
The relaxed lattice parameter of the conventional cubic cell of BAs was 4.778 \AA, which is fixed in the calculations of defected structures.   

Through this work, $\mathcal{X}$ is used to denote the defects in the BAs crystal. According to the discussed dopings, $\mathcal{X}$ is categorized into three subsets, the dopants $\mathcal{D}$, other impurities or intrinsic defects $\mathcal{I}$, and their complex $\mathcal{D}\mhyphen\mathcal{I}$. Additionally, $\mathrm{Y}_\mathrm{S_Y}$
is used to represent a single defect, a substitutional Y atom (or V for a vacancy) in the $\mathrm{S_Y}$ site, and $\mathrm{Y}_\mathrm{S_Y}$$\mathrm{Z}_\mathrm{S_Z}$ is used to represent a defect pair of coupled $\mathrm{Y}_\mathrm{S_Y}$ and $\mathrm{Z}_\mathrm{S_Z}$ single defects. The defect sites, $\mathrm{S_Y}$ and $\mathrm{S_Z}$, can only be B or As in BAs.

The key properties computed are the formation energies, electronic density of states (DOS), and the isosurfaces of charge density. The formation energy values quantified the ease of defect formation, while the isosurfaces of charge density were employed to visualize the spatial localization of impurity states, as an indicator of shallow doping state.
The formation energy $E_f$ of defect $\mathcal{X}$ is calculated as:

\begin{equation}
    E_f(\mathcal{X}) = E(\mathcal{X}) - E_0 - \sum_i{n_i^\mathcal{X}(E_i+\mu_i)}
\end{equation}
where $E_0$ is the total energy of the pristine system, $E(\mathcal{X})$ is the total energy of the defect system, $n_i^\mathcal{X}$ is the change in the number of atoms of species $i$ due to defect $\mathcal{X}$ (positive if atoms are added, negative if atoms are removed), $E_i$ is the energy per atom in the elemental phase, and $\mu_i$ is the chemical potential of species $i$.
The elemental phases of species B, As, C, O, Si, Be, and Se were considered by using $\alpha$-B, $\alpha$-As, graphite, oxygen molecule, Si in the diamond structure, Be in the hexagonal close-packed structure, and $\alpha$-Se, respectively. Additionally for oxygen molecules, due to the well-known self-interaction error using DFT \cite{droghetti_predicting_2008}, an energy correction is applied by using the suggested value from the fitted elemental-phase reference energies \cite{stevanovic_correcting_2012}.
For all species $i$, $\mu_i \le 0$. Furthermore, for all impurity atoms and dopants, $\mu_i$ was approximated as zero. The values of $\mu_B$ and $\mu_{As}$ depend on the growth condition involved: in the As-rich condition, $\mu_{As}=0$; otherwise,  $\mu_{B}=0$ in the B-rich condition. Additionally, $\mu_B$ and $\mu_{As}$ are restricted by the equilibrium condition $\mu_B + \mu_{As} = \Delta H_{BAs} $, where $\Delta H_{BAs}$ is the formation enthalpy of BAs. Throughout this work, we use the As-rich condition, following Ref.~\cite{zhengAntisitePairsSuppress2018}. Nevertheless, the PBEsol-predicted value of $\Delta H_{BAs}$ is merely $-0.056$ eV, meaning that the differences in $\mu_B$ (or $\mu_{As}$) between As- and B-rich conditions only total $-0.056$ eV, which is trivial. 
Assuming a defect $\mathcal{I}$ is already present in the system, the formation energy of a dopant $\mathcal{D}$ can be determined using the formula $E_f^\mathcal{I}(\mathcal{D})=E_f(\mathcal{D}\mhyphen \mathcal{I})-E_f(\mathcal{I})$ (see Sec.  I of SM  for more details). Comparing $E_f^\mathcal{I}(\mathcal{D})$ against $E_f(\mathcal{D})$ affords insight into whether the defect $\mathcal{I}$ is advantageous for the formation of the dopant $\mathcal{D}$.  
For DOS, $p$ orbitals of each element were plotted as solid curves throughout this work, and $s$ orbitals were plotted only when it had a higher contribution than $p$ orbitals near the Fermi level $E_F$, by using dashed curves. The isosurfaces of charge density were visualized using the VESTA package \cite{mommaVESTAThreedimensionalVisualization2011}, depicted for each band at $4\times10^{-5} \AA^{-3}$, unless for strongly localized bands, which is depicted at $1\times10^{-3} \AA^{-3}$ and will be explicitly pointed out.

\section{\label{sec:result}Results}

\subsection{\label{sec:doping_result}Shallow Dopants}

\begin{figure}[t]
\includegraphics[width=\columnwidth]{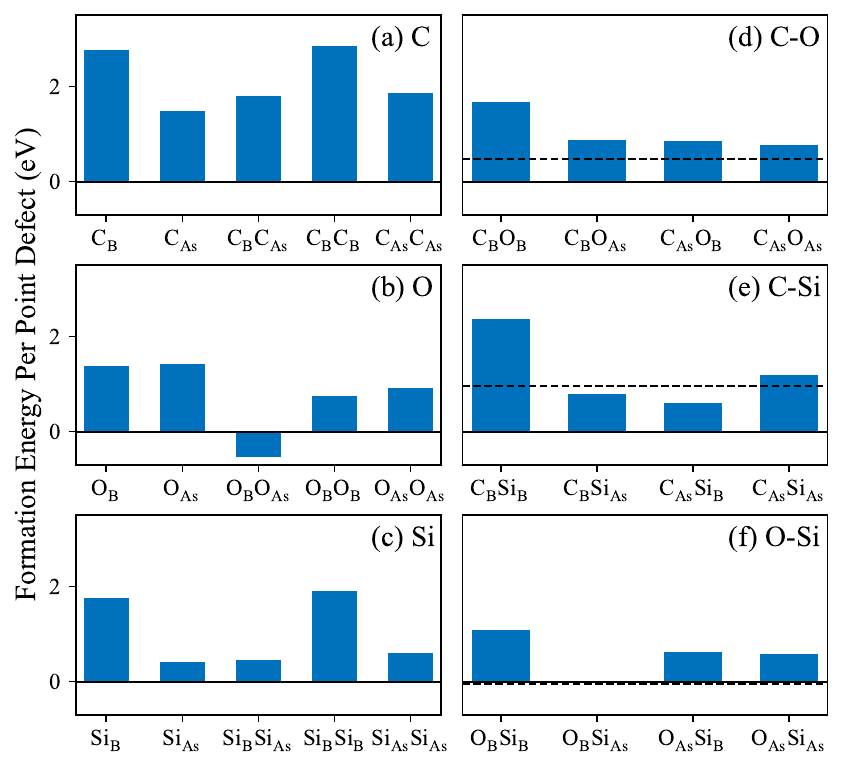}
\caption{\label{fig:defect_energy} 
Calculated formation energy per point defect formed by (a) C, (b) O, (c) Si, (d) C and O, (e) C and Si, and (f) Si and O, in pristine BAs. The dashed lines in (d), (e), and (f) indicate the mean value of the formation energy per point defect of isolated (d) C$_\text{As}$ and O$_\text{B}$O$_\text{As}$, (e) C$_\text{As}$ and Si$_\text{As}$, and (f) O$_\text{B}$O$_\text{As}$ and Si$_\text{As}$, respectively. }
\end{figure}

Here we begin by identifying the dopants. Previous studies~\cite{chaePointDefectsDopants2018, lyonsImpurityderivedPtypeConductivity2018} suggest that p-type dopants Be$_\text{B}$ and Si$_\text{As}$ and n-type dopants Se$_\text{As}$ and Si$_\text{B}$ are shallow dopants with low formation \notice{energies}. 
We confirmed that the formation energies of Be$_\text{B}$ ($0.15$ eV), Si$_\text{As}$ ($0.42$ eV), Si$_\text{B}$ ($1.76$ eV), and Se$_\text{As}$ ($1.47$ eV) are the lowest among the computed dopants (see Sec.  II of SM for more details). Note that for p-type doping (Be$_\text{B}$ and Si$_\text{As}$), the formation energy is substantially lower than that for n-type doping (Se$_\text{As}$ and Si$_\text{B}$), thus supporting the notion that p-type doping occurs more readily in BAs than does n-type doping \cite{chaePointDefectsDopants2018, lyonsImpurityderivedPtypeConductivity2018}. In addition, the DOS and isosurface of charge density were computed for Be$_\text{B}$, Si$_\text{As}$, Si$_\text{B}$, and Se$_\text{As}$ (see Fig.~\ref{fig:dop_dos0}), as well as the electronic band structures (see Sec.  V of SM). 
For Be$_\text{B}$ and Si$_\text{As}$, while the Fermi level $E_F$ slightly shifts into the valence bands, there are three-fold degenerated bands at the valence band maximum (VBM), and their isosurfaces of charge density all exhibit strong spatially delocalized nature from $p$ orbitals, indicating the shallow acceptor nature of Be$_\text{B}$ and Si$_\text{As}$. For Se$_\text{As}$ and Si$_\text{B}$, the Fermi level $E_F$ slightly shifts into the conduction bands (see Fig.~\ref{fig:dop_dos0}(c) and (d)). Two bands are observed near $E_F$, and strong localized $s$-orbital contribution from the impurities, Si and Se, are observed, both from the isosurfaces and DOS.
Only one band shows moderately delocalized behavior (left insets),
indicating the shallow donor nature of Se$_\text{As}$ and Si$_\text{B}$. The $p$ orbital DOS of B is at least 60\% larger than that of As near $E_F$, thus the spatial delocalization is mainly contributed by the $p$ orbitals of B. In summary, we use Be$_\text{B}$, Si$_\text{As}$, Si$_\text{B}$, and Se$_\text{As}$ in further investigations regarding the influence of point defects, while DOS and isosurfaces of charge density work as indicators of the shallow state nature.

\subsection{\label{sec:defect_result}Point defects}

Given the multitude of configurations involving different species and atom sites for point defects, it is imperative to select the most representative ones. First, we considered C, O, and Si as the impurity elements, due to their high concentration as substitutional impurities, as reported in Ref.~\cite{chenEffectsImpuritiesThermal2021}. As a reference, the covalent radius, electronegativity, and common charges for defect elements are summarized in Sec.  III of SM. Secondly, we restricted our consideration to single point defects and point defect pairs in the first nearest neighbor. All considered defect structures are presented using Kröger-Vink Notation \cite{kroger_relations_1956} in Sec. IV of SM. For each situation, we only picked the point defect with the lowest formation energy (see Fig.~\ref{fig:defect_energy}). Panels (a), (b), and (c) reflect situations in which individual C, O, and Si impurities existed, respectively, whereas panels (d), (e), and (f) reflect situations in which combinations of C and O, C and Si, and O and Si impurities coexisted, respectively. Note that, for a direct comparison, the y-axis value is the formation energy per point defect, meaning that the formation energy of defect pairs is divided by two. Per Fig.~\ref{fig:defect_energy}(a)--(c), C$_\text{As}$, O$_\text{B}$O$_\text{As}$, and Si$_\text{As}$ are the most favorable point defects when C, O, and Si are the only impurity elements, respectively. In Fig.~\ref{fig:defect_energy}(d), rather than forming defect pairs, C and O impurities are more likely to form isolated C$_\text{As}$ and O$_\text{B}$O$_\text{As}$ (indicated by the dashed line). In contrast, C and Si atoms are more likely to form C$_\text{As}$Si$_\text{B}$, as opposed to isolated C$_\text{As}$ and Si$_\text{As}$ (indicated by the dashed line). When O and Si coexist, the formation energy per point defect of O$_\text{B}$Si$_\text{As}$ is $0.04$ eV, and the mean formation energy per point defect of O$_\text{B}$O$_\text{As}$ and Si$_\text{As}$ is $-0.10$ eV (see Fig.~\ref{fig:defect_energy}(f)). Given the small difference (0.14 eV) between these two values, we exceptionally count in O$_\text{B}$Si$_\text{As}$ into further investigation. To summarize, we selected C$_\text{As}$, O$_\text{B}$O$_\text{As}$, Si$_\text{As}$, C$_\text{As}$Si$_\text{B}$, and O$_\text{B}$Si$_\text{As}$ as representative point defects of impurity. Notably, the lowest formation energy is for O$_\text{B}$O$_\text{As}$ (-0.53 eV), and the possible reason for the negative value is discussed in Sec.  II of SM.

\begin{figure}[t]
\includegraphics[width=0.91\columnwidth]{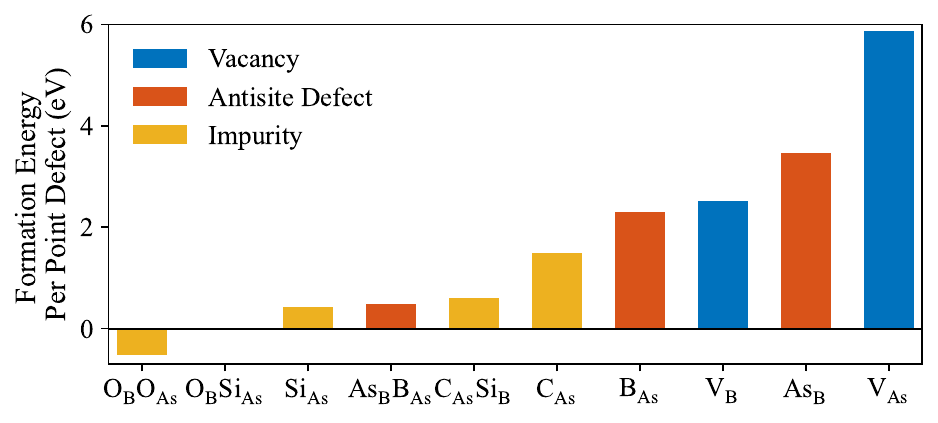}
\caption{\label{fig:defect_rank} 
Calculated formation energy of selected single point defects and point defect pairs. The obtained results are organized in ascending order.}
\end{figure}

\begin{figure*}[hbtp]
\includegraphics[width=0.91\textwidth]{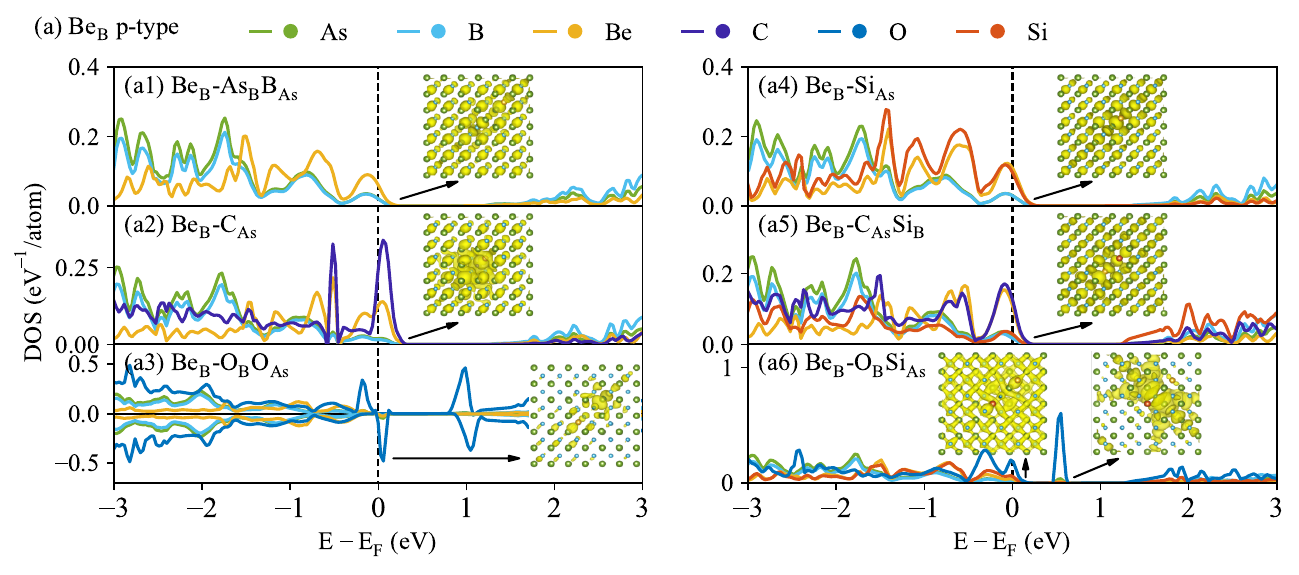}
\includegraphics[width=0.91\textwidth]{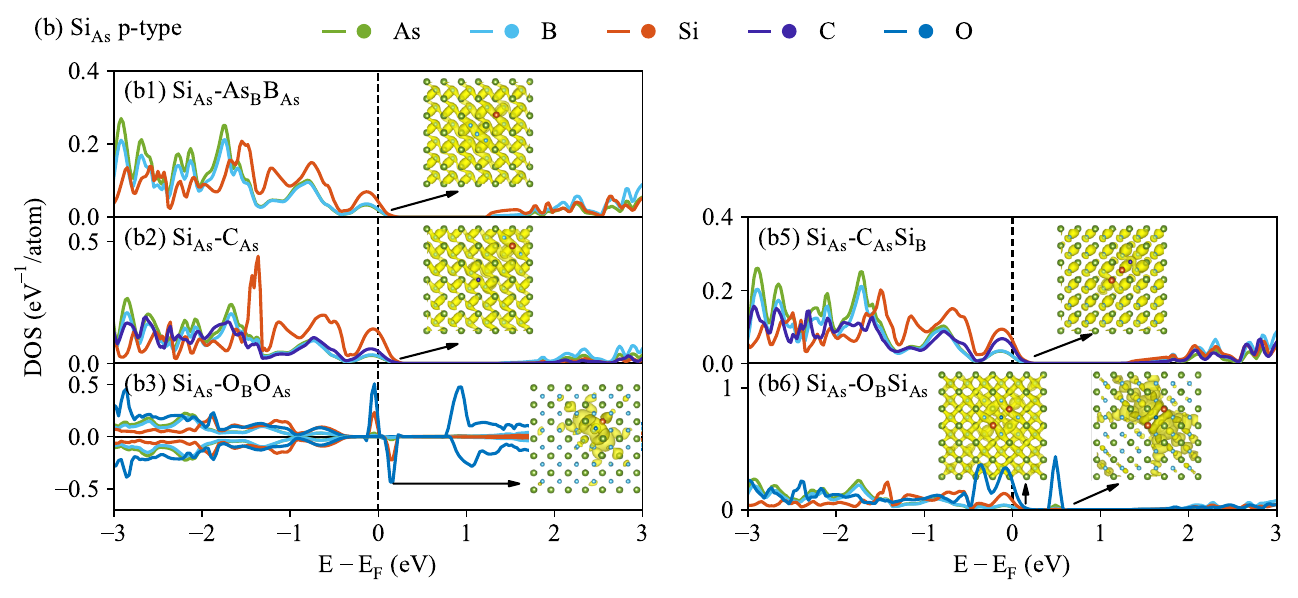}
\caption{\label{fig:dop_dos_ptype} 
Calculated electronic DOS of BAs doped by p-type dopants (a) Be$_\text{B}$ and (b) Si$_\text{As}$ in the presence of (1) As$_\text{B}$B$_\text{As}$, (2) C$_\text{As}$, (3) O$_\text{B}$O$_\text{As}$, (4) Si$_\text{As}$, (5) C$_\text{As}$Si$_\text{B}$, or (6) O$_\text{B}$Si$_\text{As}$ point \notice{defects}. The solid curves represent the DOS projection on $p$ orbitals of each element. The inset figures show the isosurfaces of charge density near VBM ($<$0.1 eV) or within the band gap, depicted at $4\times10^{-5} \AA^{-3}$ (except for O$_\text{B}$Si$_\text{As}$, which is depicted at $1\times10^{-3} \AA^{-3}$). The differently colored atoms represent different species. }
\end{figure*}

We also examined the intrinsic point defects, including antisite single defects As$_\text{B}$ and B$_\text{As}$, antisite defect pair As$_\text{B}$B$_\text{As}$, and single vacancies V$_\text{B}$ and V$_\text{As}$. The formation energy per point defect for each of these intrinsic defects is plotted alongside the selected point defects of \notice{impurity}, moving from the smallest formation energy (left) to the largest (right) (see Fig.~\ref{fig:defect_rank}). Certainly, only the antisite defect pair As$_\text{B}$B$_\text{As}$ has a formation energy comparable to that of the selected point defects of impurity. Thus, we will discuss a total of six types of point defects, (i.e., the aforementioned five selected types of impurities, plus As$_\text{B}$B$_\text{As}$), with the formation energies per point defect all being less than $1.5$ eV.

\begin{figure*}[hbtp]
\includegraphics[width=0.91\textwidth]{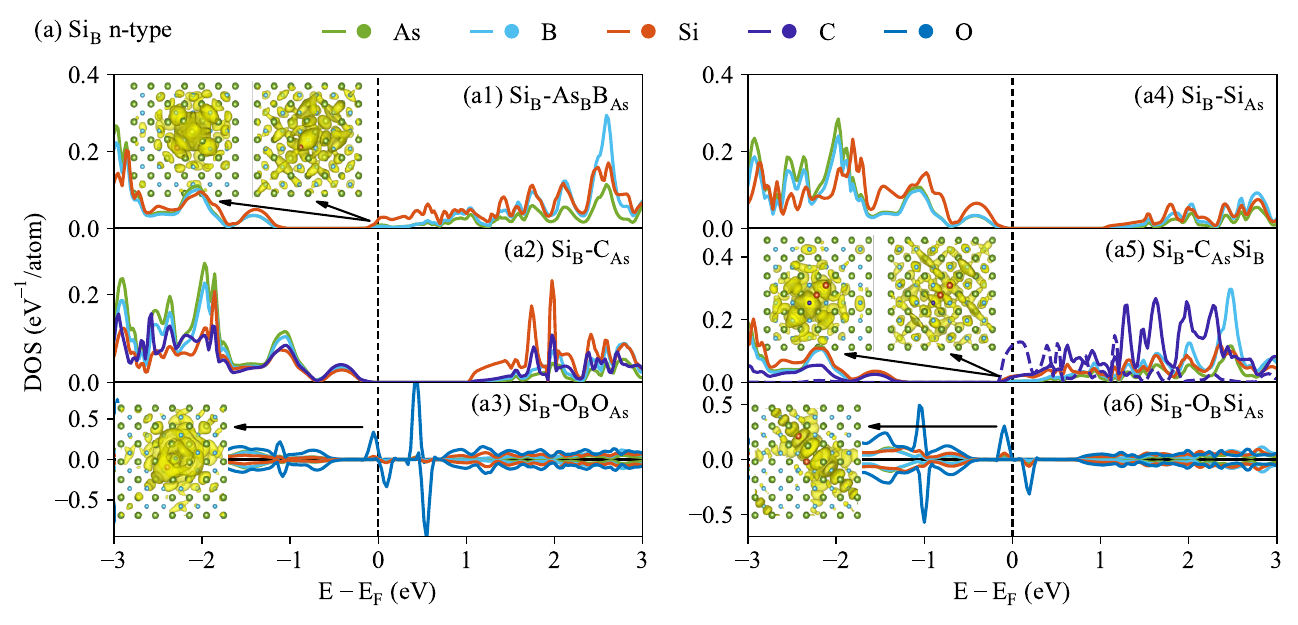}
\includegraphics[width=0.91\textwidth]{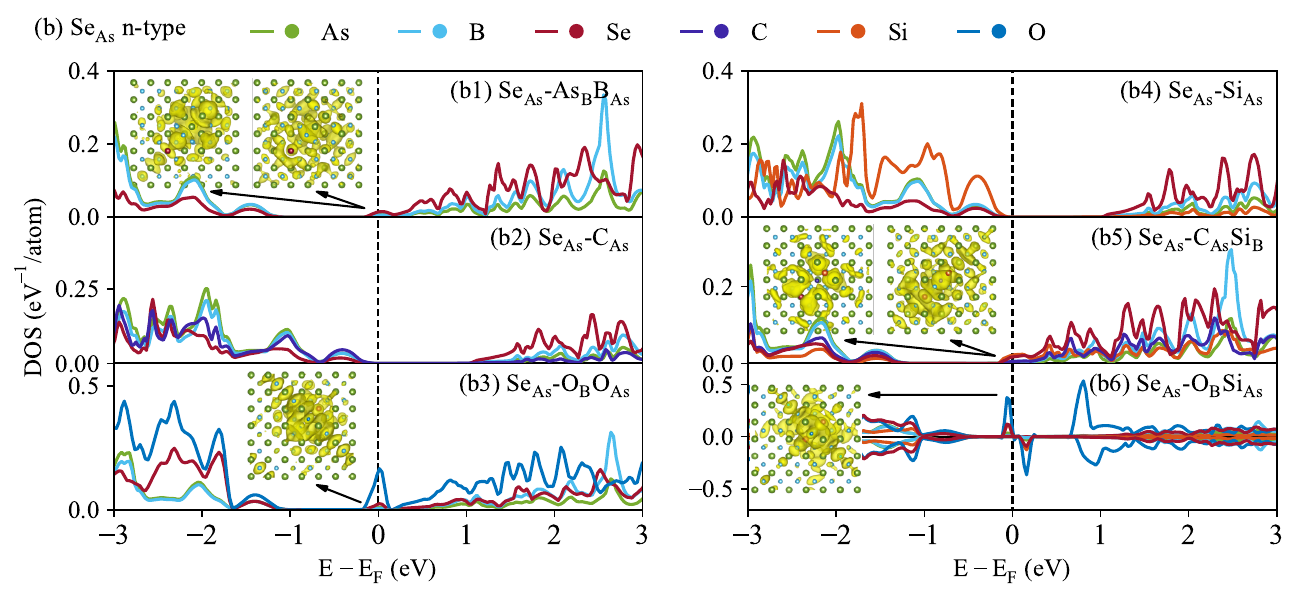}
\caption{\label{fig:dop_dos_ntype} 
Calculated electronic DOS of BAs doped by n-type dopants (a) Si$_\text{B}$ and (b) Se$_\text{As}$ in the presence of (1) As$_\text{B}$B$_\text{As}$, (2) C$_\text{As}$, (3) O$_\text{B}$O$_\text{As}$, (4) Si$_\text{As}$, (5) C$_\text{As}$Si$_\text{B}$, or (6) O$_\text{B}$Si$_\text{As}$ point \notice{defects}. The solid curves represent the DOS projection on $p$ orbitals of each element, while the dashed curves represent the $s$ orbital projection. The inset figures show the isosurfaces of charge density near CBM ($<$0.1 eV) or within the band gap, depicted at $4\times10^{-5} \AA^{-3}$. The differently colored atoms represent different species.}
\end{figure*}

\begin{figure}[t]
\includegraphics[width=\columnwidth]{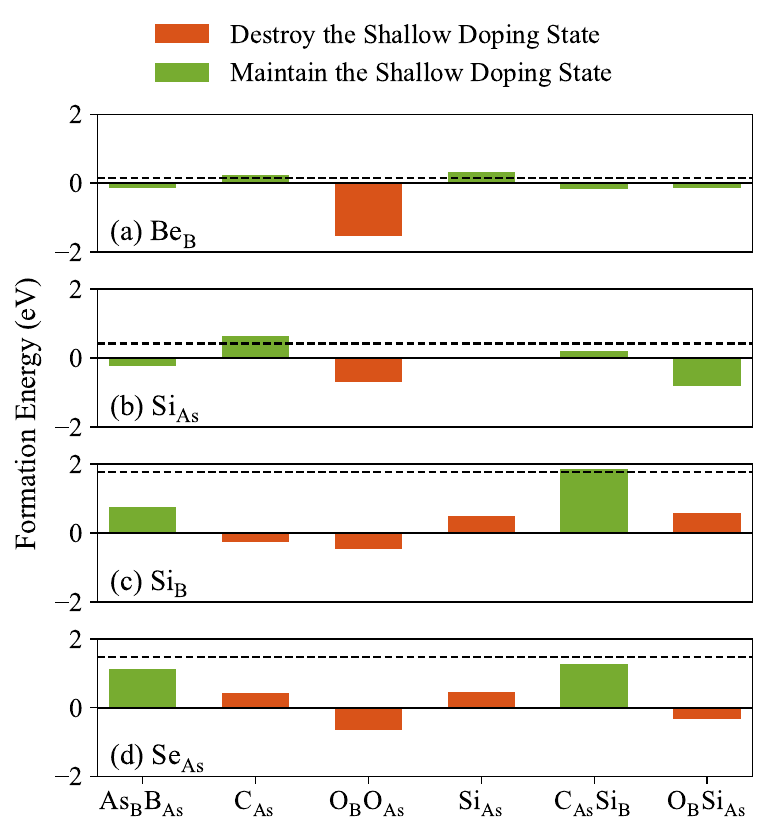}
\caption{\label{fig:doping_defect} 
Calculated formation energy of (a) Be$_\text{B}$, (b) Si$_\text{As}$, (c) Si$_\text{B}$, and (d) Se$_\text{As}$ dopants in the presence of the selected point defects. The dashed lines represent the formation energy of dopants without defects. The colors of the bars indicate whether a detrimental influence of each point defect is observed on the shallow dopant. }
\end{figure}

\subsection{\label{sec:doping_result}Effects of point defects on shallow doping}

Finally, we discuss the impact of these common point defects (As$_\text{B}$B$_\text{As}$, C$_\text{As}$, O$_\text{B}$O$_\text{As}$, Si$_\text{As}$, C$_\text{As}$, Si$_\text{B}$, and O$_\text{B}$Si$_\text{As}$) on both the formation energy and the electronic characteristics of shallow dopants Be$_\text{B}$, Si$_\text{As}$, Si$_\text{B}$, and Se$_\text{As}$. Please note that concerning Si$_\text{As}$ being regarded as a p-type dopant, Si$_\text{As}$ itself is no longer categorized as a point defect of impurity, while C$_\text{As}$Si$_\text{B}$ and O$_\text{B}$Si$_\text{As}$ continue to be considered as impurities due to the presence of C and O. In Sec.  VII of SM, we consider additional point defects As$_\text{B}$, B$_\text{As}$, and N$_\text{As}$, despite their potential lack of favorability within the BAs system. When different point defects are present, the dopants may occupy various atom sites. For each combination of point defect and dopant, all possible dopant sites within 6\AA{} of the point defects are computed. Sec.  VI of SM tabulates the detailed calculation results. Here, we only focus on the configuration with the lowest total energy and, consequently, the lowest formation energy.

We start with examining the electronic characteristics via the depiction of DOS and isosurfaces of charge density of Be$_\text{B}$ doping (see Fig.~\ref{fig:dop_dos_ptype}(a)). As mentioned above, Be$_\text{B}$ and Si$_\text{As}$ feature three-fold degenerated bands at VBM, and the presence of point defects may break the symmetry of the degenerated bands. In this case, only the highest band at VBM is shown as inset figures in Fig.~\ref{fig:dop_dos_ptype}(a) and (b), while all bands within 0.1 eV of VBM are provided in enlarged figures in Sec.  VIII of SM.
Interestingly, the p-type shallow state of Be$_\text{B}$ is maintained in the presence of all point defects, except O$_\text{B}$O$_\text{As}$, given that the Fermi level $E_F$ is still positioned slightly below VBM, and the isosurfaces of charge density near VBM continue to exhibit spatial delocalization contributed from $p$ orbitals of both B and As. This result is understandable for the charge-neutral defects As$_\text{B}$B$_\text{As}$ and C$_\text{As}$Si$_\text{B}$, as well as for the acceptor-like defects C$_\text{As}$ and Si$_\text{As}$, though the acceptor-like defects slightly shift the $E_F$ deeper into the valence band. For the donor-like defects O$_\text{B}$Si$_\text{As}$, a localized defect trap state within the band gap. Despite the narrowed band gaps, the p-type shallow state near $E_F$ is still maintained. The only detrimental case is the donor-like defects O$_\text{B}$O$_\text{As}$, where a localized defect trap state forms at the top of VBM. 
For the other p-type dopant Si$_\text{As}$, the results are similar to those for Be$_\text{B}$; namely, As$_\text{B}$B$_\text{As}$, C$_\text{As}$, C$_\text{As}$Si$_\text{B}$, and O$_\text{B}$Si$_\text{As}$ all sustain the shallow state (see Fig.~\ref{fig:dop_dos_ptype}(b)). The only difference is that O$_\text{B}$O$_\text{As}$ now forms a localized defect trap state in the middle of the band gap where $E_F$ is positioned, which is still detrimental. Even the top bands of VBM remain the shallow doping nature, Si$_\text{As}$ have to be doped beyond compensation limits to remedy the move of the Fermi level $E_F$.  

Nevertheless, in the case of n-type dopants Si$_\text{B}$ and Se$_\text{As}$, the examined point defects exert a more significant influence, owing to the acceptor-like nature of C$_\text{As}$ and Si$_\text{As}$ impurities---for which the Fermi level $E_F$ is no longer in the conduction bands---thereby destroying the n-type nature (see Fig.~\ref{fig:dop_dos_ntype}(a) and (b))). Additionally, both O$_\text{B}$O$_\text{As}$ and O$_\text{B}$Si$_\text{As}$ form localized defect trap states within the band gap where $E_F$ is located. As a result, the charge-neutral defects As$_\text{B}$B$_\text{As}$ and C$_\text{As}$Si$_\text{B}$ are the only defects that preserve the n-type shallow state, giving that $E_F$ is marginally above CBM, and one of the bands near CBM exhibits moderate spatial delocalization while the other becomes predominantly localized.

Furthermore, we now provide a synopsis regarding the formation energy of dopants in the presence of point defects. The bars in Fig~\ref{fig:doping_defect} illustrate $E_f^\mathcal{I}(\mathcal{D})$, the formation energy of the dopants $\mathcal{D}$ in the presence of each point defect $\mathcal{I}$. The dashed black lines denote $E_f(\mathcal{D})$, the formation energy of the dopants without point defects. 
Green signifies that the delocalized doping state of the original type is preserved, whereas red indicates otherwise. This color is defined by the DOS and isosurface results given in Fig.~\ref{fig:dop_dos_ptype} and \ref{fig:dop_dos_ntype}. 
Three distinct scenarios are covered: (1) a bar above the dashed line implies that the defect does not couple with the dopant, thus merely impacting the doping; (2) a red bar below the dashed line signifies that a coupled complex of the defect and dopant will likely form, potentially compromising the originally targeted doping state; \notice{and} (3) a green bar below the dashed line indicates that the coupled complex will likely form and preserve the shallow doping state. 

Following these three scenarios, the O impurity proves detrimental to all four types of dopants by the results of O$_\text{B}$O$_\text{As}$. This necessitates restriction of O's concentration. Regarding C$_\text{As}$ and Si$_\text{As}$, for p-type dopings Be$_\text{B}$ and Si$_\text{As}$, concerns over C and Si impurities are alleviated; otherwise, for n-type dopings Si$_\text{B}$ and Se$_\text{As}$, careful removal of C and Si impurities is imperative, which again emphasizes the lower n-type dopability in BAs. Most interestingly, the most favorable intrinsic defect, antisite defect pair As$_\text{B}$B$_\text{As}$, proves beneficial for all four types of dopants. Similar results can be confirmed for the other charge-neutral defect, C$_\text{As}$Si$_\text{B}$, which also is beneficial for three of the four types of dopants (excluding Si$_\text{B}$, where Si$_\text{B}$ does not couple with C$_\text{As}$Si$_\text{B}$).

\section{\label{sec:conclude}Conclusions}

In conclusion, we have computed, from first principles, the formation energies and electronic characteristics of selected dopants, point defects, and coupled complexes of dopants and point defects in BAs. The shallow doping states of p-type dopants Be$_\text{B}$ and Si$_\text{As}$, and n-type dopants Si$_\text{B}$ and Se$_\text{As}$, were directly confirmed by the delocalized charge density isosurface near the Fermi level $E_F$. The favorable single point defects and point defect pairs for C, O, and Si impurities were identified (C$_\text{As}$, O$_\text{B}$O$_\text{As}$, Si$_\text{As}$, C$_\text{As}$Si$_\text{B}$, and O$_\text{B}$Si$_\text{As}$), all with formation energies of less than $1.5$ eV. As for couplings between the selected dopants and defects, this study also identified the most favorable configurations. In terms of the influence of impurities on doping, the O impurity is detrimental to both p- and n-type doping, and C and Si impurities are detrimental to only n-type doping. Consequently, n-type doping is challenging, as it has a higher formation energy and requires the removal of C, Si, and O impurities. Interestingly, the most favorable intrinsic defect, antisite defect pair As$_\text{B}$B$_\text{As}$, positively impacts both the p- and n-type doping. These insights into the interactions between dopants and defects could potentially help expedite the further development of advanced electronics based on BAs, especially in understanding the doping limitation of n-type doping.


\section{Acknowledgements}

This work is supported through the INL Laboratory Directed Research and Development (LDRD) Program under DOE Idaho Operations Office Contract DE-AC07-05ID14517, LDRD Project ID 23A1070-064FP. This research made use of Idaho National Laboratory’s High Performance Computing systems located at the Collaborative Computing Center and supported by the Office of Nuclear Energy of the U.S. Department of Energy and the Nuclear Science User Facilities under Contract No. DE-AC07-05ID14517.

Data will be made available on request.

\bibliographystyle{apsrev4-1} 
\bibliography{bas} 

\end{document}


\raggedright{\textbf{\Large Supplemental materials to}}
\title{Impacts of Point Defects on Shallow Doping in Cubic Boron Arsenide: \\
A First Principles Study}
\author{Shuxiang Zhou$^1$, Zilong Hua$^1$, Kaustubh K. Bawane$^1$, Hao Zhou$^2$, and Tianli Feng$^2$}
\affiliation{$^1$Idaho National Laboratory, Idaho Falls, Idaho 83415, USA\\$^2$Department of Mechanical Engineering, University of Utah, Salt Lake City, Utah 84112, USA}

\maketitle

\section{\label{sec:smformation} Formation energy of dopants with the presence of point defects}
\vspace{5mm}
\justifying

The formation energy $E_f$ of a defect $\mathcal{X}$ in pristine BAs is calculated by:

\begin{equation}\label{eq:efx}
    E_f(\mathcal{X}) = E(\mathcal{X}) - E_0 - \sum_i{n_i^\mathcal{X}(E_i+\mu_i)}
\end{equation}

where $E_0$ is the total energy of the pristine system, $E(\mathcal{X})$ is the total energy of the defect system, $n_i^\mathcal{X}$ represents the number of atoms of species $i$ changed according to the defect $\mathcal{X}$ (positive if atoms are added, negatives if atoms are removed), $E_i$ is the energy per atom in the elemental phase, and $\mu_i$ is the chemical potential of species $i$. According to the discussed dopings, $\mathcal{X}$ is categorized into three subsets, the dopants $\mathcal{D}$, other impurities or intrinsic defects $\mathcal{I}$, and their complex $\mathcal{D}\mhyphen\mathcal{I}$.

Therefore, the formation energy of $\mathcal{D}\mhyphen\mathcal{I}$ is calculated by:
\begin{equation}\label{eq:efdx}
    E_f(\mathcal{D}\mhyphen\mathcal{I}) = E(\mathcal{D}\mhyphen\mathcal{I}) - E_0 - \sum_i{n_i^{\mathcal{D}\mhyphen\mathcal{I}}(E_i+\mu_i)} = E(\mathcal{D}\mhyphen\mathcal{I}) - E_0 - \sum_i{n_i^{\mathcal{D}}(E_i+\mu_i)} - \sum_i{n_i^{\mathcal{I}}(E_i+\mu_i)}
\end{equation}

As for the formation energy of a dopant $\mathcal{D}$ in the presence of a defect $\mathcal{I}$, the reference state is the defected BAs rather than pristine BAs:
\begin{equation}\label{eq:efxd}
    E_f^\mathcal{I}(\mathcal{D}) = E(\mathcal{D}\mhyphen\mathcal{I}) - E(\mathcal{I}) - \sum_i{n_i^\mathcal{D}(E_i+\mu_i)}
\end{equation}
By combining Eq.~\ref{eq:efx}-\ref{eq:efxd}, one can obtain 
\begin{equation}\label{eq:efxd2}
    E_f^\mathcal{I}(\mathcal{D})=E_f(\mathcal{D}\mhyphen \mathcal{I})-E_f(\mathcal{I})
\end{equation}

\clearpage

\section{\label{sec:smdopant} Formation energy calculation of dopants in BA\lowercase{s} }
\vspace{5mm}
\justifying

\begin{figure}[h]
\includegraphics[width=0.45\columnwidth]{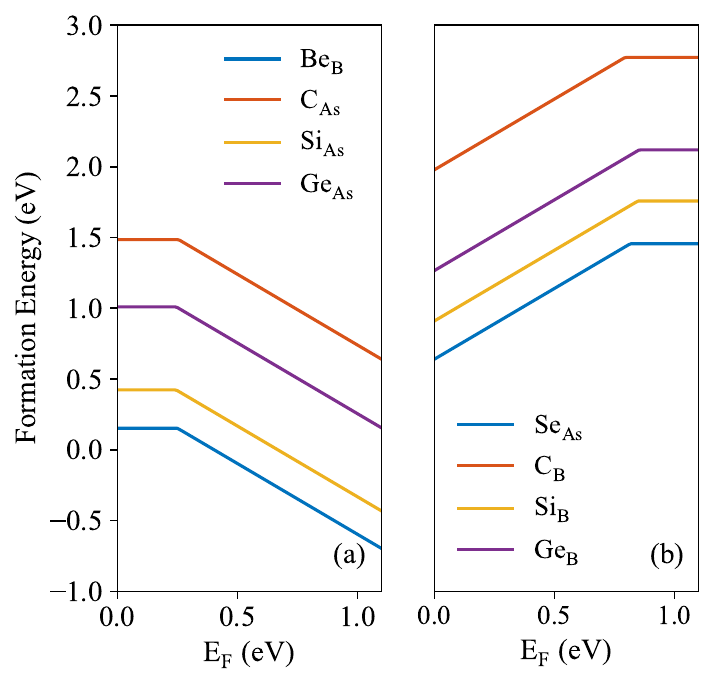}
\caption{\label{fig:sm_dop_eng} 
Calculated formation energies of (a) p-type dopants and (b) n-type dopants, as functions of Fermi energy, in which the minimum and maximum values correspond to VBM and CBM, respectively. }
\end{figure}

The formation energy of charged p-type dopants (Be$_\text{B}$, C$_\text{As}$, Si$_\text{As}$, and Ge$_\text{As}$) and n-type dopants (Se$_\text{As}$, C$_\text{B}$, Si$_\text{B}$, and Ge$_\text{B}$) as a function of the Fermi level $E_F$ was computed using PBEsol (see Fig.~\ref{fig:sm_dop_eng}). For a dopant $\mathcal{D}$ with charge $q$, the formation energy $E_f(\mathcal{D}^q)$ given by ):

\begin{equation}
    E_f(\mathcal{D}^q) = E(\mathcal{D}^q) - E_0 - \sum_i{n_i^\mathcal{D}(E_i+\mu_i)} + q(E_F+E_{VBM}) + E_{corr}(\mathcal{D}^q)
\end{equation}

The formation energy of a charged dopant depends on the Fermi level $E_F$, which is referenced to the energy of the valence band maximum (VBM), $E_{VBM}$. Specifically, $E_F$ spans from zero to the band gap $E_{gap}$, wherein $E_F=0$ signifies the Fermi level resting on the VBM, and $E_F=E_{gap}$ indicates that the Fermi level aligns with the conduction band minimum (CBM). 

In DFT calculations involving non-zero charges, it becomes necessary to simulate a neutralizing background to circumvent infinite charge issues under periodic boundary conditions. Thereby, a correction energy $E_{corr}$ is introduced to account for the nonphysical electrostatic interaction between periodic charged-defect images, which is computed using SXDEFECTALIGN package \cite{freysoldtFullyInitioFiniteSize2009}. The SXDEFECTALIGN package implements the Freysoldt, Neugebauer, and Van de Walle (FNV) method [3], which assumes the shape
of the defect charge as a spherical model, fits the spherical model parameters with the defect wave functions, and obtains the correction by comparing the calculated potential.  The charges $q$ computed in the systems varied from -2 to 2. 

From our calculations, Be$_\text{B}$ and Si$_\text{As}$ are the two p-type dopants with the lowest formation energy, while Se$_\text{As}$ and Si$_\text{B}$ are the two n-type dopants with the lowest formation energy.

Additionally, for the formation energy of O$_\text{B}$O$_\text{As}$, which is the lowest one (-0.53 eV) in this work, we re-compute its value via the exchange-correlation functional of HSEsol (0.10 eV). Please note that PBE may underestimate formation energy when compared with HSE.

\clearpage

\section{\label{sec:smdopant} covalent radius, electronegativity, and common charges }
\vspace{5mm}
\justifying

We tabulate covalent radius, electronegativity, and common charges for B, As, and defect elements (Be, Si, Se, C, O) as below. The bold text in common charges highlights the charge in BAs system, considering the electronegativity of each element.

\begin{table}[h]
\caption{ The comparison of covalent radius \cite{cordero_covalent_2008}, electronegativity \cite{allen_electronegativity_1989}, and common element charges between B, As, Be, Si, Se, C, and O.
}
\begin{tabular}{@{}r|c|c|c}
\hline\hline 
Element & Covalent Radius (pm) & Electronegativity & Common Charges\\ 
\hline\hline 
B  & 84  & 2.05 & 3-, \textbf{3+} \\
As & 119 & 2.21 & \textbf{3-}, 3+, 5+ \\
\hline 
Be & 96  & 1.58 & \textbf{2+} \\
Si & 111 & 1.92 & \textbf{4+}, 4- \\
Se & 120 & 2.42 & \textbf{2-}, 4+, 6+ \\
\hline 
C & 69-76 & 2.55 & 4+, \textbf{4-} \\
O & 66   & 3.44 & \textbf{2-}  \\
\hline\hline 
\end{tabular}
\end{table}
\normalsize
\clearpage

\section{\label{sec:smdopant} Presentation of Defected structure using Kröger-Vink Notation \cite{kroger_relations_1956}}
\vspace{5mm}
\justifying

We start with the stoichiometric equations for BAs:
\begin{equation}
    \text{B}\text{As} \rightleftharpoons \text{B}_\text{B}^\times + \text{As}_\text{As}^\times
\end{equation}
\begin{equation}
    \text{As}_\text{As}^\times \rightleftharpoons \text{V}_\text{As}^{\cdot\cdot\cdot} + 3\text{e}^\prime + \text{As}
\end{equation}
\begin{equation}
    \text{B}_\text{B}^\times \rightleftharpoons \text{V}_\text{B}^{\prime\prime\prime} + 3\text{h}^\cdot + \text{B}
\end{equation}
\begin{equation}
    \varnothing \rightleftharpoons \text{e}^\prime + \text{h}^\cdot
\end{equation}
\begin{equation}
    \varnothing \rightleftharpoons \text{V}_\text{As}^{\cdot\cdot\cdot} + \text{V}_\text{B}^{\prime\prime\prime} 
\end{equation}

For antisite defects:
\begin{equation}
    \text{As} + \text{V}_\text{B}^{\prime\prime\prime} \rightleftharpoons \text{As}_\text{B}^{\prime\prime\prime\prime\prime\prime} + 3\text{h}^\cdot
\end{equation}
\begin{equation}
    \text{B} + \text{V}_\text{As}^{\cdot\cdot\cdot} \rightleftharpoons \text{B}_\text{As}^{\cdot\cdot\cdot\cdot\cdot\cdot} + 3\text{e}^\prime
\end{equation}

For Be defects:
\begin{equation}
    \text{Be} + \text{V}_\text{B}^{\prime\prime\prime} \rightleftharpoons \text{Be}_\text{B}^\prime + 2\text{e}^\prime
\end{equation}

For Si defects:
\begin{equation}
    \text{Si} + \text{V}_\text{As}^{\cdot\cdot\cdot} \rightleftharpoons \text{Si}_\text{As}^{\cdot\cdot\cdot\cdot\cdot\cdot\cdot} + 4\text{e}^\prime
\end{equation}
\begin{equation}
    \text{Si} + \text{V}_\text{B}^{\prime\prime\prime} \rightleftharpoons \text{Si}_\text{B}^\cdot + 4\text{e}^\prime
\end{equation}

For Se defects:
\begin{equation}
    \text{Se} + \text{V}_\text{As}^{\cdot\cdot\cdot} \rightleftharpoons \text{Se}_\text{As}^\cdot + 2\text{h}^\cdot
\end{equation}

For C defect:
\begin{equation}
    \text{C} + \text{V}_\text{As}^{\cdot\cdot\cdot} \rightleftharpoons \text{C}_\text{As}^{\prime} + 4\text{h}^\cdot
\end{equation}
\begin{equation}
    \text{C} + \text{V}_\text{B}^{\prime\prime\prime} \rightleftharpoons \text{C}_\text{B}^{\prime\prime\prime\prime\prime\prime\prime} + 4\text{h}^\cdot
\end{equation}

For O defect:
\begin{equation}
    \text{O} + \text{V}_\text{As}^{\cdot\cdot\cdot} \rightleftharpoons \text{O}_\text{As}^\cdot + 2\text{h}^\cdot
\end{equation}
\begin{equation}
    \text{O} + \text{V}_\text{B}^{\prime\prime\prime} \rightleftharpoons \text{O}_\text{B}^{\prime\prime\prime\prime\prime} + 2\text{h}^\cdot
\end{equation}

Please note that the charge state is speculated using the electronegativity and common charge of each element, and is not proved by DFT. Usually, the charge transfer values from DFT are not integers. The extra electrons and holes will be re-distributed in DFT by minimizing the total energy. This notation is only used to present the defected structures, as given below. Please also note that the DFT calculations presented in the manuscript are charge-neutral. We used a supercell of $\text{B}_{108}\text{As}_{108}$, thus $x=1/108$ in the following equations.

Here we list the defected structures for the 4 dopings (as in Fig.1).

doping $\text{Be}_\text{B}$: B$_{1-x}$As$_{}$Be$_{x}$
\begin{equation}
    \text{B}\text{As} - x\text{B} + x\text{Be} \rightleftharpoons (1-x)\text{B}\text{As} + x(\text{Be}_\text{B}^\prime + \text{h}^\cdot + \text{As}_\text{As}^\times)
\end{equation}

doping $\text{Si}_\text{As}$: B$_{}$As$_{1-x}$Si$_{x}$
\begin{equation}
    \text{B}\text{As} - x\text{As} + x\text{Si} \rightleftharpoons (1-x)\text{B}\text{As} + x(\text{Si}_\text{As}^{\cdot\cdot\cdot\cdot\cdot\cdot\cdot} + 7\text{e}^\prime + \text{B}_\text{B}^\times)
\end{equation}

doping $\text{Si}_\text{B}$: B$_{1-x}$As$_{}$Si$_{x}$
\begin{equation}
    \text{B}\text{As} - x\text{B} + x\text{Si} \rightleftharpoons (1-x)\text{B}\text{As} + x(\text{Si}_\text{B}^\cdot + \text{e}^\prime + \text{As}_\text{As}^\times)
\end{equation}

doping $\text{Se}_\text{As}$: B$_{}$As$_{1-x}$Se$_{x}$
\begin{equation}
    \text{B}\text{As} - x\text{As} + x\text{Se} \rightleftharpoons (1-x)\text{B}\text{As} + x(\text{Se}_\text{As}^\cdot + \text{e}^\prime + \text{B}_\text{B}^\times)
\end{equation}

Here we list the defect structures for the 6 single defects and 22 defect pairs (as in Fig.2 and 3).

single defect $\text{V}_\text{B}$: B$_{1-x}$As$_{}$
\begin{equation}
    \text{B}\text{As} - x\text{B}  
    \rightleftharpoons (1-x)\text{B}\text{As} + 
    x(\text{V}_\text{B}^{\prime\prime\prime} + 3\text{h}^\cdot + \text{As}_\text{As}^\times)
\end{equation}

single defect $\text{V}_\text{As}$: B$_{}$As$_{1-x}$
\begin{equation}
    \text{B}\text{As} - x\text{As}  
    \rightleftharpoons (1-x)\text{B}\text{As} + 
    x(\text{V}_\text{As}^{\cdot\cdot\cdot} + 3\text{e}^\prime + \text{B}_\text{B}^\times)
\end{equation}

single defect $\text{C}_\text{B}$: B$_{1-x}$As$_{}$C$_{x}$
\begin{equation}
    \text{B}\text{As} - x\text{B} + x\text{C} 
    \rightleftharpoons (1-x)\text{B}\text{As} + 
    x(\text{C}_\text{B}^{\prime\prime\prime\prime\prime\prime\prime} + 7\text{h}^\cdot + \text{As}_\text{As}^\times)
\end{equation}

single defect $\text{C}_\text{As}$: B$_{}$As$_{1-x}$C$_{x}$
\begin{equation}
    \text{B}\text{As} - x\text{As} + x\text{C} 
    \rightleftharpoons (1-x)\text{B}\text{As} + 
    x(\text{C}_\text{As}^{\prime} + \text{h}^\cdot + \text{B}_\text{B}^\times)
\end{equation}

single defect $\text{O}_\text{B}$: B$_{1-x}$As$_{}$O$_{x}$
\begin{equation}
    \text{B}\text{As} - x\text{B} + x\text{O} 
    \rightleftharpoons (1-x)\text{B}\text{As} + 
    x(\text{O}_\text{B}^{\prime\prime\prime\prime\prime} + 5\text{h}^\cdot + \text{As}_\text{As}^\times)
\end{equation}

single defect $\text{O}_\text{As}$: B$_{}$As$_{1-x}$O$_{x}$
\begin{equation}
    \text{B}\text{As} - x\text{As} + x\text{O} 
    \rightleftharpoons (1-x)\text{B}\text{As} + 
    x(\text{O}_\text{As}^\cdot + \text{e}^\prime + \text{B}_\text{B}^\times)
\end{equation}

defect pair $\text{C}_\text{B}$$\text{C}_\text{As}$: B$_{1-x}$As$_{1-x}$C$_{2x}$
\begin{equation}
    \text{B}\text{As} - x\text{B} - x\text{As} + 2x\text{C} 
    \rightleftharpoons (1-x)\text{B}\text{As} + 
    x(\text{C}_\text{B}^{\prime\prime\prime\prime\prime\prime\prime} + \text{C}_\text{As}^{\prime} + 8\text{h}^\cdot)
\end{equation}

defect pair $\text{C}_\text{B}$$\text{C}_\text{B}$: B$_{1-2x}$As$_{}$C$_{2x}$
\begin{equation}
    \text{B}\text{As} - 2x\text{B} + 2x\text{C} 
    \rightleftharpoons (1-2x)\text{B}\text{As} + 
    2x(\text{C}_\text{B}^{\prime\prime\prime\prime\prime\prime\prime} + 7\text{h}^\cdot + \text{As}_\text{As}^\times)
\end{equation}

defect pair $\text{C}_\text{As}$$\text{C}_\text{As}$: B$_{}$As$_{1-2x}$C$_{2x}$
\begin{equation}
    \text{B}\text{As} - 2x\text{As} + 2x\text{C} 
    \rightleftharpoons (1-2x)\text{B}\text{As} + 
    2x(\text{C}_\text{As}^{\prime} + \text{h}^\cdot + \text{B}_\text{B}^\times)
\end{equation}

defect pair $\text{O}_\text{B}$$\text{O}_\text{As}$: B$_{1-x}$As$_{1-x}$O$_{2x}$
\begin{equation}
    \text{B}\text{As} - x\text{B} - x\text{As} + 2x\text{O} 
    \rightleftharpoons (1-x)\text{B}\text{As} + 
    x(\text{O}_\text{B}^{\prime\prime\prime\prime\prime} + \text{O}_\text{As}^{\cdot} + 4\text{h}^\cdot)
\end{equation}

defect pair $\text{O}_\text{B}$$\text{O}_\text{B}$: B$_{1-2x}$As$_{}$O$_{2x}$
\begin{equation}
    \text{B}\text{As} - 2x\text{B} + 2x\text{O} 
    \rightleftharpoons (1-2x)\text{B}\text{As} + 
    2x(\text{O}_\text{B}^{\prime\prime\prime\prime\prime} + 5\text{h}^\cdot + \text{As}_\text{As}^\times)
\end{equation}

defect pair $\text{O}_\text{As}$$\text{O}_\text{As}$: B$_{}$As$_{1-2x}$O$_{2x}$
\begin{equation}
    \text{B}\text{As} - 2x\text{As} + 2x\text{O} 
    \rightleftharpoons (1-2x)\text{B}\text{As} + 
    2x(\text{O}_\text{As}^\cdot + \text{e}^\prime + \text{B}_\text{B}^\times)
\end{equation}

defect pair $\text{Si}_\text{B}$$\text{Si}_\text{As}$: B$_{1-x}$As$_{1-x}$Si$_{2x}$
\begin{equation}
    \text{B}\text{As} - x\text{B} - x\text{As} + 2x\text{Si} 
    \rightleftharpoons (1-x)\text{B}\text{As} + 
    x(\text{Si}_\text{B}^\cdot + \text{Si}_\text{As}^{\cdot\cdot\cdot\cdot\cdot\cdot\cdot}+8\text{e}^\prime)
\end{equation}

defect pair $\text{Si}_\text{B}$$\text{Si}_\text{B}$: B$_{1-2x}$As$_{}$Si$_{2x}$
\begin{equation}
    \text{B}\text{As} - 2x\text{B} + 2x\text{Si} \rightleftharpoons (1-2x)\text{B}\text{As} + 2x(\text{Si}_\text{B}^\cdot + \text{e}^\prime + \text{As}_\text{As}^\times)
\end{equation}

defect pair $\text{Si}_\text{As}$$\text{Si}_\text{As}$: B$_{}$As$_{1-2x}$Si$_{2x}$
\begin{equation}
    \text{B}\text{As} - 2x\text{As} + 2x\text{Si} \rightleftharpoons (1-2x)\text{B}\text{As} + 2x(\text{Si}_\text{As}^{\cdot\cdot\cdot\cdot\cdot\cdot\cdot} + 7\text{e}^\prime + \text{B}_\text{B}^\times)
\end{equation}

defect pair $\text{C}_\text{B}$$\text{O}_\text{B}$: B$_{1-2x}$As$_{}$C$_{x}$O$_{x}$
\begin{equation}
    \text{B}\text{As} - 2x\text{B} + x\text{C} + x\text{O} 
    \rightleftharpoons (1-2x)\text{B}\text{As} + 
    x(\text{C}_\text{B}^{\prime\prime\prime\prime\prime\prime\prime} + \text{O}_\text{B}^{\prime\prime\prime\prime\prime} + 12\text{h}^\cdot + 2\text{As}_\text{As}^\times)
\end{equation}

defect pair $\text{C}_\text{B}$$\text{O}_\text{As}$: B$_{1-x}$As$_{1-x}$C$_{x}$O$_{x}$
\begin{equation}
    \text{B}\text{As} - x\text{B} - x\text{As} + x\text{C} + x\text{O} 
    \rightleftharpoons (1-x)\text{B}\text{As} + 
    x(\text{C}_\text{B}^{\prime\prime\prime\prime\prime\prime\prime} + \text{O}_\text{As}^\cdot + 6\text{h}^\cdot)
\end{equation}

defect pair $\text{C}_\text{As}$$\text{O}_\text{B}$:  B$_{1-x}$As$_{1-x}$C$_{x}$O$_{x}$
\begin{equation}
    \text{B}\text{As} - x\text{B} - x\text{As} + x\text{C} + x\text{O} 
    \rightleftharpoons (1-x)\text{B}\text{As} + 
    x(\text{C}_\text{As}^{\prime} + \text{O}_\text{B}^{\prime\prime\prime\prime\prime} + 6\text{h}^\cdot)
\end{equation}

defect pair $\text{C}_\text{As}$$\text{O}_\text{As}$: B$_{}$As$_{1-2x}$C$_{x}$O$_{x}$
\begin{equation}
    \text{B}\text{As} - 2x\text{As} + x\text{C} + x\text{O} 
    \rightleftharpoons (1-2x)\text{B}\text{As} + 
    x(\text{C}_\text{As}^{\prime} + \text{O}_\text{As}^\cdot +  2\text{B}_\text{B}^\times)
\end{equation}

defect pair $\text{C}_\text{B}$$\text{Si}_\text{B}$:  B$_{1-2x}$As$_{}$C$_{x}$Si$_{x}$
\begin{equation}
    \text{B}\text{As} - 2x\text{B} + x\text{C} + x\text{Si} 
    \rightleftharpoons (1-2x)\text{B}\text{As} + 
    x(\text{C}_\text{B}^{\prime\prime\prime\prime\prime\prime\prime} + \text{Si}_\text{B}^{\cdot} + 6\text{h}^\cdot + 2\text{As}_\text{As}^\times)
\end{equation}

defect pair $\text{C}_\text{B}$$\text{Si}_\text{As}$:  B$_{1-x}$As$_{1-x}$C$_{x}$Si$_{x}$
\begin{equation}
    \text{B}\text{As} - x\text{B} - x\text{As} + x\text{C} + x\text{Si} 
    \rightleftharpoons (1-x)\text{B}\text{As} + 
    x(\text{C}_\text{B}^{\prime\prime\prime\prime\prime\prime\prime} + \text{Si}_\text{As}^{\cdot\cdot\cdot\cdot\cdot\cdot\cdot} )
\end{equation}

defect pair $\text{C}_\text{As}$$\text{Si}_\text{B}$: B$_{1-x}$As$_{1-x}$C$_{x}$Si$_{x}$
\begin{equation}
    \text{B}\text{As} - x\text{B} - x\text{As} + x\text{C} + x\text{Si} 
    \rightleftharpoons (1-x)\text{B}\text{As} + 
    x(\text{C}_\text{As}^{\prime} + \text{Si}_\text{B}^{\cdot} )
\end{equation}

defect pair $\text{C}_\text{As}$$\text{Si}_\text{As}$: B$_{}$As$_{1-2x}$C$_{x}$O$_{x}$
\begin{equation}
    \text{B}\text{As} - 2x\text{As} + x\text{C} + x\text{Si} 
    \rightleftharpoons (1-2x)\text{B}\text{As} + 
    x(\text{C}_\text{As}^{\prime} + \text{Si}_\text{As}^{\cdot\cdot\cdot\cdot\cdot\cdot\cdot} + 6\text{e}^\prime + 2\text{B}_\text{B}^\times)
\end{equation}

defect pair $\text{O}_\text{B}$$\text{Si}_\text{B}$: B$_{1-2x}$As$_{}$O$_{x}$Si$_{x}$
\begin{equation}
    \text{B}\text{As} - 2x\text{B} + x\text{O} + x\text{Si} 
    \rightleftharpoons (1-2x)\text{B}\text{As} + 
    x(\text{O}_\text{B}^{\prime\prime\prime\prime\prime} + \text{Si}_\text{B}^{\cdot} + 4\text{h}^\cdot + 2\text{As}_\text{As}^\times)
\end{equation}

defect pair $\text{O}_\text{B}$$\text{Si}_\text{As}$: B$_{1-x}$As$_{1-x}$O$_{x}$Si$_{x}$
\begin{equation}
    \text{B}\text{As} - x\text{B} - x\text{As} + x\text{O} + x\text{Si} 
    \rightleftharpoons (1-x)\text{B}\text{As} + 
    x(\text{O}_\text{B}^{\prime\prime\prime\prime\prime} + \text{Si}_\text{As}^{\cdot\cdot\cdot\cdot\cdot\cdot\cdot} + 2\text{e}^\prime)
\end{equation}

defect pair $\text{O}_\text{As}$$\text{Si}_\text{B}$: B$_{1-x}$As$_{1-x}$O$_{x}$Si$_{x}$
\begin{equation}
    \text{B}\text{As} - x\text{B} - x\text{As} + x\text{O} + x\text{Si} 
    \rightleftharpoons (1-x)\text{B}\text{As} + 
    x(\text{O}_\text{As}^\cdot + \text{Si}_\text{B}^{\cdot} + 2\text{e}^\prime)
\end{equation}

defect pair $\text{O}_\text{As}$$\text{Si}_\text{As}$: B$_{}$As$_{1-2x}$O$_{x}$Si$_{x}$
\begin{equation}
    \text{B}\text{As} - 2x\text{As} + x\text{O} + x\text{Si} 
    \rightleftharpoons (1-2x)\text{B}\text{As} + 
    x(\text{O}_\text{As}^\cdot + \text{Si}_\text{As}^{\cdot\cdot\cdot\cdot\cdot\cdot\cdot}  + 8\text{e}^\prime + 2\text{B}_\text{B}^\times)
\end{equation}

defect pair $\text{B}_\text{As}$$\text{As}_\text{B}$: BAs
\begin{equation}
    \text{B}\text{As} 
    \rightleftharpoons (1-x)\text{B}\text{As} + 
    x(\text{B}_\text{As}^{\cdot\cdot\cdot\cdot\cdot\cdot} + \text{As}_\text{B}^{\prime\prime\prime\prime\prime\prime})
\end{equation}

\clearpage
\section{\label{sec:smdopant} Electronic band structures of doped BA\lowercase{s} }
\vspace{5mm}
\justifying

The electronic band structures of pristine BAs and BAs doped with Be$_\text{B}$, Si$_\text{As}$,  Si$_\text{B}$, and Se$_\text{As}$ are presented in this section. The colored dots represent the contributions from the dopant elements in the corresponding Bloch wave functions, where the dot size is proportional to the value of the contribution. 

\begin{figure}[h]
\includegraphics[width=0.2\textwidth]{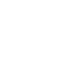}
\includegraphics[width=0.38\textwidth]{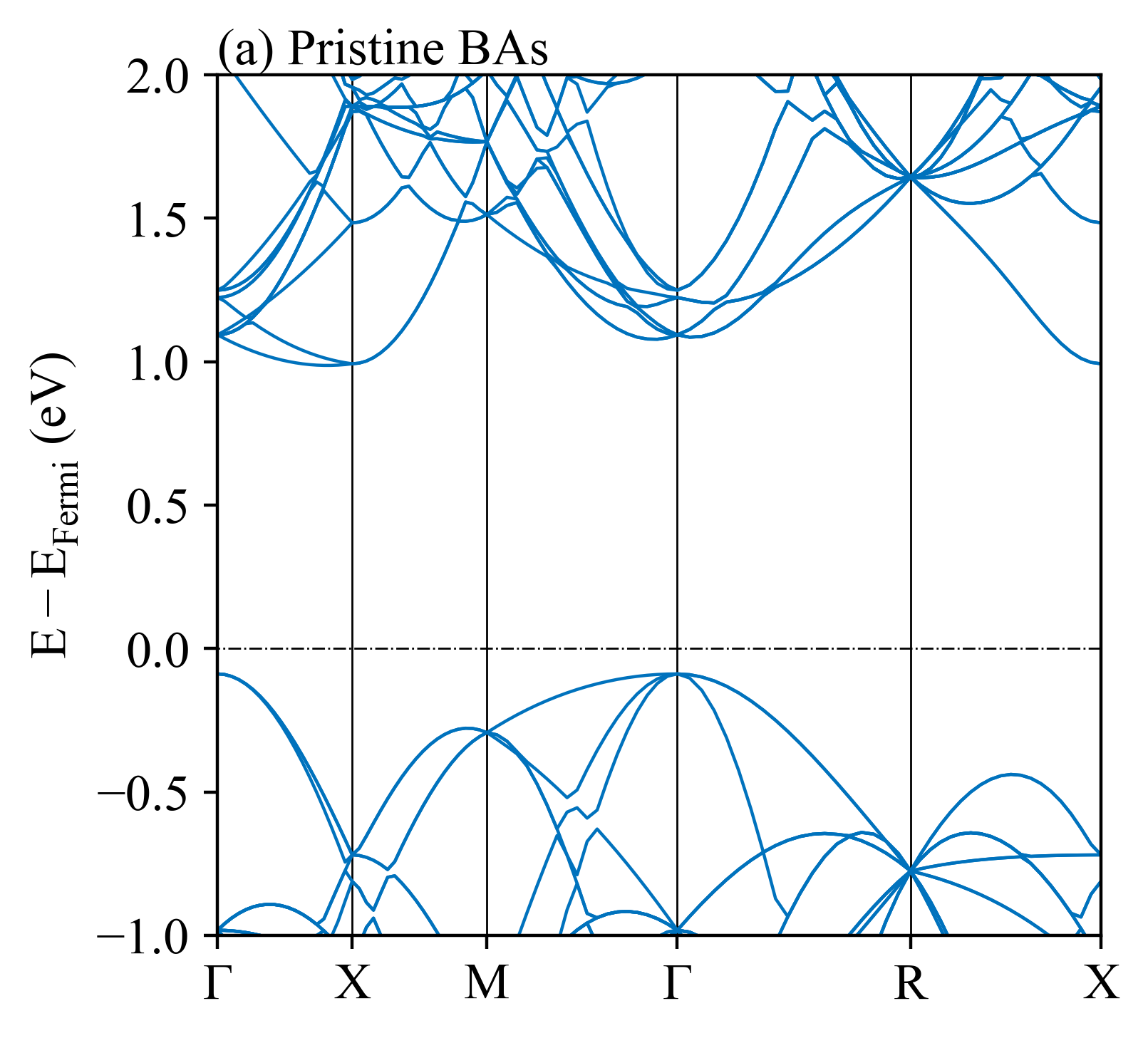}
\includegraphics[width=0.2\textwidth]{blank.png}
\includegraphics[width=0.38\columnwidth]{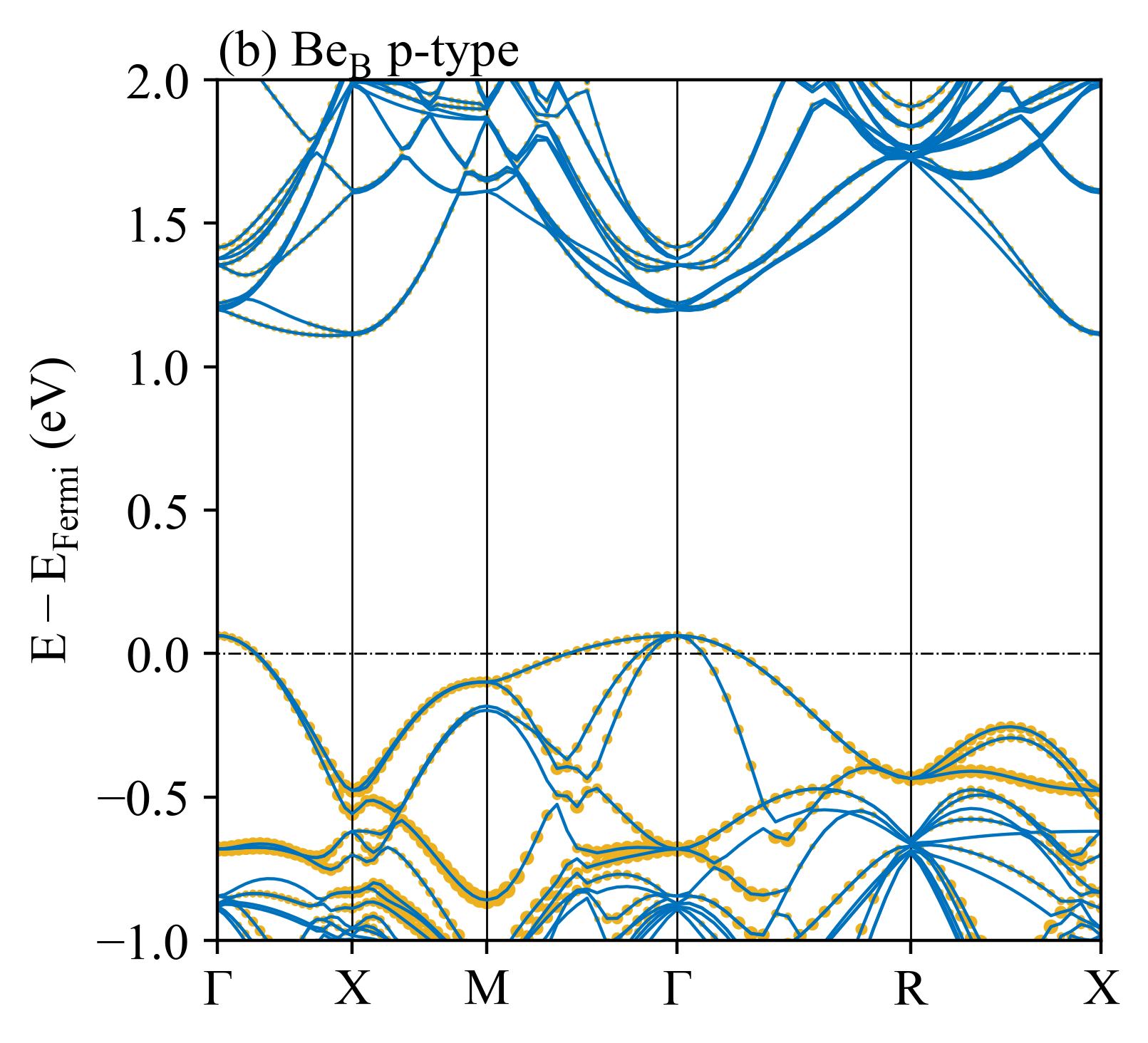}
\includegraphics[width=0.38\columnwidth]{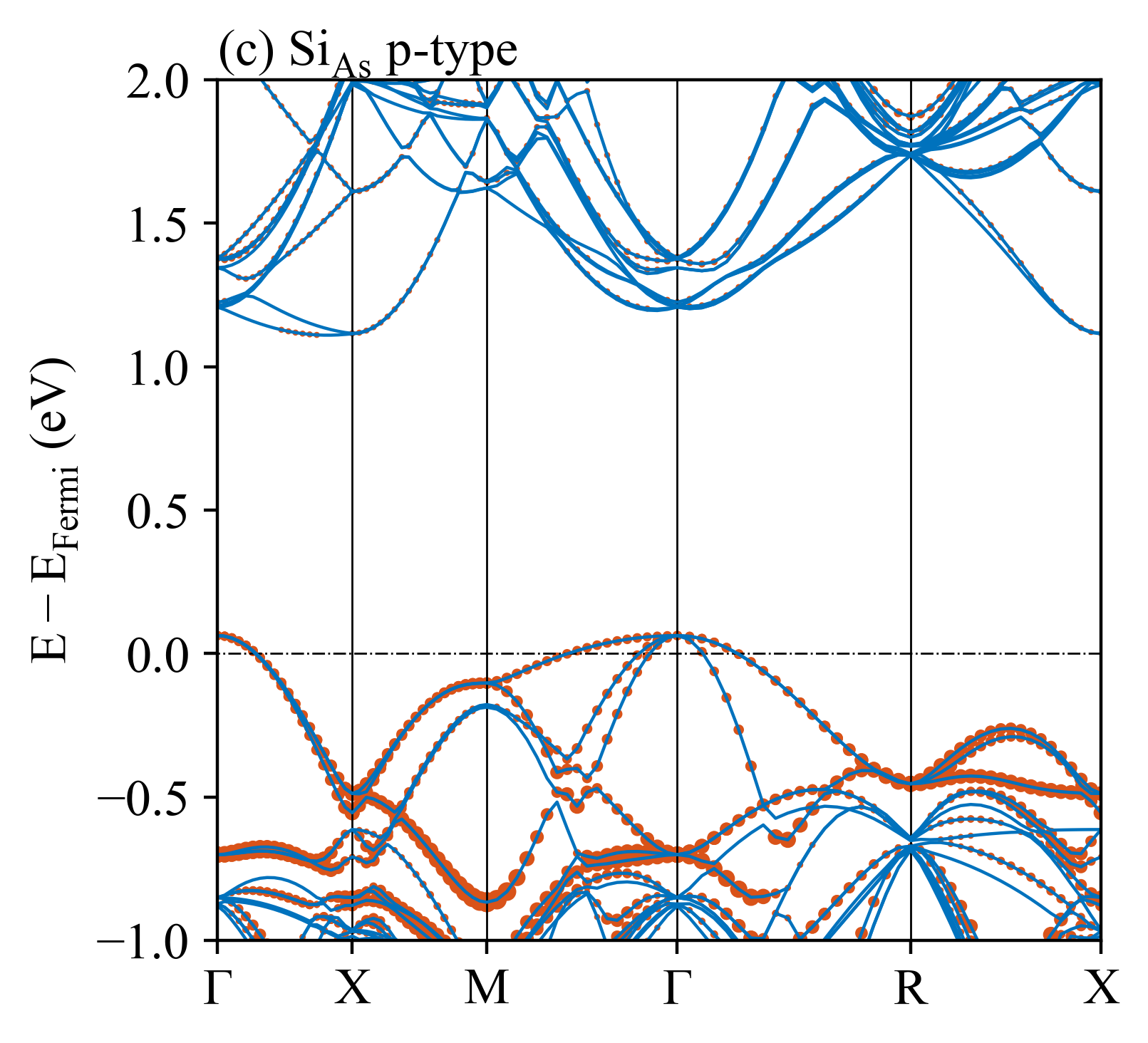}
\includegraphics[width=0.38\columnwidth]{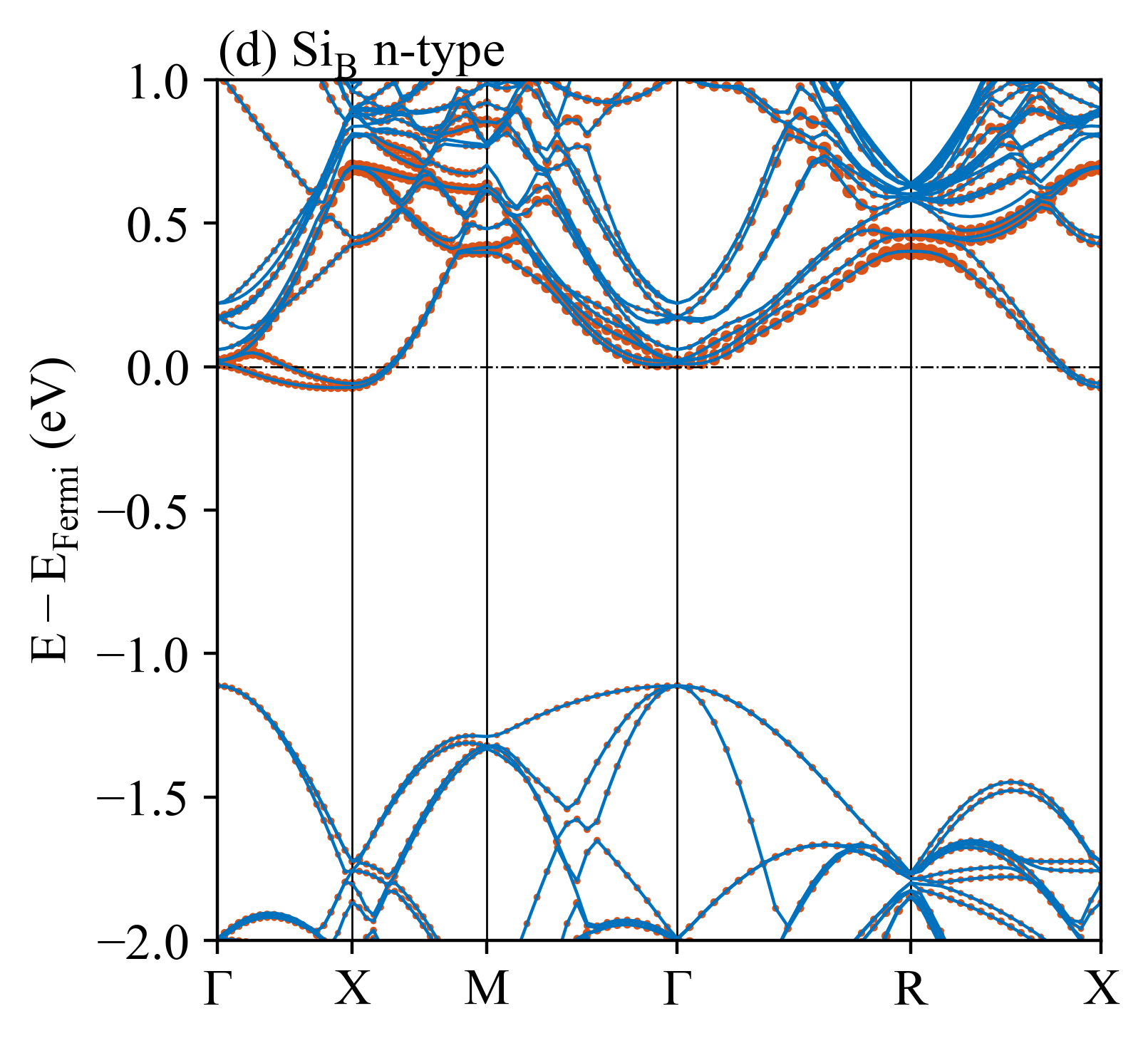}
\includegraphics[width=0.38\columnwidth]{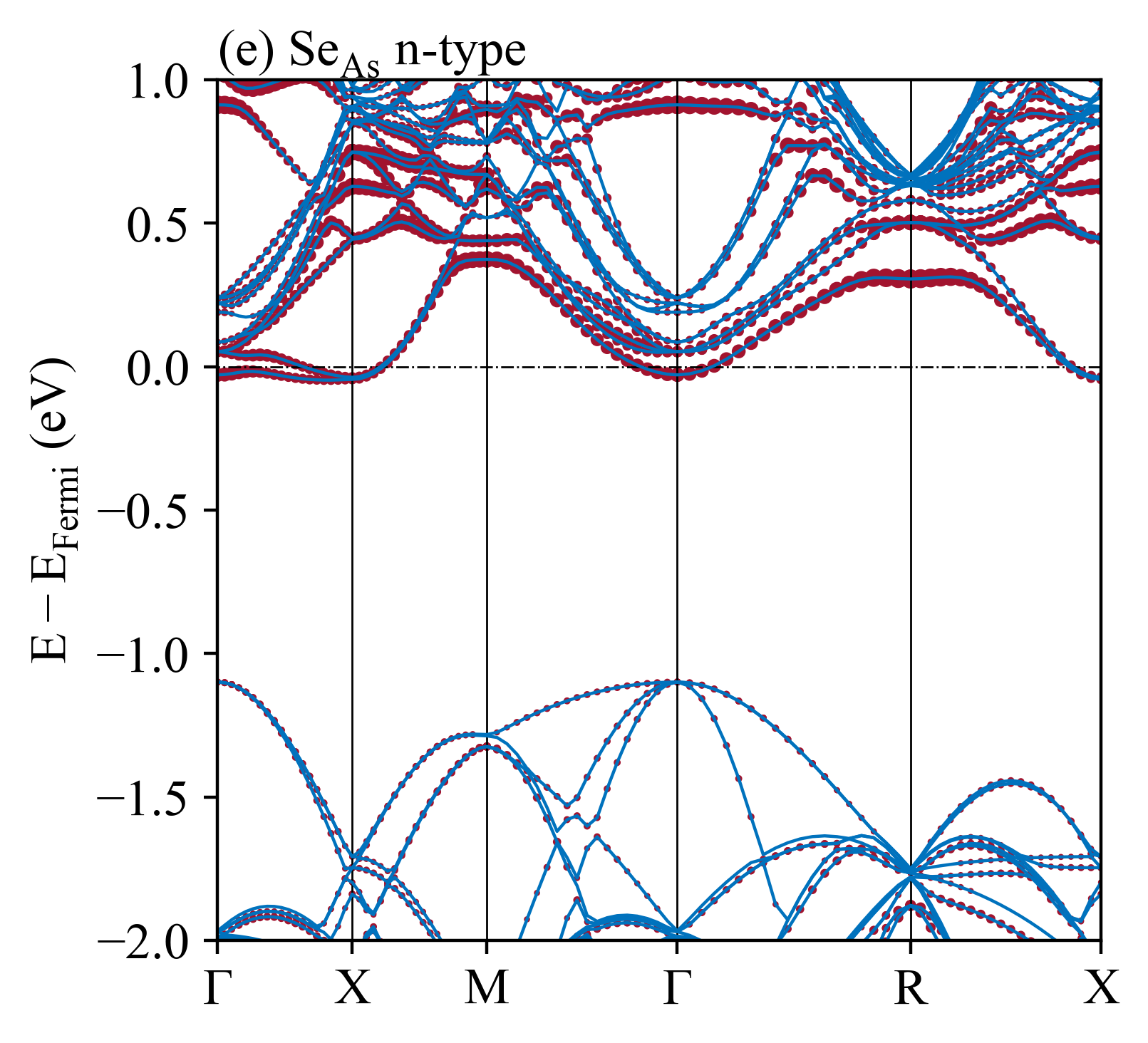}
\caption{\label{fig:sm_band} 
Calculated electronic band structures of (a) pristine BAs and BAs doped with (b) Be$_\text{B}$, (c) Si$_\text{As}$,  (d) Si$_\text{B}$, and (e) Se$_\text{As}$. The colored dots represent the contributions from the dopant elements in the corresponding Bloch wave functions, where the dot size is proportional to the value of the contribution. }
\end{figure} 

\clearpage

\section{\label{sec:smconfiguration} Configurations of coupled complex of dopants and point defects }
\vspace{5mm}
\justifying

For each coupled complex of dopants and point defects, we compute multiple configurations, where the dopants and point defects occupy different sites and, thus have different distances. We first tabulate the fractional coordinates of each site (see Table~\ref{table:sm_sites}), then provide the total energy of each configuration (see Table~\ref{table:sm_complex1}-\ref{table:sm_complex6}). For each coupled complex, the lowest total energy is highlighted in bold text, to indicate the specific configuration used in the DOS and formation energy calculation. 

In this work, we discuss four dopants: Be$_\text{B}$, Si$_\text{As}$, Se$_\text{As}$, and Si$_\text{B}$. According to the defect site, they are divided into two categories: D$_\text{B}$  and D$_\text{As}$. 
And we discuss 9 types of point defects: C$_\text{As}$, O$_\text{B}$O$_\text{As}$, Si$_\text{As}$, C$_\text{As}$Si$_\text{B}$, O$_\text{B}$Si$_\text{As}$, As$_\text{B}$B$_\text{As}$, As$_\text{B}$, B$_\text{As}$, and N$_\text{As}$. Similarly, they are divided into three categories according to the defect site, single defect Y$_\text{B}$ and Y$_\text{As}$, and defect pair Y$_\text{B}$Z$_\text{As}$.
Therefore, if each dopant is coupled with each defect, we have $2\times3=6$ types of couplings in total, as listed below: 
\begin{itemize}
    \item D$_\text{B}$-Y$_\text{As}$
    \item D$_\text{As}$-Y$_\text{As}$
    \item D$_\text{B}$-Y$_\text{B}$
    \item D$_\text{As}$-Y$_\text{B}$
    \item D$_\text{B}$-Y$_\text{B}$Z$_\text{As}$
    \item D$_\text{As}$-Y$_\text{B}$Z$_\text{As}$
\end{itemize}



\begin{longtable}{lc|lc|lc|lc}
\caption{\label{table:sm_sites}The fractional coordinates of each B and As sites in the $3\times3\times3$ supercell of the conventional cubic cell of BAs. } 
\\
\hline\hline\\[-1.0em]

Sites & Fractional coordinates & Sites & Fractional coordinates & Sites & Fractional coordinates & Sites & Fractional coordinates \\
\hline\\[-1.0em]
\endhead

\hline\hline
\endfoot

As1 & (0, 0, 0) & 
As55 & (0.5, 0, 0.1667) & 
B1 & (0.0833, 0.0833, 0.0833) & 
B55 & (0.5833, 0.0833, 0.25) \\ 
As2 & (0, 0, 0.3333) & 
As56 & (0.5, 0, 0.5) & 
B2 & (0.0833, 0.0833, 0.4167) & 
B56 & (0.5833, 0.0833, 0.5833) \\ 
As3 & (0, 0, 0.6667) & 
As57 & (0.5, 0, 0.8333) & 
B3 & (0.0833, 0.0833, 0.75) & 
B57 & (0.5833, 0.0833, 0.9167) \\ 
As4 & (0, 0.1667, 0.1667) & 
As58 & (0.5, 0.1667, 0) & 
B4 & (0.0833, 0.25, 0.25) & 
B58 & (0.5833, 0.25, 0.0833) \\ 
As5 & (0, 0.1667, 0.5) & 
As59 & (0.5, 0.1667, 0.3333) & 
B5 & (0.0833, 0.25, 0.5833) & 
B59 & (0.5833, 0.25, 0.4167) \\ 
As6 & (0, 0.1667, 0.8333) & 
As60 & (0.5, 0.1667, 0.6667) & 
B6 & (0.0833, 0.25, 0.9167) & 
B60 & (0.5833, 0.25, 0.75) \\ 
As7 & (0, 0.3333, 0) & 
As61 & (0.5, 0.3333, 0.1667) & 
B7 & (0.0833, 0.4167, 0.0833) & 
B61 & (0.5833, 0.4167, 0.25) \\ 
As8 & (0, 0.3333, 0.3333) & 
As62 & (0.5, 0.3333, 0.5) & 
B8 & (0.0833, 0.4167, 0.4167) & 
B62 & (0.5833, 0.4167, 0.5833) \\ 
As9 & (0, 0.3333, 0.6667) & 
As63 & (0.5, 0.3333, 0.8333) & 
B9 & (0.0833, 0.4167, 0.75) & 
B63 & (0.5833, 0.4167, 0.9167) \\ 
As10 & (0, 0.5, 0.1667) & 
As64 & (0.5, 0.5, 0) & 
B10 & (0.0833, 0.5833, 0.25) & 
B64 & (0.5833, 0.5833, 0.0833) \\ 
As11 & (0, 0.5, 0.5) & 
As65 & (0.5, 0.5, 0.3333) & 
B11 & (0.0833, 0.5833, 0.5833) & 
B65 & (0.5833, 0.5833, 0.4167) \\ 
As12 & (0, 0.5, 0.8333) & 
As66 & (0.5, 0.5, 0.6667) & 
B12 & (0.0833, 0.5833, 0.9167) & 
B66 & (0.5833, 0.5833, 0.75) \\ 
As13 & (0, 0.6667, 0) & 
As67 & (0.5, 0.6667, 0.1667) & 
B13 & (0.0833, 0.75, 0.0833) & 
B67 & (0.5833, 0.75, 0.25) \\ 
As14 & (0, 0.6667, 0.3333) & 
As68 & (0.5, 0.6667, 0.5) & 
B14 & (0.0833, 0.75, 0.4167) & 
B68 & (0.5833, 0.75, 0.5833) \\ 
As15 & (0, 0.6667, 0.6667) & 
As69 & (0.5, 0.6667, 0.8333) & 
B15 & (0.0833, 0.75, 0.75) & 
B69 & (0.5833, 0.75, 0.9167) \\ 
As16 & (0, 0.8333, 0.1667) & 
As70 & (0.5, 0.8333, 0) & 
B16 & (0.0833, 0.9167, 0.25) & 
B70 & (0.5833, 0.9167, 0.0833) \\ 
As17 & (0, 0.8333, 0.5) & 
As71 & (0.5, 0.8333, 0.3333) & 
B17 & (0.0833, 0.9167, 0.5833) & 
B71 & (0.5833, 0.9167, 0.4167) \\ 
As18 & (0, 0.8333, 0.8333) & 
As72 & (0.5, 0.8333, 0.6667) & 
B18 & (0.0833, 0.9167, 0.9167) & 
B72 & (0.5833, 0.9167, 0.75) \\ 
As19 & (0.1667, 0, 0.1667) & 
As73 & (0.6667, 0, 0) & 
B19 & (0.25, 0.0833, 0.25) & 
B73 & (0.75, 0.0833, 0.0833) \\ 
As20 & (0.1667, 0, 0.5) & 
As74 & (0.6667, 0, 0.3333) & 
B20 & (0.25, 0.0833, 0.5833) & 
B74 & (0.75, 0.0833, 0.4167) \\ 
As21 & (0.1667, 0, 0.8333) & 
As75 & (0.6667, 0, 0.6667) & 
B21 & (0.25, 0.0833, 0.9167) & 
B75 & (0.75, 0.0833, 0.75) \\ 
As22 & (0.1667, 0.1667, 0) & 
As76 & (0.6667, 0.1667, 0.1667) & 
B22 & (0.25, 0.25, 0.0833) & 
B76 & (0.75, 0.25, 0.25) \\ 
As23 & (0.1667, 0.1667, 0.3333) & 
As77 & (0.6667, 0.1667, 0.5) & 
B23 & (0.25, 0.25, 0.4167) & 
B77 & (0.75, 0.25, 0.5833) \\ 
As24 & (0.1667, 0.1667, 0.6667) & 
As78 & (0.6667, 0.1667, 0.8333) & 
B24 & (0.25, 0.25, 0.75) & 
B78 & (0.75, 0.25, 0.9167) \\ 
As25 & (0.1667, 0.3333, 0.1667) & 
As79 & (0.6667, 0.3333, 0) & 
B25 & (0.25, 0.4167, 0.25) & 
B79 & (0.75, 0.4167, 0.0833) \\ 
As26 & (0.1667, 0.3333, 0.5) & 
As80 & (0.6667, 0.3333, 0.3333) & 
B26 & (0.25, 0.4167, 0.5833) & 
B80 & (0.75, 0.4167, 0.4167) \\ 
As27 & (0.1667, 0.3333, 0.8333) & 
As81 & (0.6667, 0.3333, 0.6667) & 
B27 & (0.25, 0.4167, 0.9167) & 
B81 & (0.75, 0.4167, 0.75) \\ 
As28 & (0.1667, 0.5, 0) & 
As82 & (0.6667, 0.5, 0.1667) & 
B28 & (0.25, 0.5833, 0.0833) & 
B82 & (0.75, 0.5833, 0.25) \\ 
As29 & (0.1667, 0.5, 0.3333) & 
As83 & (0.6667, 0.5, 0.5) & 
B29 & (0.25, 0.5833, 0.4167) & 
B83 & (0.75, 0.5833, 0.5833) \\ 
As30 & (0.1667, 0.5, 0.6667) & 
As84 & (0.6667, 0.5, 0.8333) & 
B30 & (0.25, 0.5833, 0.75) & 
B84 & (0.75, 0.5833, 0.9167) \\ 
As31 & (0.1667, 0.6667, 0.1667) & 
As85 & (0.6667, 0.6667, 0) & 
B31 & (0.25, 0.75, 0.25) & 
B85 & (0.75, 0.75, 0.0833) \\ 
As32 & (0.1667, 0.6667, 0.5) & 
As86 & (0.6667, 0.6667, 0.3333) & 
B32 & (0.25, 0.75, 0.5833) & 
B86 & (0.75, 0.75, 0.4167) \\ 
As33 & (0.1667, 0.6667, 0.8333) & 
As87 & (0.6667, 0.6667, 0.6667) & 
B33 & (0.25, 0.75, 0.9167) & 
B87 & (0.75, 0.75, 0.75) \\ 
As34 & (0.1667, 0.8333, 0) & 
As88 & (0.6667, 0.8333, 0.1667) & 
B34 & (0.25, 0.9167, 0.0833) & 
B88 & (0.75, 0.9167, 0.25) \\ 
As35 & (0.1667, 0.8333, 0.3333) & 
As89 & (0.6667, 0.8333, 0.5) & 
B35 & (0.25, 0.9167, 0.4167) & 
B89 & (0.75, 0.9167, 0.5833) \\ 
As36 & (0.1667, 0.8333, 0.6667) & 
As90 & (0.6667, 0.8333, 0.8333) & 
B36 & (0.25, 0.9167, 0.75) & 
B90 & (0.75, 0.9167, 0.9167) \\ 
As37 & (0.3333, 0, 0) & 
As91 & (0.8333, 0, 0.1667) & 
B37 & (0.4167, 0.0833, 0.0833) & 
B91 & (0.9167, 0.0833, 0.25) \\ 
As38 & (0.3333, 0, 0.3333) & 
As92 & (0.8333, 0, 0.5) & 
B38 & (0.4167, 0.0833, 0.4167) & 
B92 & (0.9167, 0.0833, 0.5833) \\ 
As39 & (0.3333, 0, 0.6667) & 
As93 & (0.8333, 0, 0.8333) & 
B39 & (0.4167, 0.0833, 0.75) & 
B93 & (0.9167, 0.0833, 0.9167) \\ 
As40 & (0.3333, 0.1667, 0.1667) & 
As94 & (0.8333, 0.1667, 0) & 
B40 & (0.4167, 0.25, 0.25) & 
B94 & (0.9167, 0.25, 0.0833) \\ 
As41 & (0.3333, 0.1667, 0.5) & 
As95 & (0.8333, 0.1667, 0.3333) & 
B41 & (0.4167, 0.25, 0.5833) & 
B95 & (0.9167, 0.25, 0.4167) \\ 
As42 & (0.3333, 0.1667, 0.8333) & 
As96 & (0.8333, 0.1667, 0.6667) & 
B42 & (0.4167, 0.25, 0.9167) & 
B96 & (0.9167, 0.25, 0.75) \\ 
As43 & (0.3333, 0.3333, 0) & 
As97 & (0.8333, 0.3333, 0.1667) & 
B43 & (0.4167, 0.4167, 0.0833) & 
B97 & (0.9167, 0.4167, 0.25) \\ 
As44 & (0.3333, 0.3333, 0.3333) & 
As98 & (0.8333, 0.3333, 0.5) & 
B44 & (0.4167, 0.4167, 0.4167) & 
B98 & (0.9167, 0.4167, 0.5833) \\ 
As45 & (0.3333, 0.3333, 0.6667) & 
As99 & (0.8333, 0.3333, 0.8333) & 
B45 & (0.4167, 0.4167, 0.75) & 
B99 & (0.9167, 0.4167, 0.9167) \\ 
As46 & (0.3333, 0.5, 0.1667) & 
As100 & (0.8333, 0.5, 0) & 
B46 & (0.4167, 0.5833, 0.25) & 
B100 & (0.9167, 0.5833, 0.0833) \\ 
As47 & (0.3333, 0.5, 0.5) & 
As101 & (0.8333, 0.5, 0.3333) & 
B47 & (0.4167, 0.5833, 0.5833) & 
B101 & (0.9167, 0.5833, 0.4167) \\ 
As48 & (0.3333, 0.5, 0.8333) & 
As102 & (0.8333, 0.5, 0.6667) & 
B48 & (0.4167, 0.5833, 0.9167) & 
B102 & (0.9167, 0.5833, 0.75) \\ 
As49 & (0.3333, 0.6667, 0) & 
As103 & (0.8333, 0.6667, 0.1667) & 
B49 & (0.4167, 0.75, 0.0833) & 
B103 & (0.9167, 0.75, 0.25) \\ 
As50 & (0.3333, 0.6667, 0.3333) & 
As104 & (0.8333, 0.6667, 0.5) & 
B50 & (0.4167, 0.75, 0.4167) & 
B104 & (0.9167, 0.75, 0.5833) \\ 
As51 & (0.3333, 0.6667, 0.6667) & 
As105 & (0.8333, 0.6667, 0.8333) & 
B51 & (0.4167, 0.75, 0.75) & 
B105 & (0.9167, 0.75, 0.9167) \\ 
As52 & (0.3333, 0.8333, 0.1667) & 
As106 & (0.8333, 0.8333, 0) & 
B52 & (0.4167, 0.9167, 0.25) & 
B106 & (0.9167, 0.9167, 0.0833) \\ 
As53 & (0.3333, 0.8333, 0.5) & 
As107 & (0.8333, 0.8333, 0.3333) & 
B53 & (0.4167, 0.9167, 0.5833) & 
B107 & (0.9167, 0.9167, 0.4167) \\ 
As54 & (0.3333, 0.8333, 0.8333) & 
As108 & (0.8333, 0.8333, 0.6667) & 
B54 & (0.4167, 0.9167, 0.9167) & 
B108 & (0.9167, 0.9167, 0.75) \\

\end{longtable}


\begin{table}[h]

\caption{\label{table:sm_complex1} For the coupled complex of dopant and defect D$_\text{B}$-Y$_\text{As}$, the total energy $E$ and the distance $d$ between D$_\text{B}$ and Y$_\text{As}$ as functions of the site of D$_\text{B}$. The site of Y$_\text{As}$ is at As1.} 
\begin{tabular}{@{}l|l|cccc|cccc}
\hline\hline
\begin{tabular}{@{}c@{}}Site of \\ D$_\text{B}$\end{tabular} & 
\begin{tabular}{@{}c@{}}$d$ \\ (\AA)\end{tabular}  &
\begin{tabular}{@{}c@{}}$E($Be$_\text{B}$-C$_\text{As})$ \\ (eV)\end{tabular}  & 
\begin{tabular}{@{}c@{}}$E($Be$_\text{B}$-Si$_\text{As})$ \\ (eV)\end{tabular}  & 
\begin{tabular}{@{}c@{}}$E($Be$_\text{B}$-B$_\text{As})$ \\ (eV)\end{tabular}  & 
\begin{tabular}{@{}c@{}}$E($Be$_\text{B}$-N$_\text{As})$ \\ (eV)\end{tabular}  & 
\begin{tabular}{@{}c@{}}$E($Si$_\text{B}$-C$_\text{As})$ \\ (eV)\end{tabular}  & 
\begin{tabular}{@{}c@{}}$E($Si$_\text{B}$-Si$_\text{As})$ \\ (eV)\end{tabular}  & 
\begin{tabular}{@{}c@{}}$E($Si$_\text{B}$-B$_\text{As})$ \\ (eV)\end{tabular}  & 
\begin{tabular}{@{}c@{}}$E($Si$_\text{B}$-N$_\text{As})$ \\ (eV)\end{tabular}  \\
\hline

B1  & 2.07 & \textbf{-1317.62} & -1314.11 & \textbf{-1314.81} & \textbf{-1318.44} & \textbf{-1319.89} & \textbf{-1316.28} & \textbf{-1317.07} & \textbf{-1318.34}
\\
B21 & 3.96 & -1317.33 & -1314.60 & -1313.99 & -1317.18 & -1318.77 & -1315.97 & -1315.70 & -1317.40
\\
B22 & 5.21 & -1317.54 & -1314.62 & -1314.22 & -1317.39 & -1318.84 & -1315.92 & -1315.68 & -1317.57
\\
B2  & 6.21 & -1317.47 & \textbf{-1314.69} & -1314.14 & -1317.25 & -1318.70 & -1315.87 & -1315.53 & -1317.44
\\

\hline\hline
\end{tabular}
\end{table}


\begin{table}[h]

\caption{\label{table:sm_complex2} For the coupled complex of dopant and defect D$_\text{As}$-Y$_\text{As}$, the total energy $E$ and the distance $d$ between D$_\text{As}$ and Y$_\text{As}$ as functions of the site of D$_\text{As}$. The site of Y$_\text{As}$ is at As1.} 
\begin{tabular}{@{}l|l|ccc|cccc}
\hline\hline
\begin{tabular}{@{}c@{}}Site of \\ D$_\text{As}$\end{tabular} & 
\begin{tabular}{@{}c@{}}$d$ \\ (\AA)\end{tabular}  &
\begin{tabular}{@{}c@{}}$E($Si$_\text{As}$-C$_\text{As})$ \\ (eV)\end{tabular}  & 
\begin{tabular}{@{}c@{}}$E($Si$_\text{As}$-B$_\text{As})$ \\ (eV)\end{tabular}  & 
\begin{tabular}{@{}c@{}}$E($Si$_\text{As}$-N$_\text{As})$ \\ (eV)\end{tabular}  & 
\begin{tabular}{@{}c@{}}$E($Se$_\text{As}$-C$_\text{As})$ \\ (eV)\end{tabular}  & 
\begin{tabular}{@{}c@{}}$E($Se$_\text{As}$-Si$_\text{As})$ \\ (eV)\end{tabular}  & 
\begin{tabular}{@{}c@{}}$E($Se$_\text{As}$-B$_\text{As})$ \\ (eV)\end{tabular}  & 
\begin{tabular}{@{}c@{}}$E($Se$_\text{As}$-N$_\text{As})$ \\ (eV)\end{tabular}  \\
\hline

As21 & 3.38 & -1320.74 & -1317.49 & -1320.72 & \textbf{-1319.26} & \textbf{-1316.38} & \textbf{-1316.23} & \textbf{-1317.82}
\\
As2  & 4.78 & -1320.90 & -1317.56 & -1320.72 & -1319.07 & -1316.24 & -1315.93 & -1317.78
\\
As23 & 5.85 & -1320.93 & -1317.54 & -1320.72 & -1319.03 & -1316.21 & -1315.87 & -1317.77
\\
As38 & 6.76 & \textbf{-1320.96} & \textbf{-1317.58} & \textbf{-1320.74} & -1319.00 & -1316.20 & -1315.83 & -1317.75
\\

\hline\hline
\end{tabular}
\end{table}


\begin{table}[h]
\caption{\label{table:sm_complex5} For the coupled complex of dopant and defect D$_\text{B}$-As$_\text{B}$, the total energy $E$ and the distance $d$ between D$_\text{B}$ and As$_\text{B}$ as functions of the site of D$_\text{B}$. The site of As$_\text{B}$ is at B1.} 
\begin{tabular}{@{}l|l|c|c}
\hline\hline
\begin{tabular}{@{}c@{}}Site of \\ D$_\text{B}$\end{tabular} & 
\begin{tabular}{@{}c@{}}$d$ \\ (\AA)\end{tabular}  &
\begin{tabular}{@{}c@{}}$E($Be$_\text{B}$-As$_\text{B})$ \\ (eV)\end{tabular}  & 
\begin{tabular}{@{}c@{}}$E($Si$_\text{B}$-As$_\text{B})$ \\ (eV)\end{tabular}   \\
\hline

B22  & 3.38 & \textbf{-1310.16} & -1309.24
\\
B2 & 4.78 & -1309.97 & -1309.36
\\
B23 & 5.85 & -1309.95 & \textbf{-1309.37}
\\
B38  & 6.76 & -1309.88 & -1309.29
\\

\hline\hline
\end{tabular}
\end{table}


\begin{table}[h]
\caption{\label{table:sm_complex4} For the coupled complex of dopant and defect D$_\text{As}$-As$_\text{B}$, the total energy $E$ and the distance $d$ between D$_\text{As}$ and As$_\text{B}$ as functions of the site of D$_\text{As}$. The site of As$_\text{B}$ is at B1.} 
\begin{tabular}{@{}l|l|c|c}
\hline\hline
\begin{tabular}{@{}c@{}}Site of \\ D$_\text{As}$\end{tabular} & 
\begin{tabular}{@{}c@{}}$d$ \\ (\AA)\end{tabular}  &
\begin{tabular}{@{}c@{}}$E($Si$_\text{As}$-As$_\text{B})$ \\ (eV)\end{tabular}  & 
\begin{tabular}{@{}c@{}}$E($Se$_\text{As}$-As$_\text{B})$ \\ (eV)\end{tabular}  \\
\hline

As1 & 2.07 & \textbf{-1314.18} & \textbf{-1310.50}
\\
As2 & 3.96 & -1313.54 & -1309.64
\\
As38 & 5.21 & -1313.50 & -1309.73
\\
As44 & 6.21 & -1313.38 & -1309.69
\\

\hline\hline
\end{tabular}
\end{table}




\begin{sidewaystable}[h]
\caption{\label{table:sm_complex3} For the coupled complex of dopant and defect D$_\text{B}$-Y$_\text{B}$Z$_\text{As}$, the total energy $E$, the distance $d_1$ between D$_\text{B}$ and Y$_\text{B}$, and the distance $d_2$ between D$_\text{B}$ and Z$_\text{As}$, as functions of the site of D$_\text{B}$. The sites of Y$_\text{B}$ and Z$_\text{As}$ are at B1 and As1, respectively.} 
\begin{tabular}{@{}l|l|l|cccc|cccc}
\hline\hline
\begin{tabular}{@{}c@{}}Site of \\ D$_\text{B}$\end{tabular} & 
\begin{tabular}{@{}c@{}}$d_1$ \\ (\AA)\end{tabular}  &
\begin{tabular}{@{}c@{}}$d_2$ \\ (\AA)\end{tabular}  &
\begin{tabular}{@{}c@{}}$E($Be$_\text{B}$-O$_\text{B}$O$_\text{As})$ \\ (eV)\end{tabular}  & 
\begin{tabular}{@{}c@{}}$E($Be$_\text{B}$-Si$_\text{B}$C$_\text{As})$ \\ (eV)\end{tabular}  & 
\begin{tabular}{@{}c@{}}$E($Be$_\text{B}$-O$_\text{B}$Si$_\text{As})$ \\ (eV)\end{tabular}  & 
\begin{tabular}{@{}c@{}}$E($Be$_\text{B}$-As$_\text{B}$B$_\text{As})$ \\ (eV)\end{tabular}  & 
\begin{tabular}{@{}c@{}}$E($Si$_\text{B}$-O$_\text{B}$O$_\text{As})$ \\ (eV)\end{tabular}  & 
\begin{tabular}{@{}c@{}}$E($Si$_\text{B}$-Si$_\text{B}$C$_\text{As})$ \\ (eV)\end{tabular}  & 
\begin{tabular}{@{}c@{}}$E($Si$_\text{B}$-O$_\text{B}$Si$_\text{As})$ \\ (eV)\end{tabular}  & 
\begin{tabular}{@{}c@{}}$E($Si$_\text{B}$-As$_\text{B}$B$_\text{As})$ \\ (eV)\end{tabular}  \\
\hline

B106 & 3.38 & 2.07 & \textbf{-1315.45} & \textbf{-1316.95} & -1312.81 & -1312.29 & \textbf{-1316.14} & \textbf{-1316.68} & -1314.12 & \textbf{-1313.84}
\\
B94 & 3.38 & 3.96 & -1314.11  & -1316.66 & -1313.34 & -1312.81 & -1314.40 & -1316.02 & \textbf{-1314.63} & -1312.81
\\
B3 & 4.78 & 3.96 & -1313.77 & -1316.52 & -1313.21 & -1312.57 & -1313.83 & -1316.12 & -1313.98 & -1312.99
\\
B4 & 3.38 & 5.21 & -1313.87 & -1316.83 & \textbf{-1313.58} & \textbf{-1312.96} & -1314.37 & -1315.99 & -1314.19 & -1312.70
\\
B108 & 5.85 & 3.96 & -1313.74 & -1316.50 & -1313.18 & -1312.54 & -1313.80 & -1316.13 & -1314.18 & -1313.02
\\
B88 & 5.85 & 5.21 & -1314.05 & -1316.70 & -1313.24 & -1312.69 & -1314.19 & -1316.12 & -1314.23 & -1313.03
\\
B2 & 4.78 & 6.21 & -1313.85 & -1316.67 & -1313.24 & -1312.78 & -1313.92 & -1316.04 & -1314.17 & -1312.85
\\

\hline\hline
\end{tabular}

\bigskip\bigskip



\caption{\label{table:sm_complex6} For the coupled complex of dopant and defect D$_\text{As}$-Y$_\text{B}$Z$_\text{As}$, the total energy $E$, the distance $d_1$ between D$_\text{As}$ and Y$_\text{B}$, and the distance $d_2$ between D$_\text{As}$ and Z$_\text{As}$, as functions of the site of D$_\text{As}$. The sites of Y$_\text{B}$ and Z$_\text{As}$ are at B1 and As1, respectively.} 
\begin{tabular}{@{}l|l|l|cccc|cccc}
\hline\hline
\begin{tabular}{@{}c@{}}Site of \\ D$_\text{As}$\end{tabular} & 
\begin{tabular}{@{}c@{}}$d_1$ \\ (\AA)\end{tabular}  &
\begin{tabular}{@{}c@{}}$d_2$ \\ (\AA)\end{tabular}  &
\begin{tabular}{@{}c@{}}$E($Si$_\text{As}$-O$_\text{B}$O$_\text{As})$ \\ (eV)\end{tabular}  & 
\begin{tabular}{@{}c@{}}$E($Si$_\text{As}$-Si$_\text{B}$C$_\text{As})$ \\ (eV)\end{tabular}  & 
\begin{tabular}{@{}c@{}}$E($Si$_\text{As}$-O$_\text{B}$Si$_\text{As})$ \\ (eV)\end{tabular}  & 
\begin{tabular}{@{}c@{}}$E($Si$_\text{As}$-As$_\text{B}$B$_\text{As})$ \\ (eV)\end{tabular}  & 
\begin{tabular}{@{}c@{}}$E($Se$_\text{As}$-O$_\text{B}$O$_\text{As})$ \\ (eV)\end{tabular}  & 
\begin{tabular}{@{}c@{}}$E($Se$_\text{As}$-Si$_\text{B}$C$_\text{As})$ \\ (eV)\end{tabular}  & 
\begin{tabular}{@{}c@{}}$E($Se$_\text{As}$-O$_\text{B}$Si$_\text{As})$ \\ (eV)\end{tabular}  & 
\begin{tabular}{@{}c@{}}$E($Se$_\text{As}$-As$_\text{B}$B$_\text{As})$ \\ (eV)\end{tabular}  \\
\hline

As19 & 2.07 & 3.38 &\textbf{ -1318.36} & \textbf{-1320.33} & \textbf{-1318.00} & \textbf{-1316.81} & \textbf{-1315.76} & -1317.08 & \textbf{-1315.58} &-1313.28
\\
As21 & 3.96 & 3.38 & -1317.40 & -1319.18 & -1316.66 & -1316.00 & -1314.40 & -1317.26 & -1314.67 &-1313.42
\\
As2 & 3.96 & 4.78 & -1317.37 & -1319.35 & -1316.72 & -1316.27 & -1314.25 & -1317.09 & -1314.59 &-1313.16
\\
As93 & 5.21 & 3.38 & -1317.28 & -1319.29 & -1316.65 & -1315.96 & -1314.28 & \textbf{-1317.31} & -1314.34 &\textbf{-1313.50}
\\
As23 & 3.96 & 5.85 & -1317.36 & -1319.39 & -1316.76 & -1316.35 & -1314.24 & -1317.05 & -1314.55 &-1313.10
\\
As3 & 6.21 & 4.78 & -1317.29 & -1319.29 & -1316.70 & -1316.10 & -1314.22 & -1317.17 & -1314.21 &-1313.30
\\
As42 & 5.21 & 5.85 & -1317.42 & -1319.47 & -1316.83 & -1316.25 & -1314.46 & -1317.13 & -1314.59 &-1313.21
\\

\hline\hline
\end{tabular}
\end{sidewaystable}

\clearpage

\clearpage

\section{\label{sec:smother} Influence of other defects on the dopants}
\vspace{5mm}
\justifying

\begin{figure}[h]
\includegraphics[width=0.5\columnwidth]{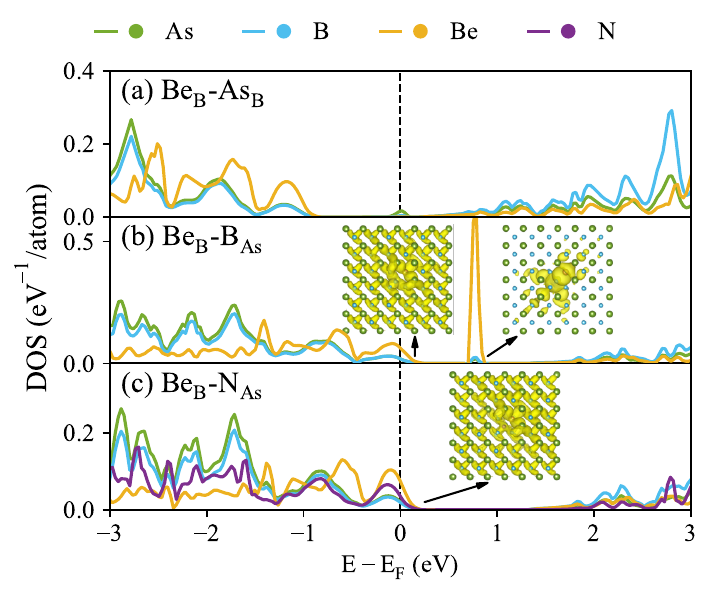}
\caption{\label{fig:sm_dop_dos_beb} 
Calculated electronic DOS of Be$_\text{B}$-doped BAs in the presence of (a) As$_\text{B}$, (b) B$_\text{As}$, and (c) N$_\text{As}$ point defect. The inset figures show the isosurfaces of charge density near VBM ($<$0.1eV) or within the band gap. }
\end{figure}

\begin{figure}[h]
\includegraphics[width=0.5\columnwidth]{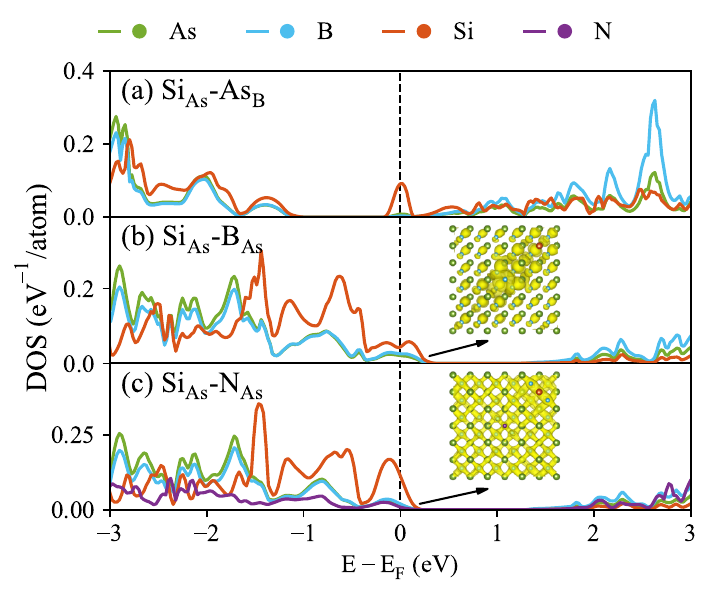}
\caption{\label{fig:sm_dop_dos_sias} 
Calculated electronic DOS of Si$_\text{As}$-doped BAs in the presence of (a) As$_\text{B}$, (b) B$_\text{As}$, and (c) N$_\text{As}$ point defect. The inset figures show the isosurfaces of charge density near VBM ($<$0.1eV) or within the band gap. }
\end{figure}

\begin{figure}[h]
\includegraphics[width=0.5\columnwidth]{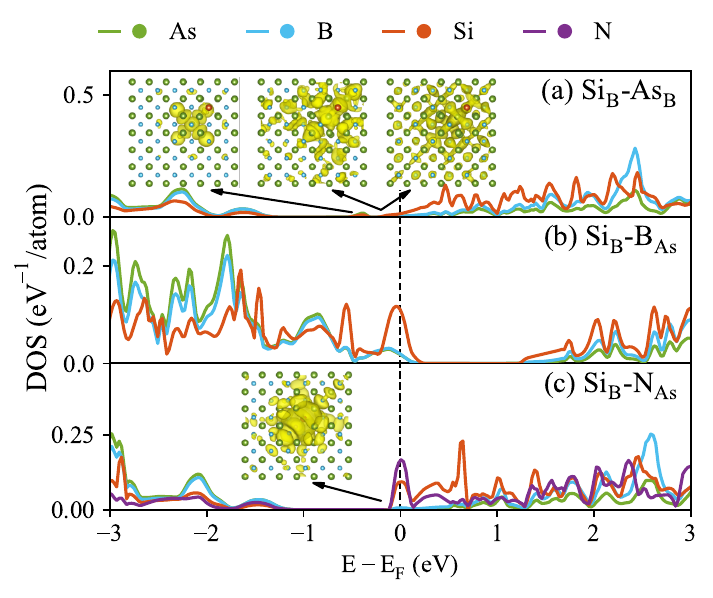}
\caption{\label{fig:sm_dop_dos_sib} 
Calculated electronic DOS of Si$_\text{B}$-doped BAs in the presence of (a) As$_\text{B}$, (b) B$_\text{As}$, and (c) N$_\text{As}$ point defect. The inset figures show the isosurfaces of charge density near CBM ($<$0.1eV) or within the band gap. }
\end{figure}

\begin{figure}[h]
\includegraphics[width=0.5\columnwidth]{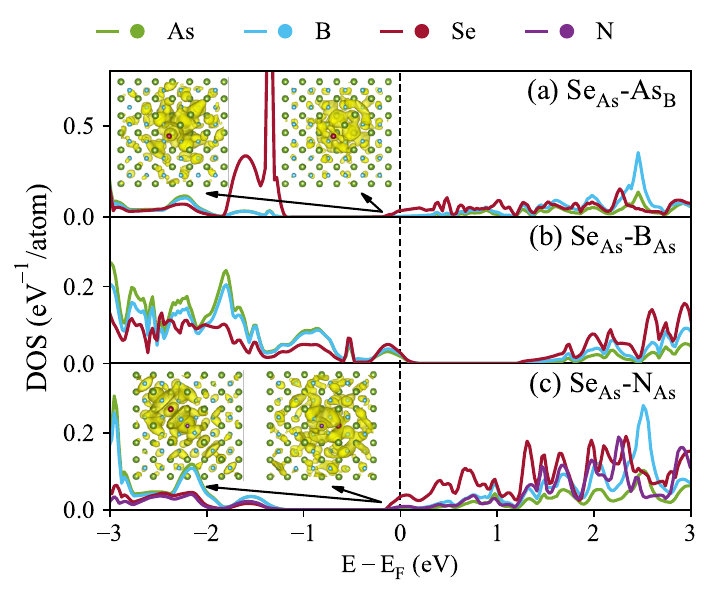}
\caption{\label{fig:sm_dop_dos_seas} 
Calculated electronic DOS of Se$_\text{As}$-doped BAs in the presence of (a) As$_\text{B}$, (b) B$_\text{As}$, and (c) N$_\text{As}$ point defect. The inset figures show the isosurfaces of charge density near CBM ($<$0.1eV) or within the band gap. }
\end{figure}

\begin{figure}[h]
\includegraphics[width=0.3\columnwidth]{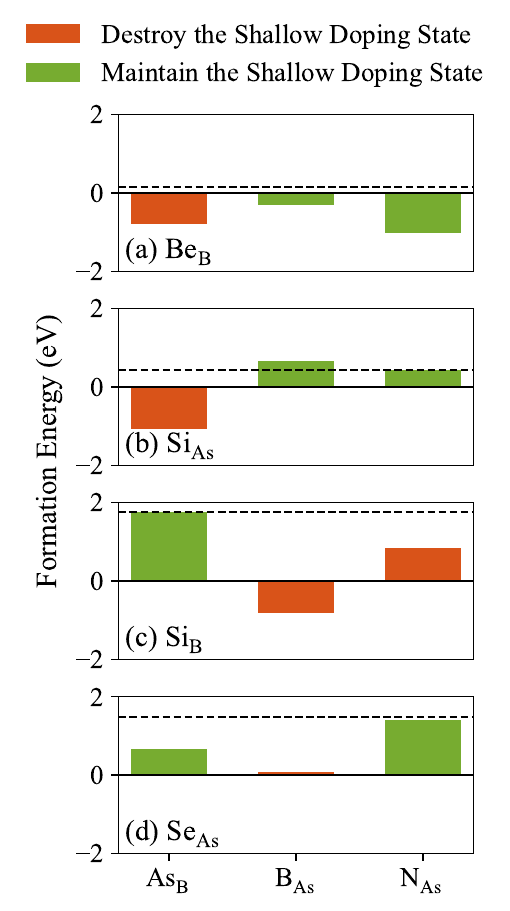}
\caption{\label{fig:sm_doping_defect} 
Calculated formation energy of (a) Be$_\text{B}$, (b) Si$_\text{As}$, (c) Si$_\text{B}$, and (d) Se$_\text{As}$ dopants in the presence of selected point defects. The dashed lines represent the formation energy of dopants without point defects. The green color signifies that the shallow doping state is preserved as the dopants in pristine BAs, whereas the red color indicates otherwise. }
\end{figure}

In this section we present the results for defects As$_\text{B}$, B$_\text{As}$, and N$_\text{As}$. As$_\text{B}$ and B$_\text{As}$ are common antisite defects, while N$_\text{As}$ is a charge-neutral impurity, as N is in the same group of As. In N single point defects and the point defect pairs in the first nearest neighbors, the formation energies per point defect are 2.56, 0.82, 0.90, 2.59, 1.00 eV for N$_\text{B}$, N$_\text{As}$, N$_\text{B}$N$_\text{As}$, N$_\text{B}$N$_\text{B}$, and N$_\text{As}$N$_\text{As}$, respectively. Therefore, N$_\text{As}$ is the most favorable among these defects. The DOS and isosurface results are presented in Fig.~\ref{fig:sm_dop_dos_beb}-\ref{fig:sm_dop_dos_seas}, and the formation energy is in Fig.~\ref{fig:sm_doping_defect}. As$_\text{B}$ is harmful to p-type dopants, and is beneficial to Se$_\text{As}$; B$_\text{As}$ is harmful to n-type dopants, and is beneficial to Be$_\text{B}$. While N$_\text{As}$ does not couple with As-site dopants Si$_\text{As}$ and Se$_\text{As}$, it is beneficial to Be$_\text{B}$, as a p-type B-site dopant, and is harmful to Si$_\text{B}$, as a n-type B-site dopant. 

\clearpage

\section{\label{sec:smother} Isosurfaces of charge density}
\vspace{5mm}
\justifying

In this section, we provide enlarged figures of isosurfaces of charge density, for all bands near CBM or VBM ($<$0.1 eV) and within the band gap. In each case, from left to right, the energies of electronic bands are arranged from small to large values, indicating that the band indexes also increase from small to large values. The isosurface of charge density is depicted for each band at $4\times10^{-5} \AA^{-3}$, unless for strongly localized bands, which is depicted at $1\times10^{-3} \AA^{-3}$ and will be explicitly pointed out.

\subsection{Dopants}
Be$_\text{B}$: A three-fold degenerated band at VBM
\begin{figure}[h]
\includegraphics[width=0.23\columnwidth]{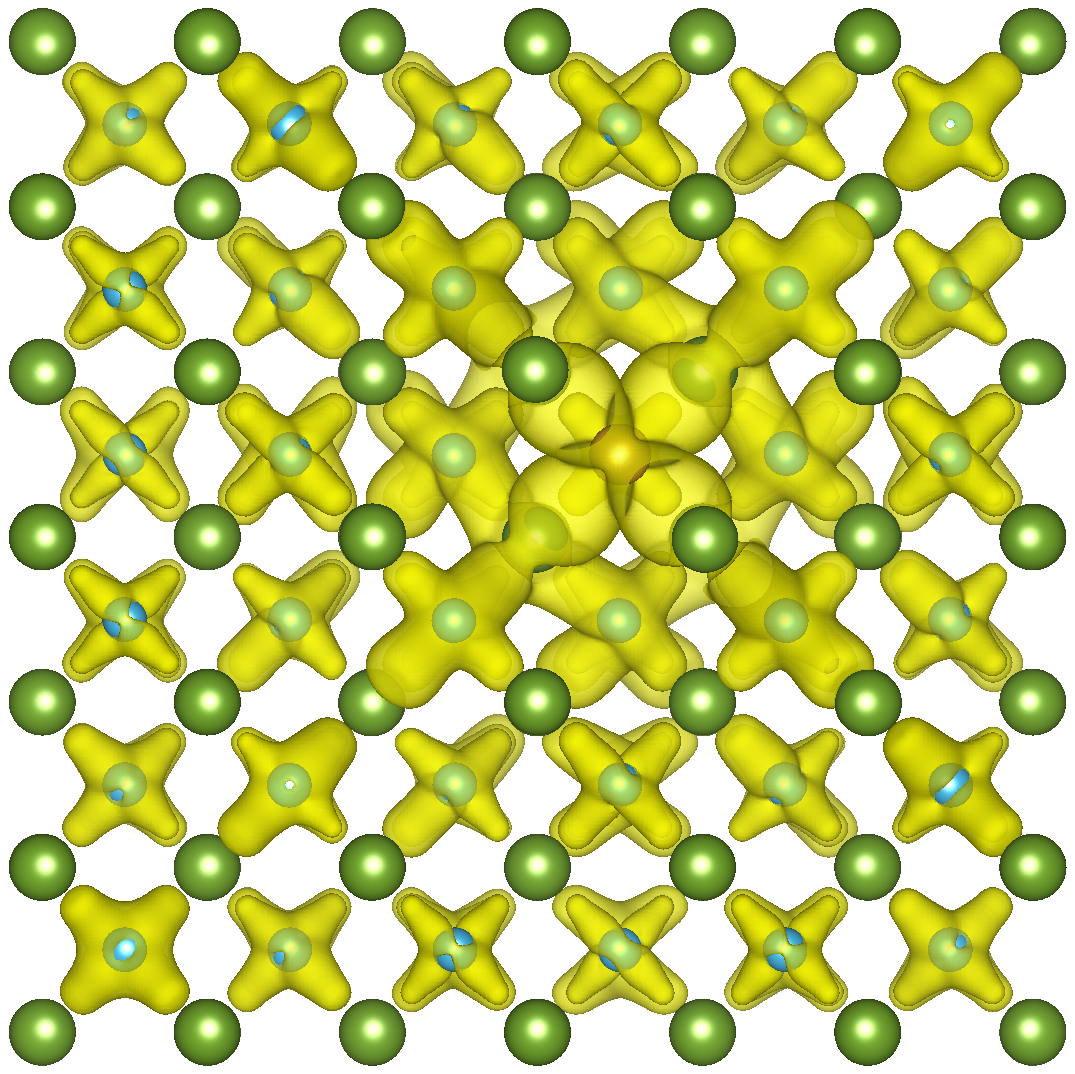}
\end{figure} 

Si$_\text{As}$: A three-fold degenerated band at VBM
\begin{figure}[h]
\includegraphics[width=0.23\columnwidth]{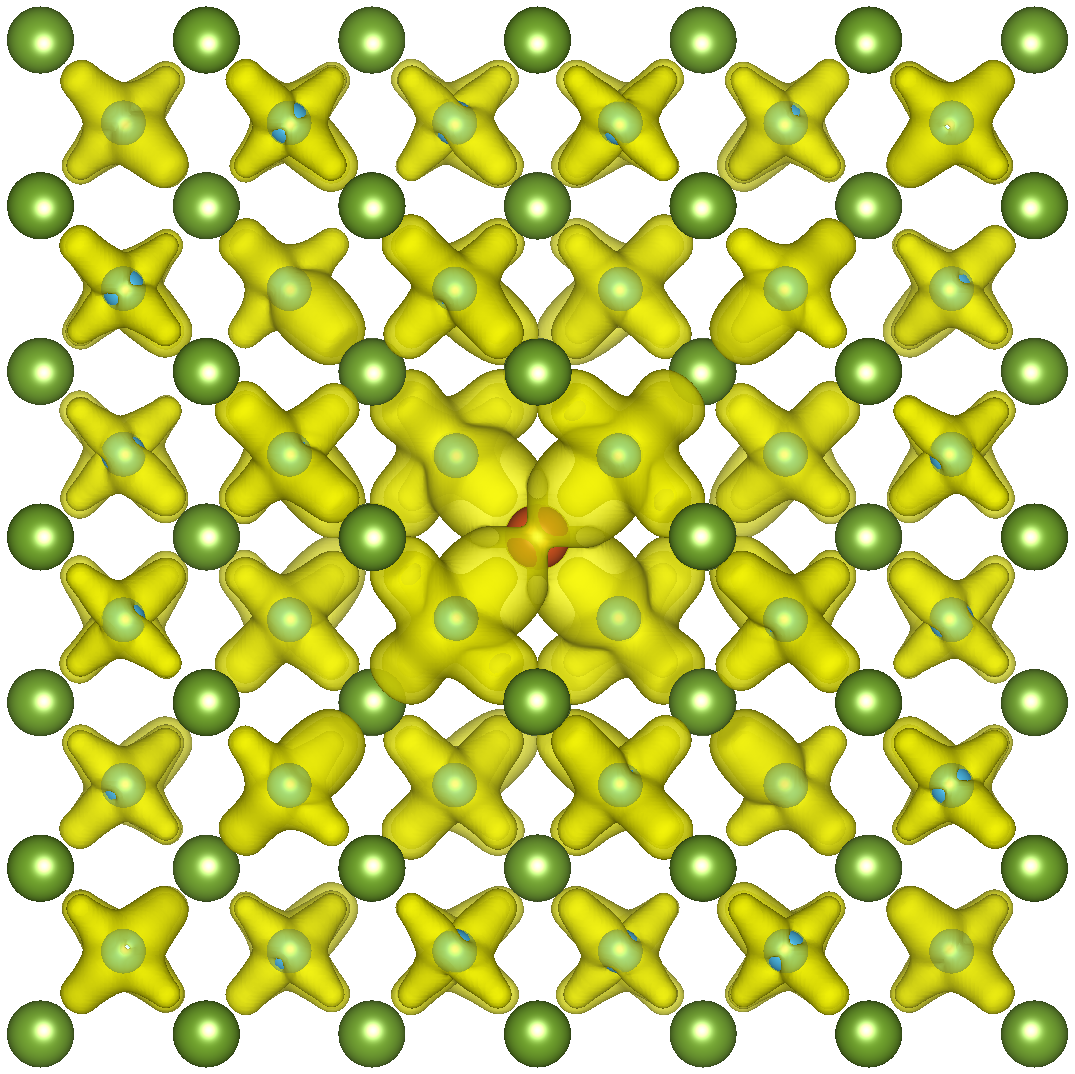}
\end{figure} 

Si$_\text{B}$: Two bands near CBM
\begin{figure}[h]
\includegraphics[width=0.23\columnwidth]{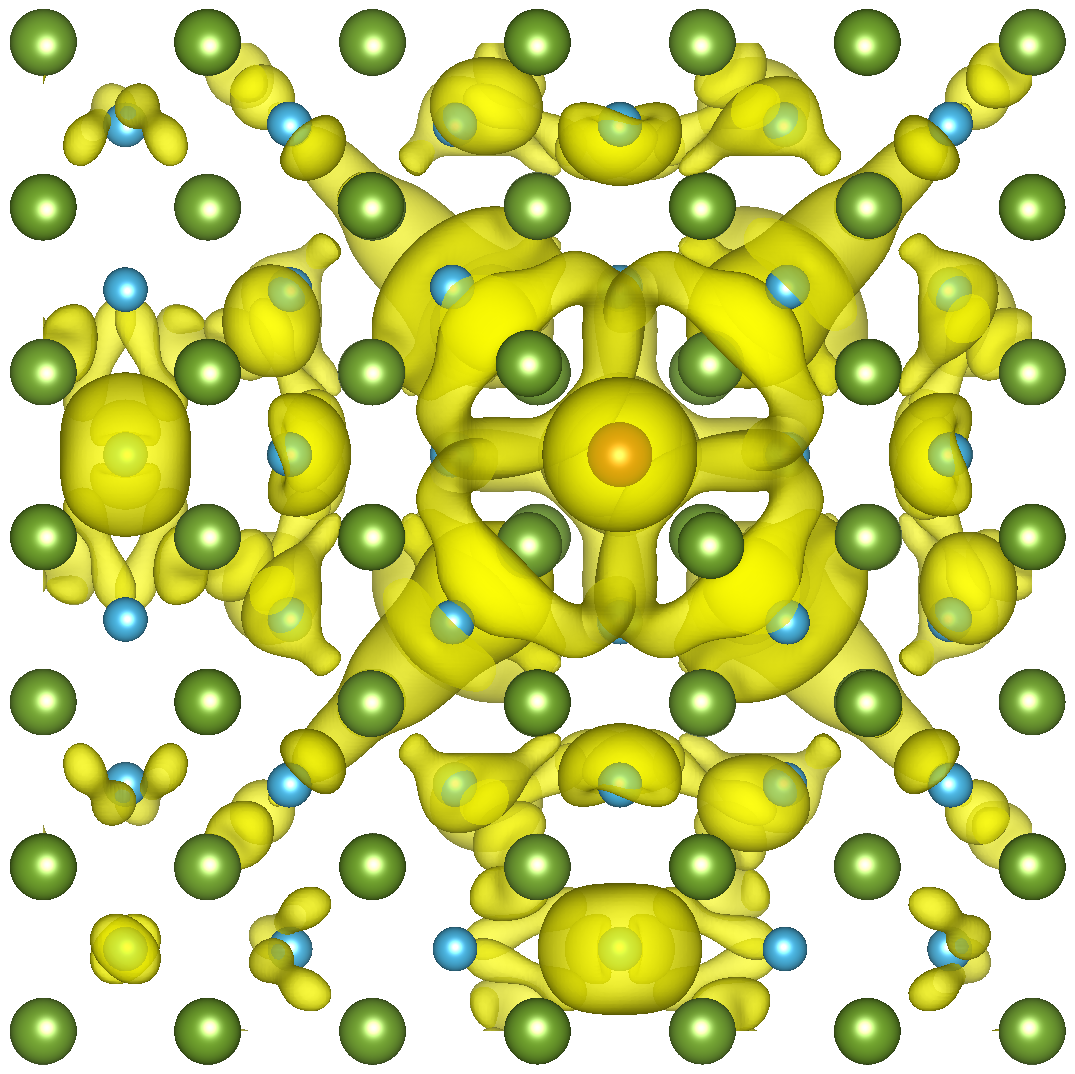}
\hspace{1 cm}
\includegraphics[width=0.23\columnwidth]{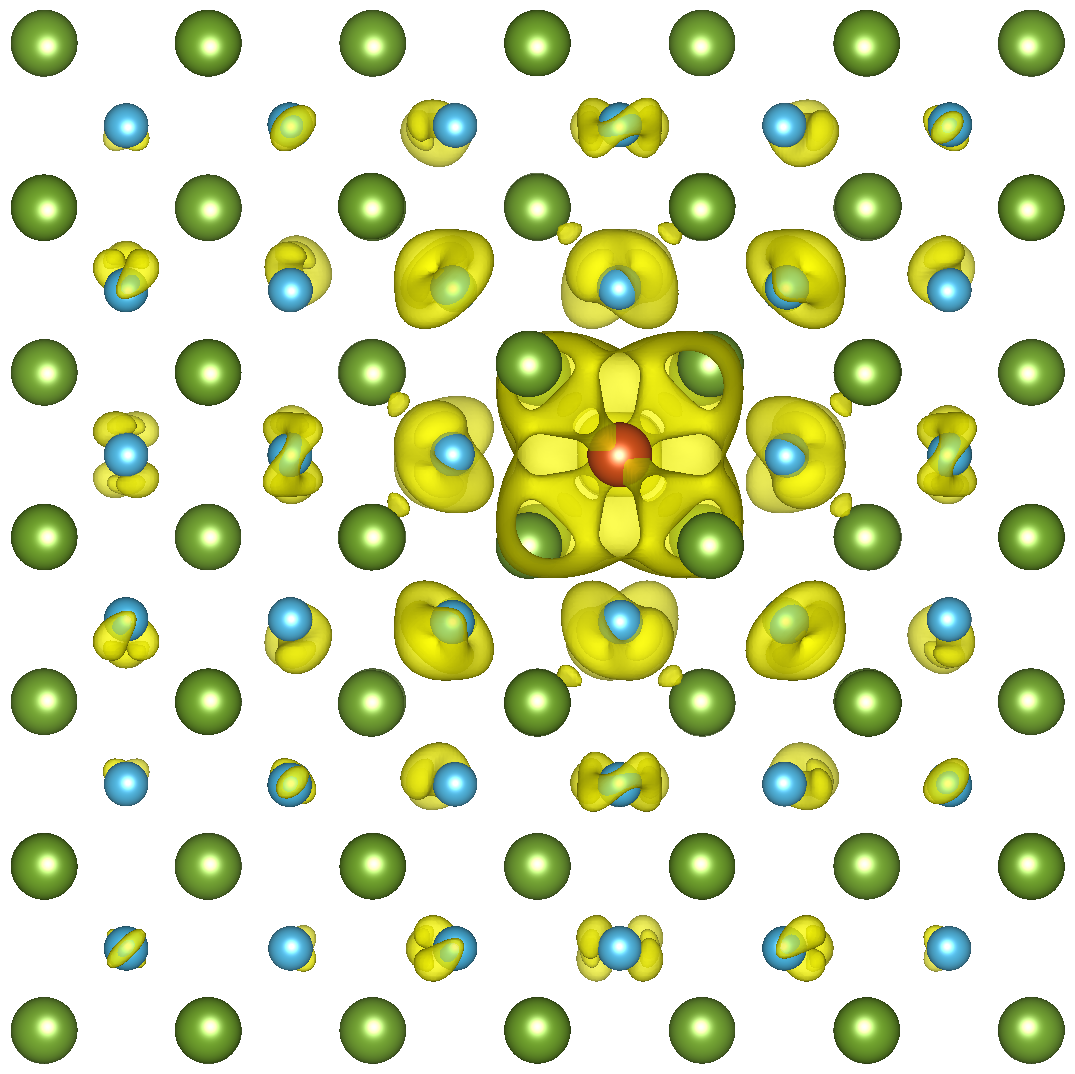}
\end{figure} 

\clearpage
Se$_\text{As}$: Two bands near CBM
\begin{figure}[h]
\includegraphics[width=0.23\columnwidth]{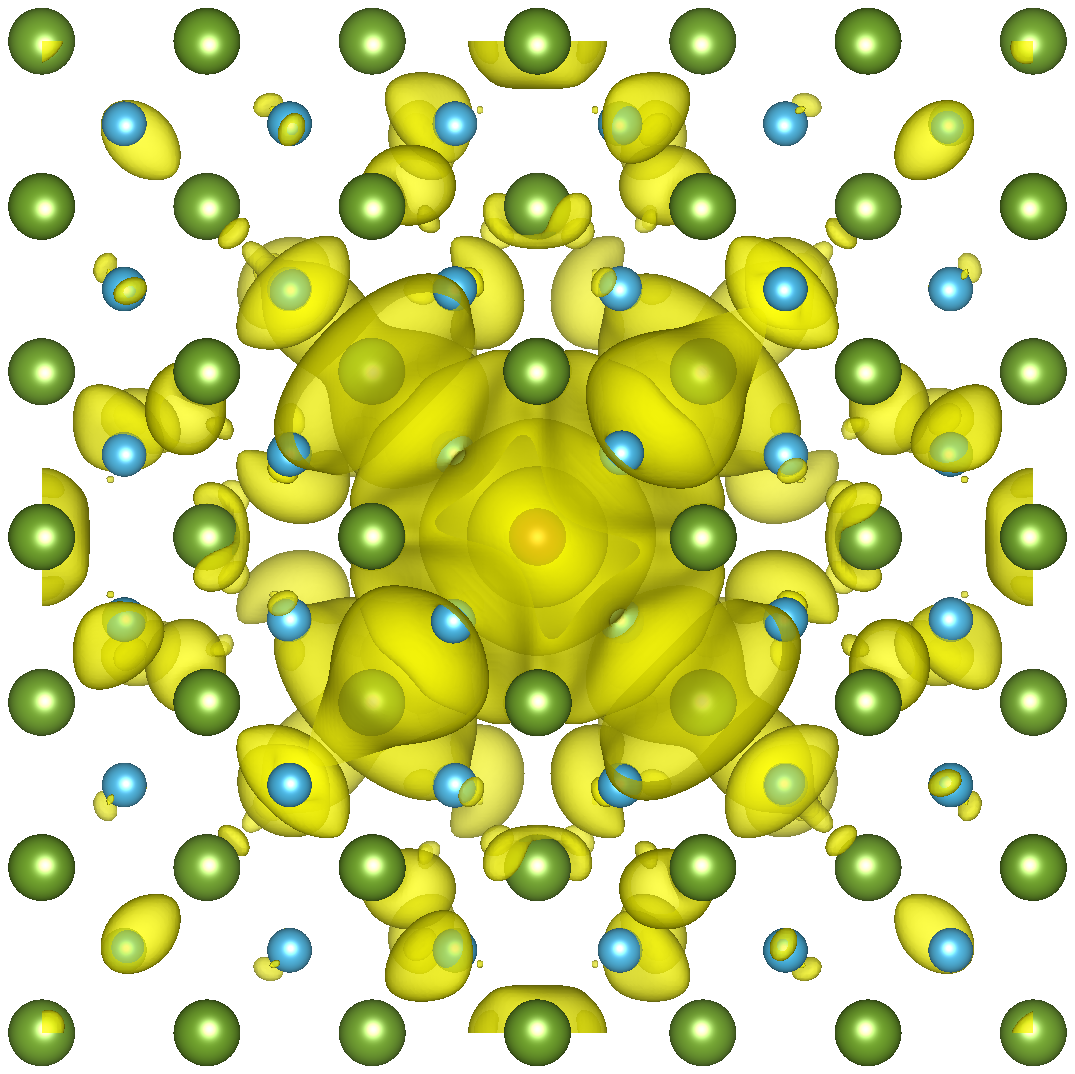}
\hspace{1 cm}
\includegraphics[width=0.23\columnwidth]{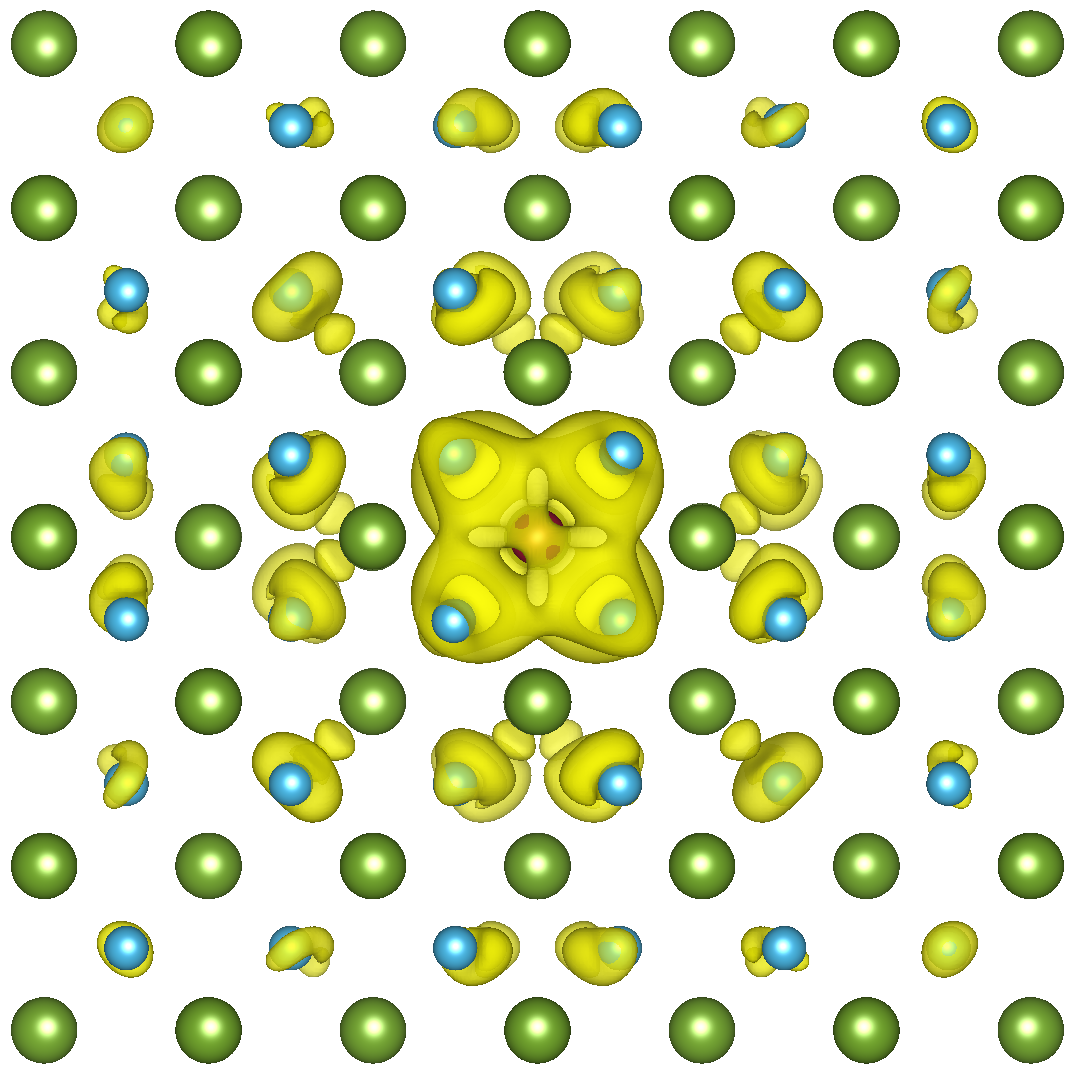}
\end{figure} 

\subsection{Dopants with point defects}
Be$_\text{B}$-As$_\text{B}$B$_\text{As}$: Three bands near VBM
\begin{figure}[h]
\includegraphics[width=0.23\columnwidth]
{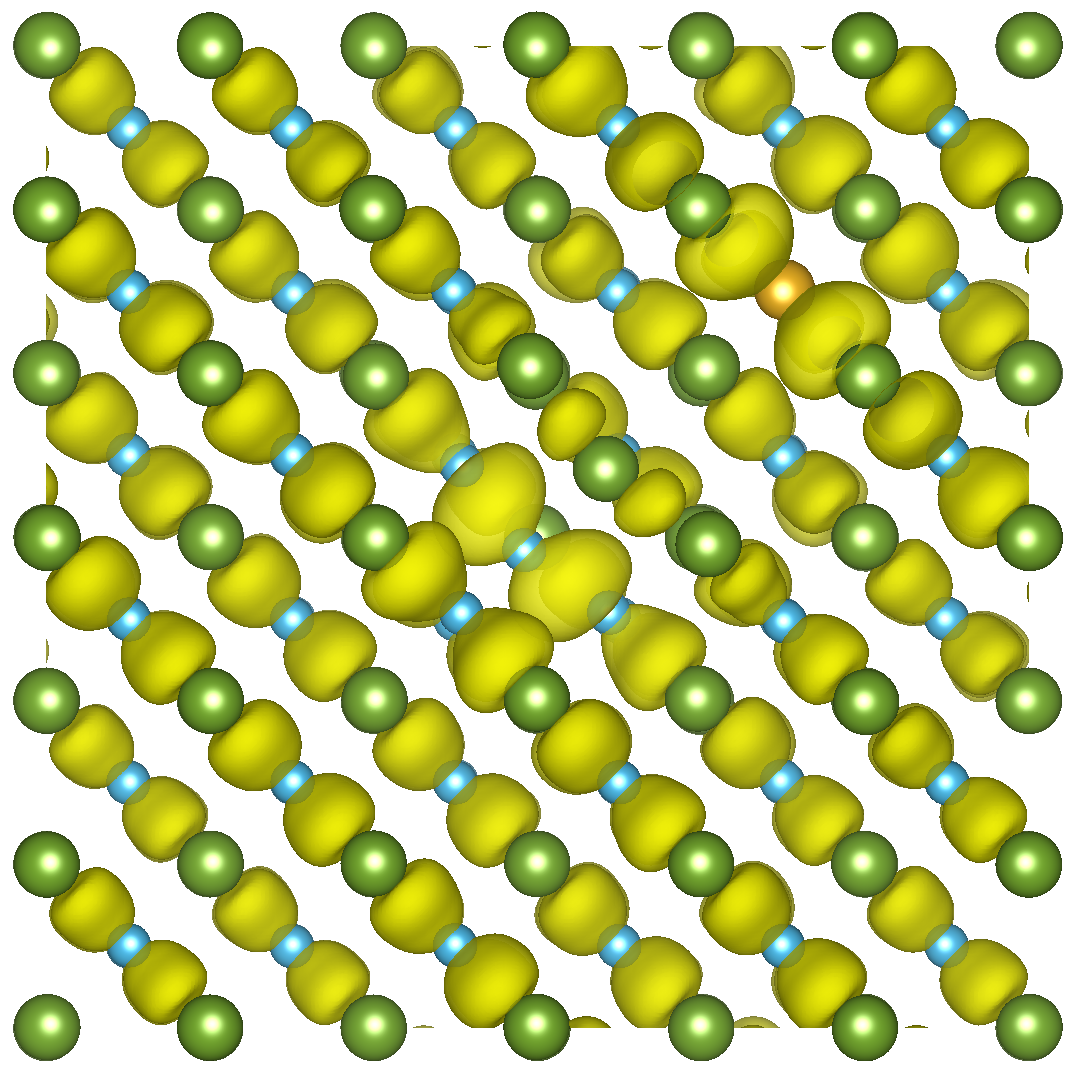}
\hspace{.5 cm}
\includegraphics[width=0.23\columnwidth]
{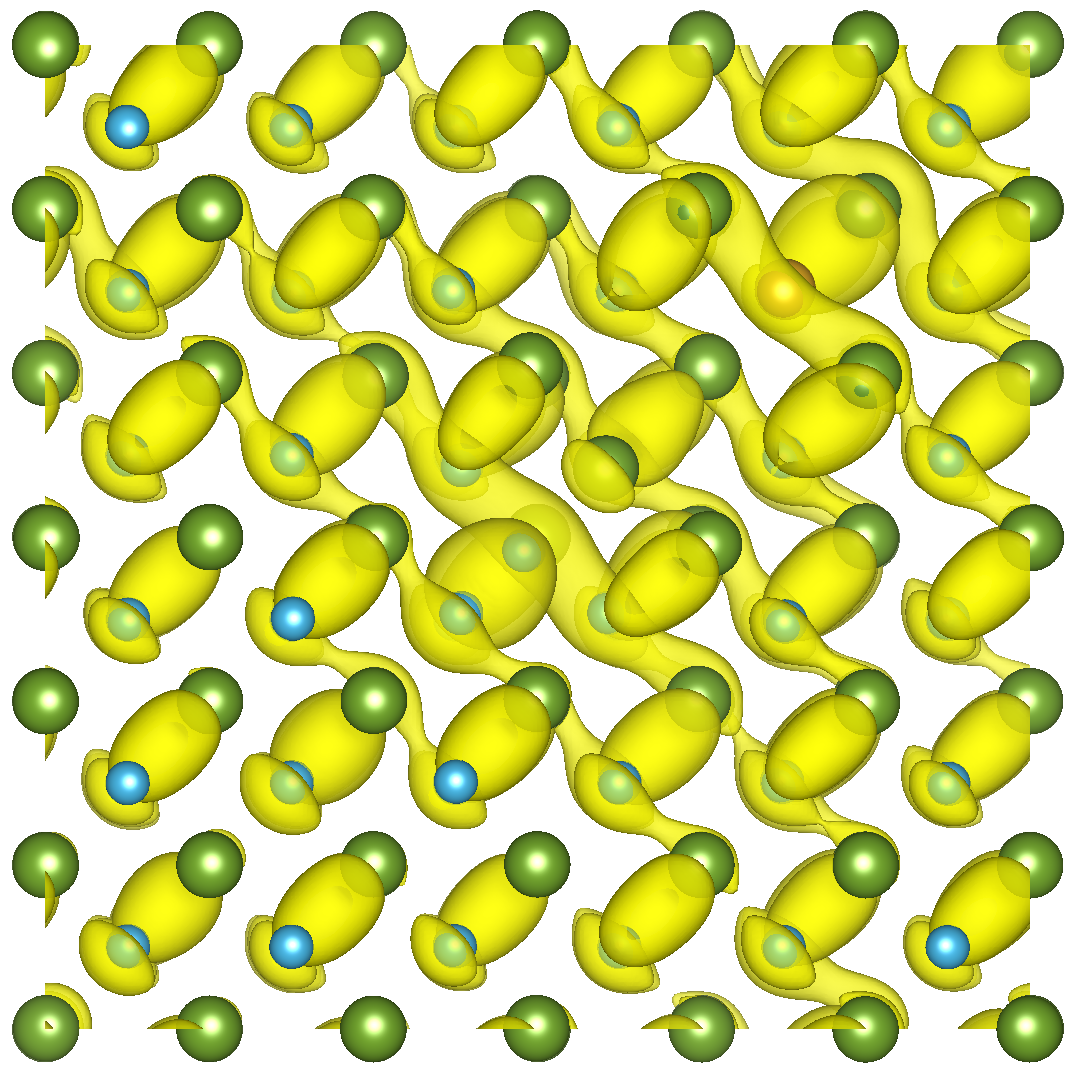}
\hspace{.5 cm}
\includegraphics[width=0.23\columnwidth]
{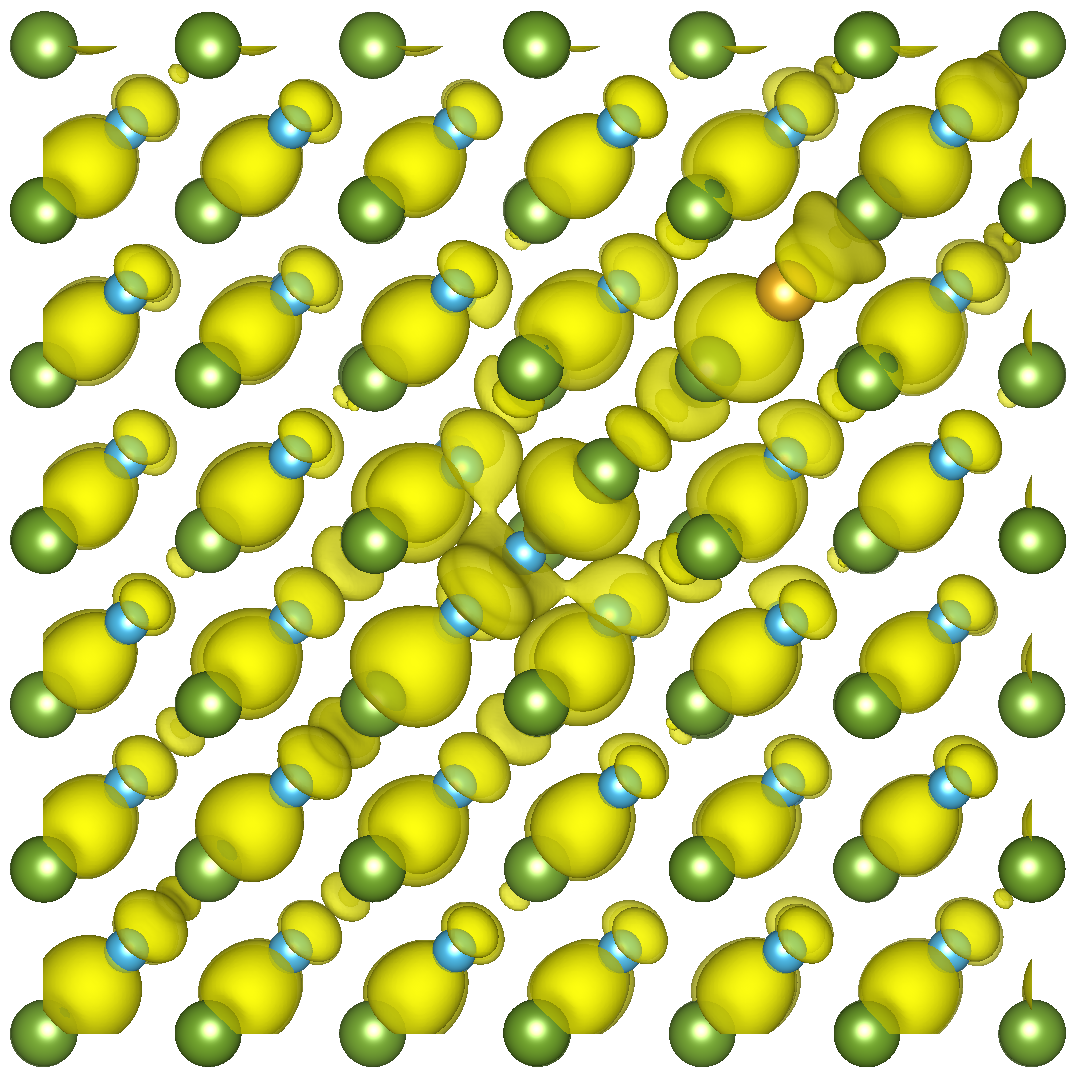}
\end{figure} 

Be$_\text{B}$-C$_\text{As}$: One band near VBM:
\begin{figure}[h]
\includegraphics[width=0.23\columnwidth]{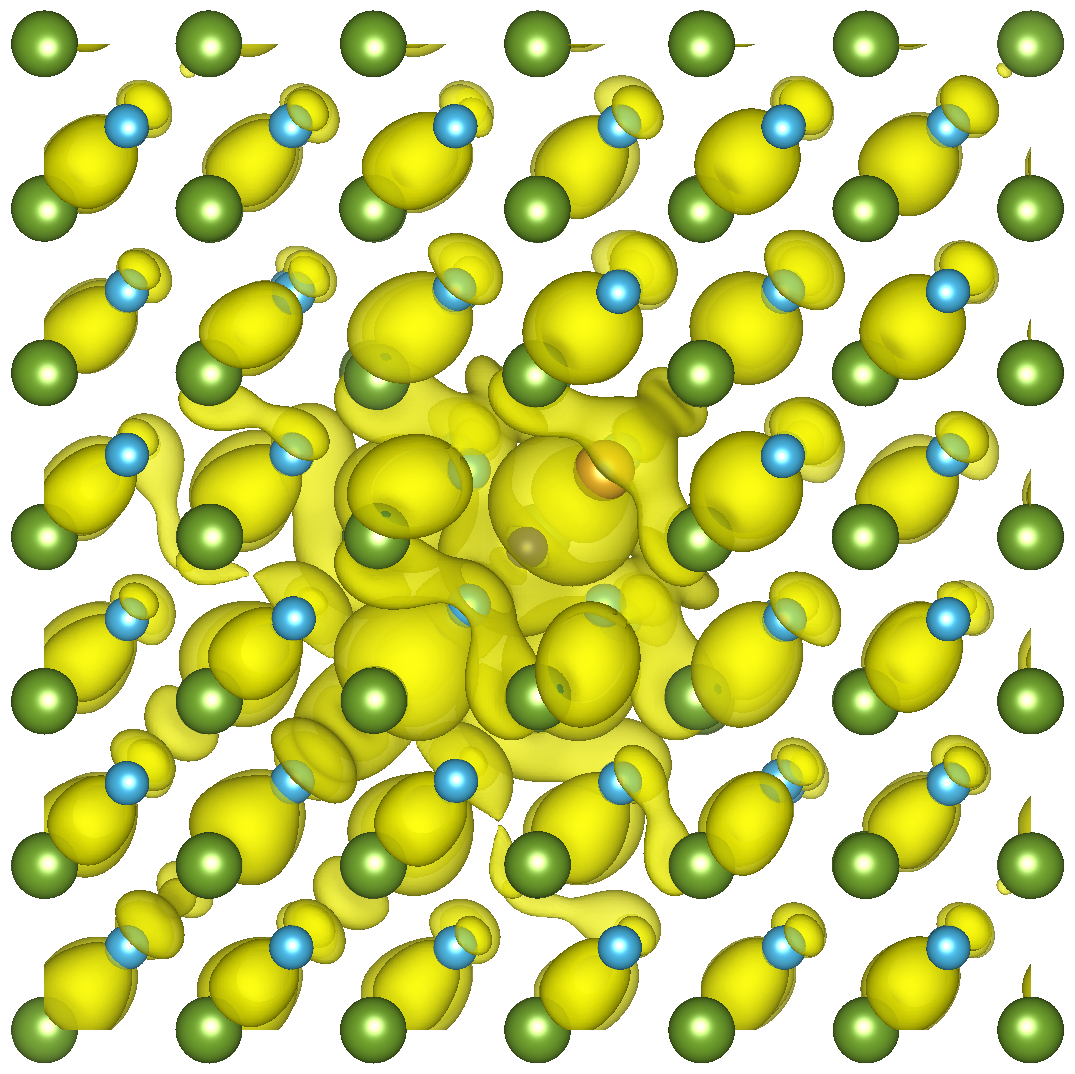}
\end{figure} 

Be$_\text{B}$-O$_\text{B}$O$_\text{As}$: One band at VBM, and one band at CBM (both are strongly localized and depicted at $1\times10^{-3} \AA^{-3}$)
\begin{figure}[h]
\includegraphics[width=0.23\columnwidth]
{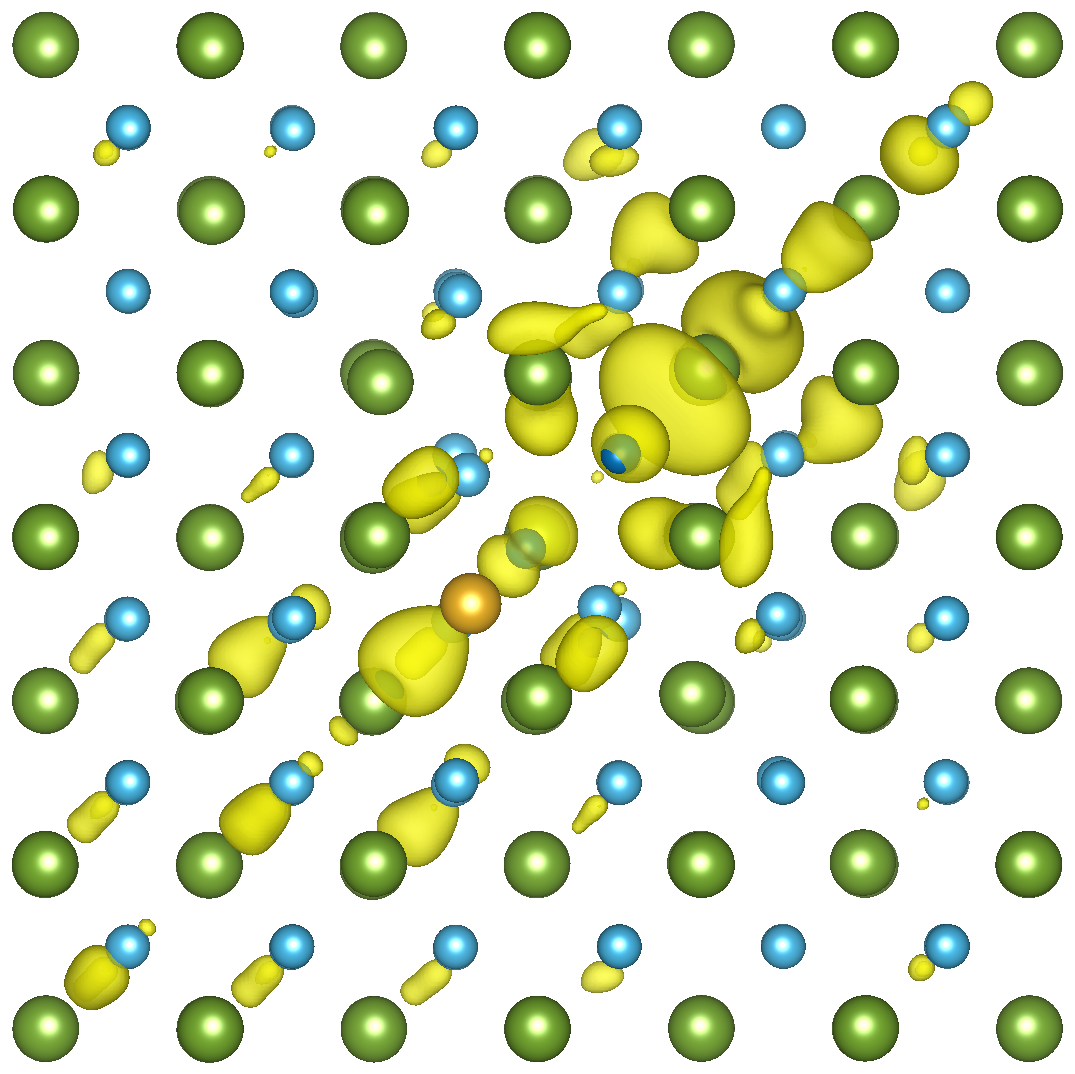}
\hspace{1 cm}
\includegraphics[width=0.23\columnwidth]
{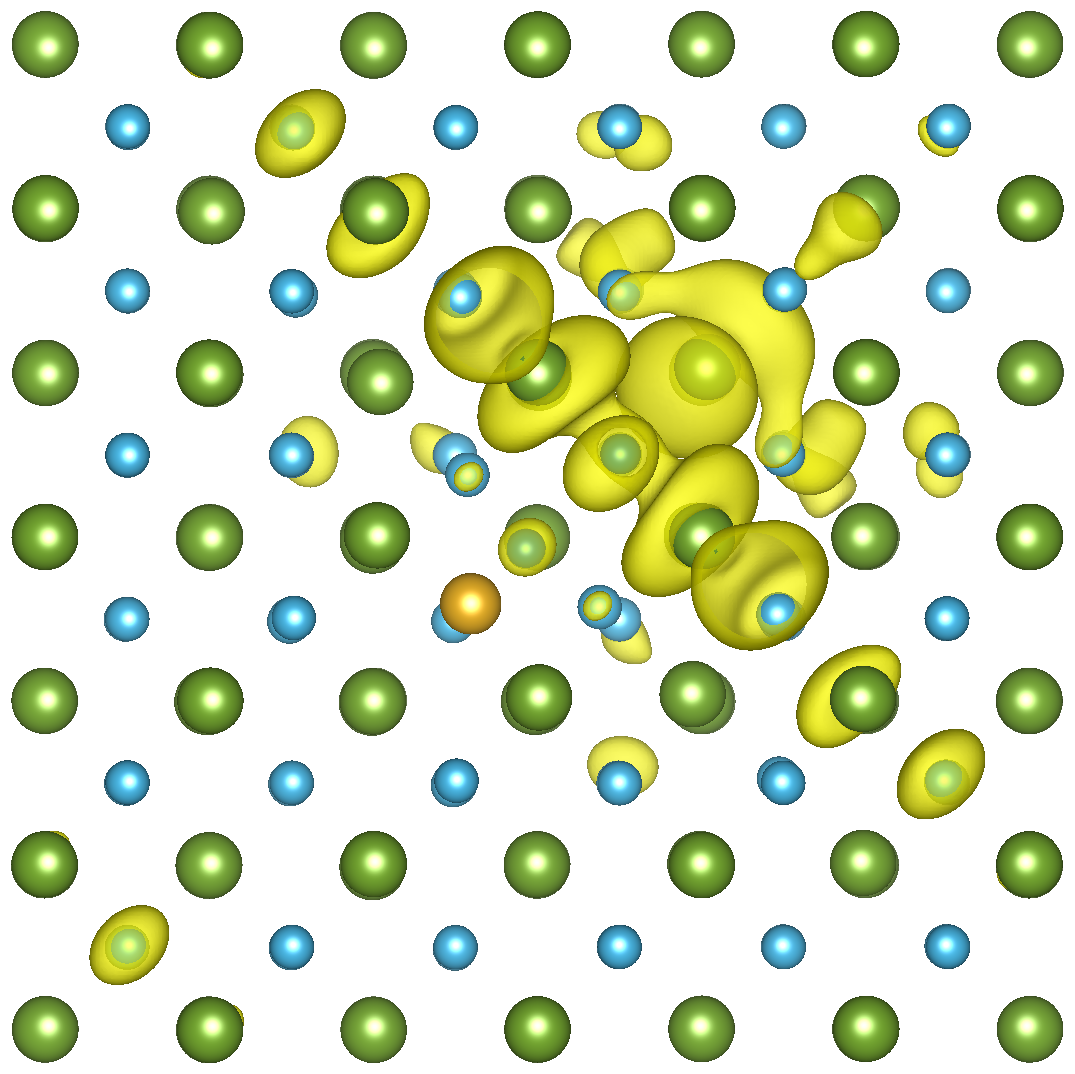}
\end{figure} 

\clearpage
Be$_\text{B}$-Si$_\text{As}$: Three bands near VBM
\begin{figure}[h]
\includegraphics[width=0.23\columnwidth]
{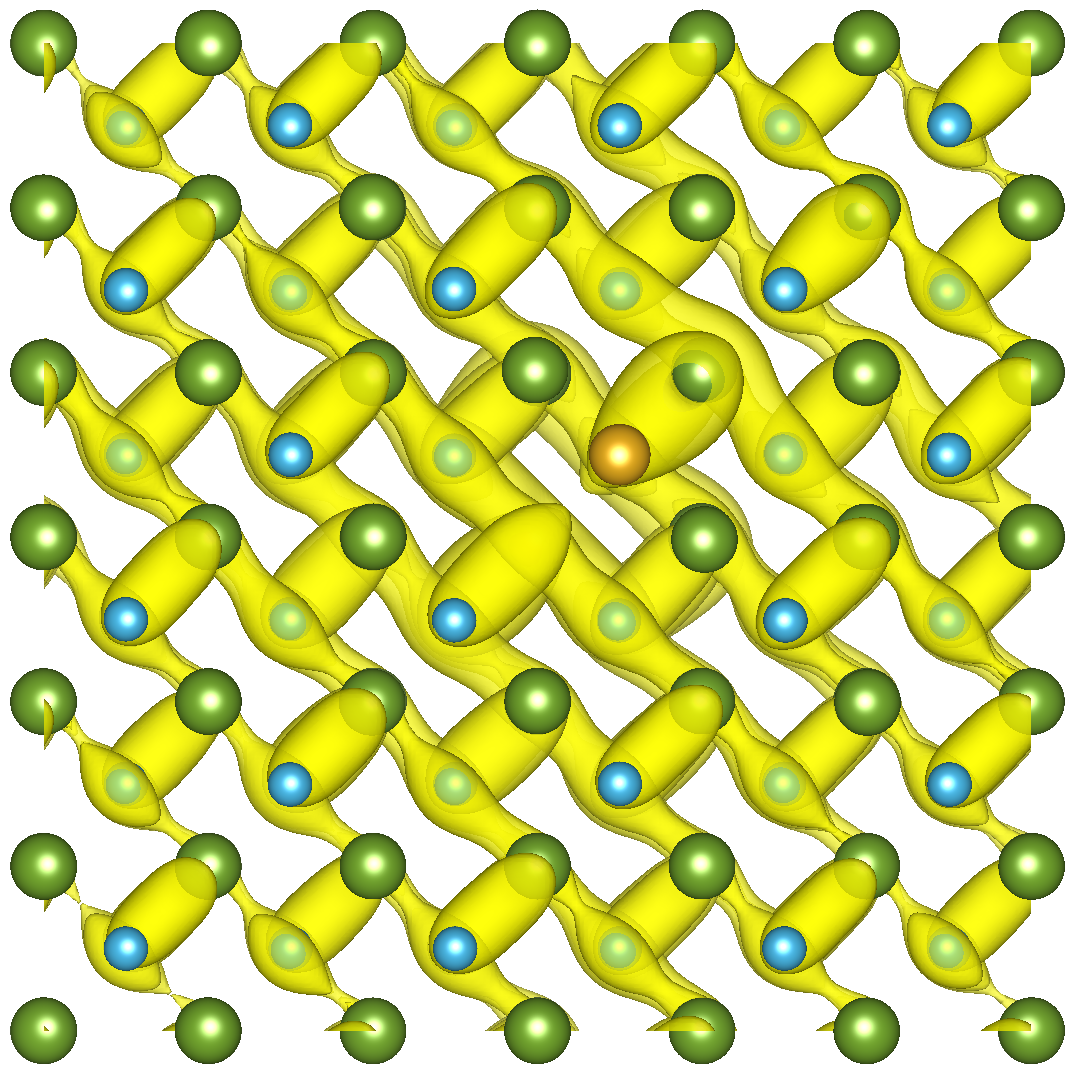}
\hspace{.5 cm}
\includegraphics[width=0.23\columnwidth]
{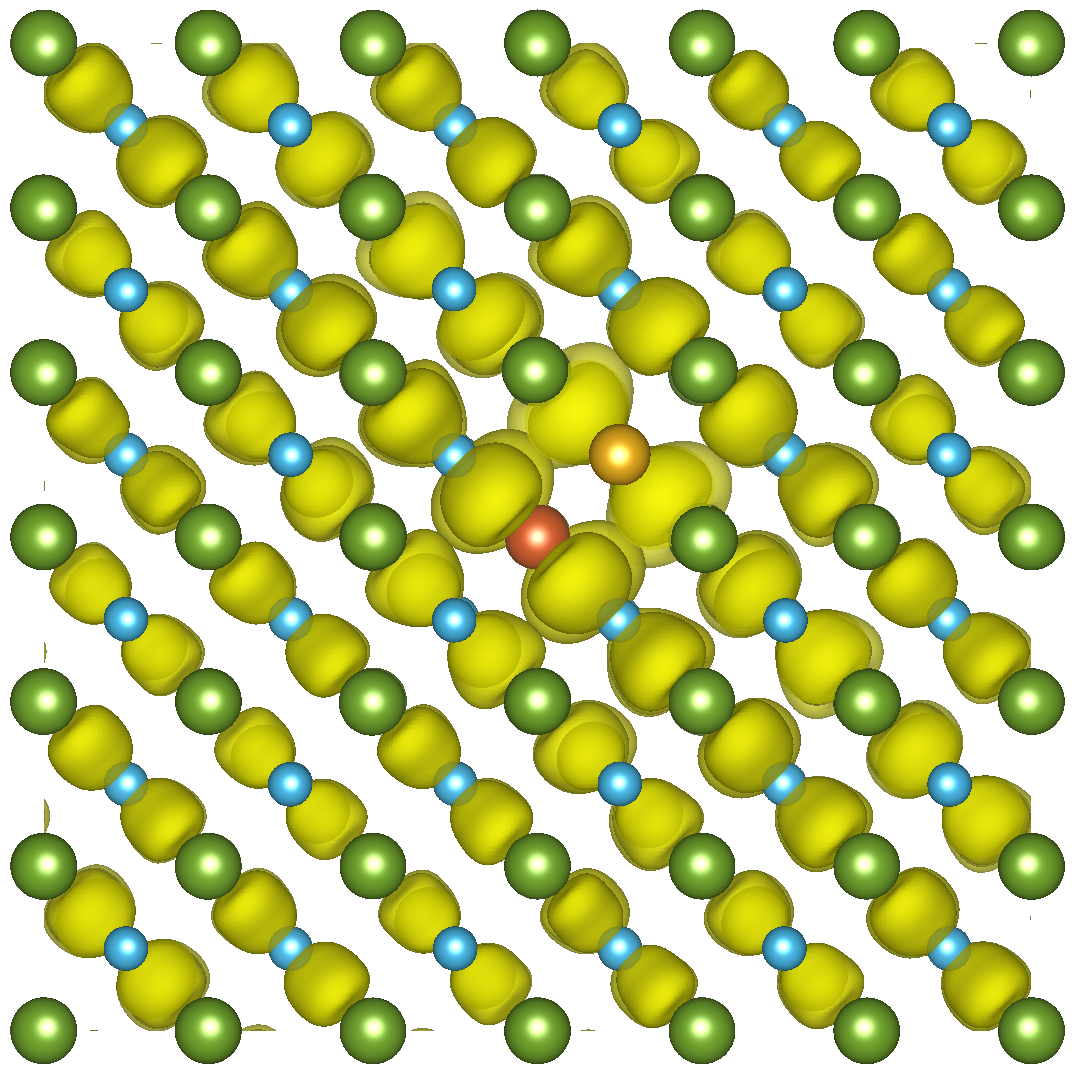}
\hspace{.5 cm}
\includegraphics[width=0.23\columnwidth]
{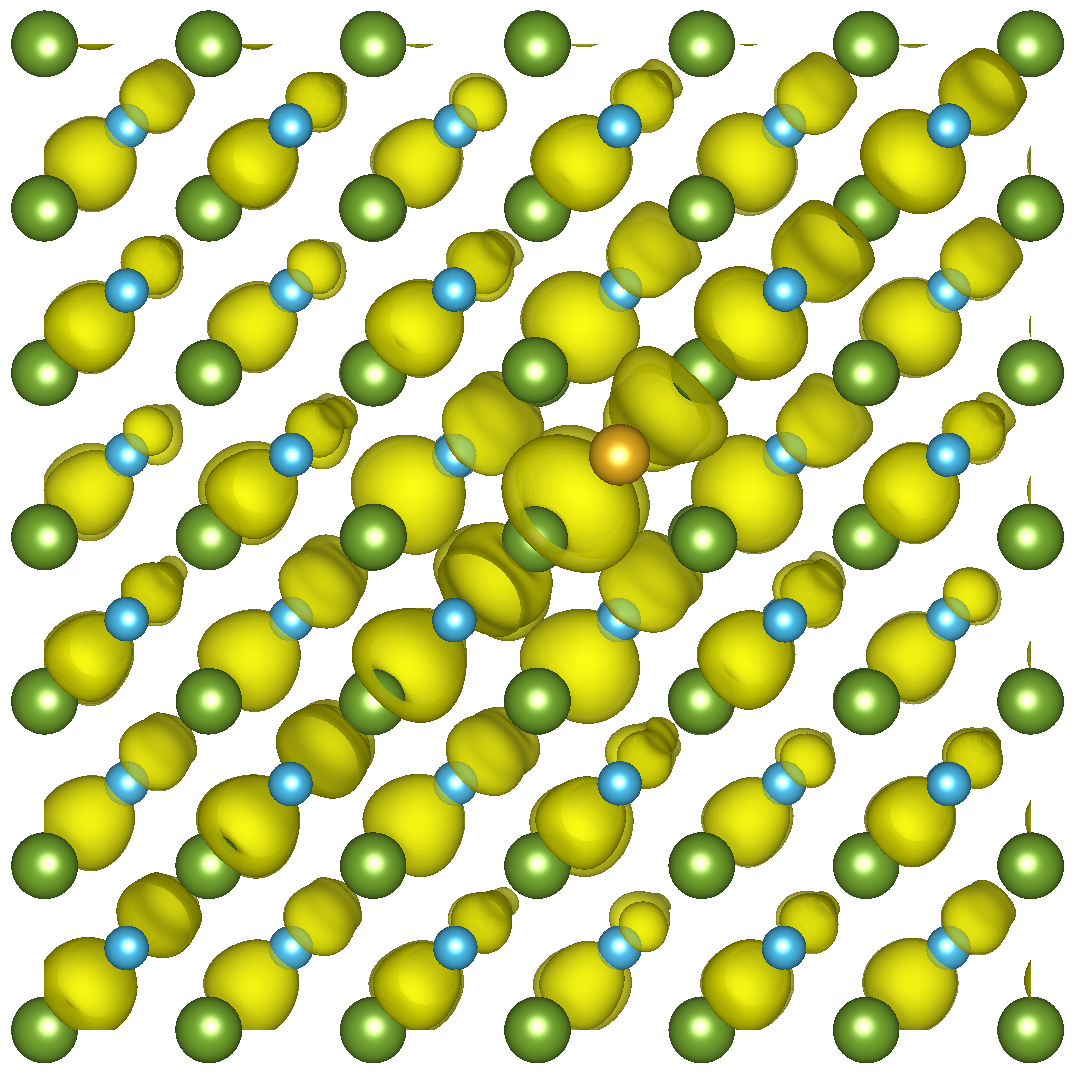}
\end{figure} 

Be$_\text{B}$-C$_\text{As}$Si$_\text{B}$: Three bands near VBM
\begin{figure}[h]
\includegraphics[width=0.23\columnwidth]
{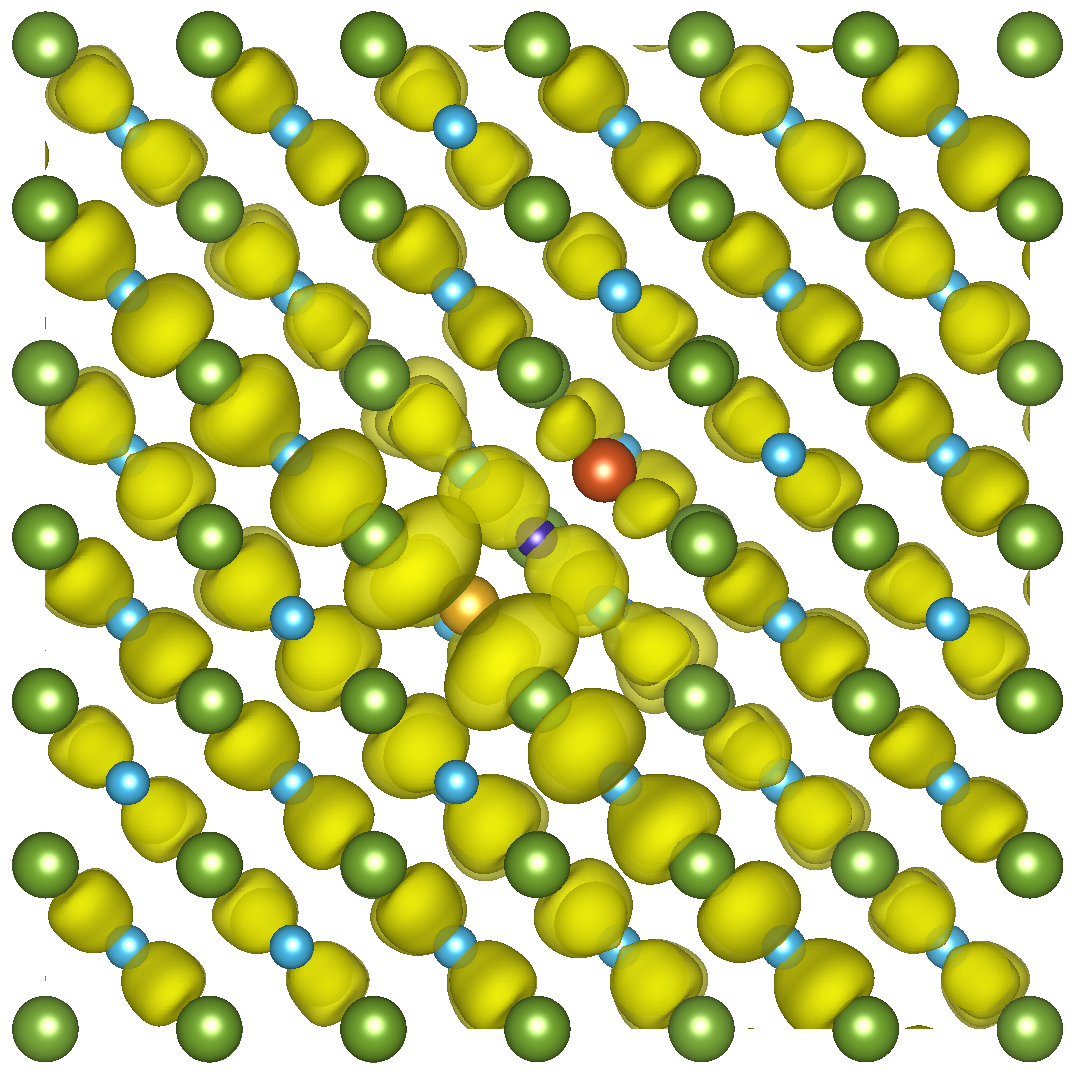}
\hspace{.5 cm}
\includegraphics[width=0.23\columnwidth]
{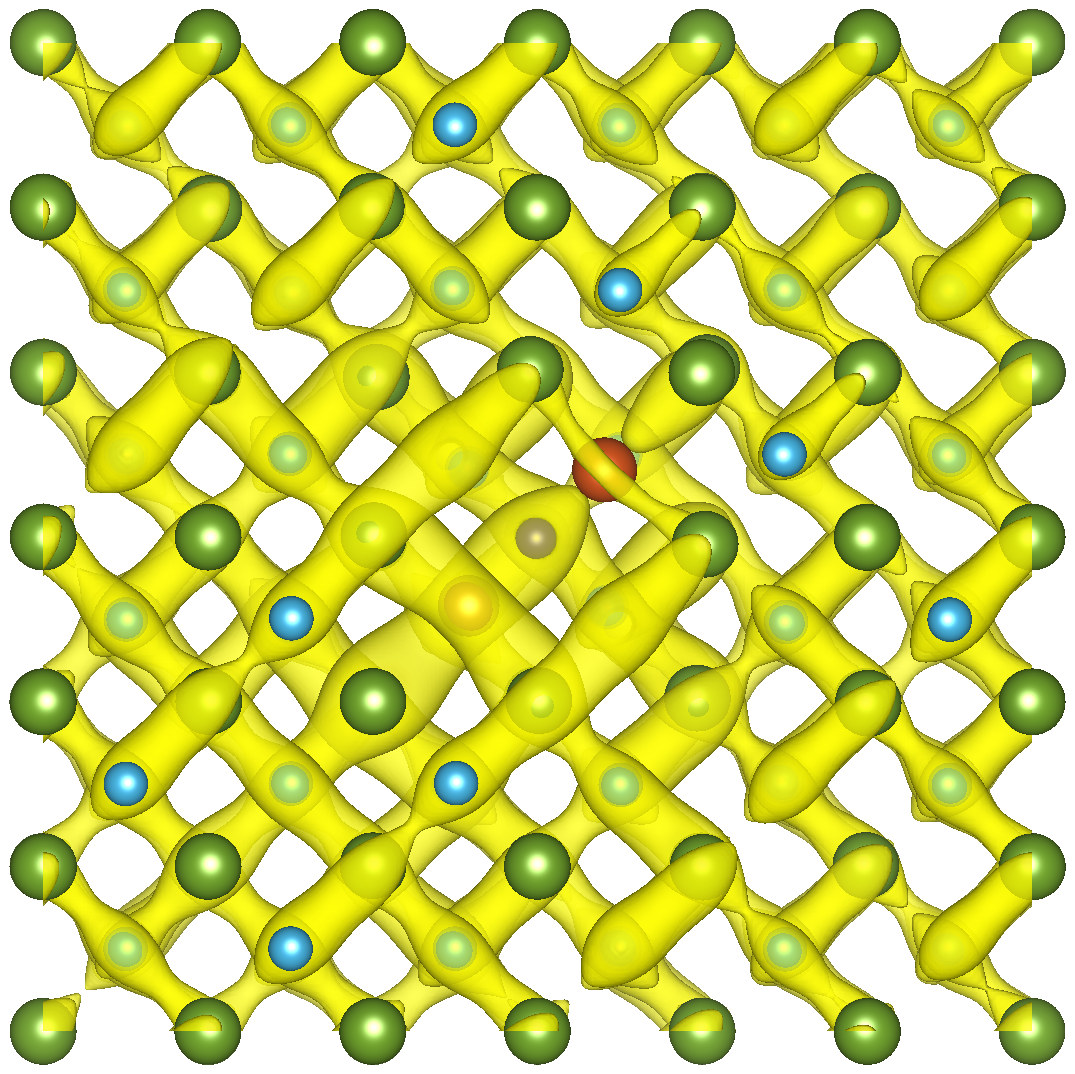}
\hspace{.5 cm}
\includegraphics[width=0.23\columnwidth]
{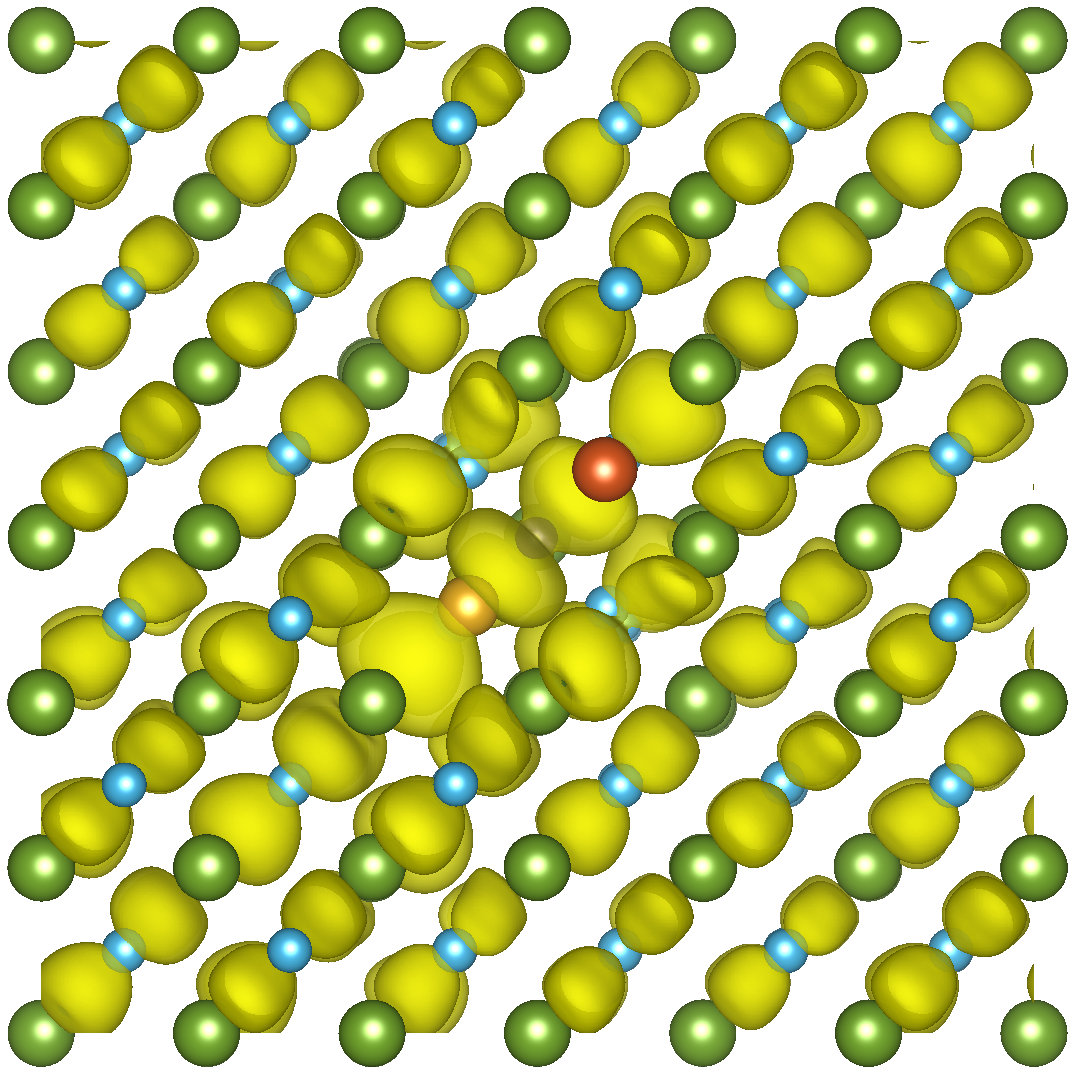}
\end{figure} 

Be$_\text{B}$-O$_\text{B}$Si$_\text{As}$: One band at VBM, and one band in the band gap
\begin{figure}[h]
\includegraphics[width=0.23\columnwidth]
{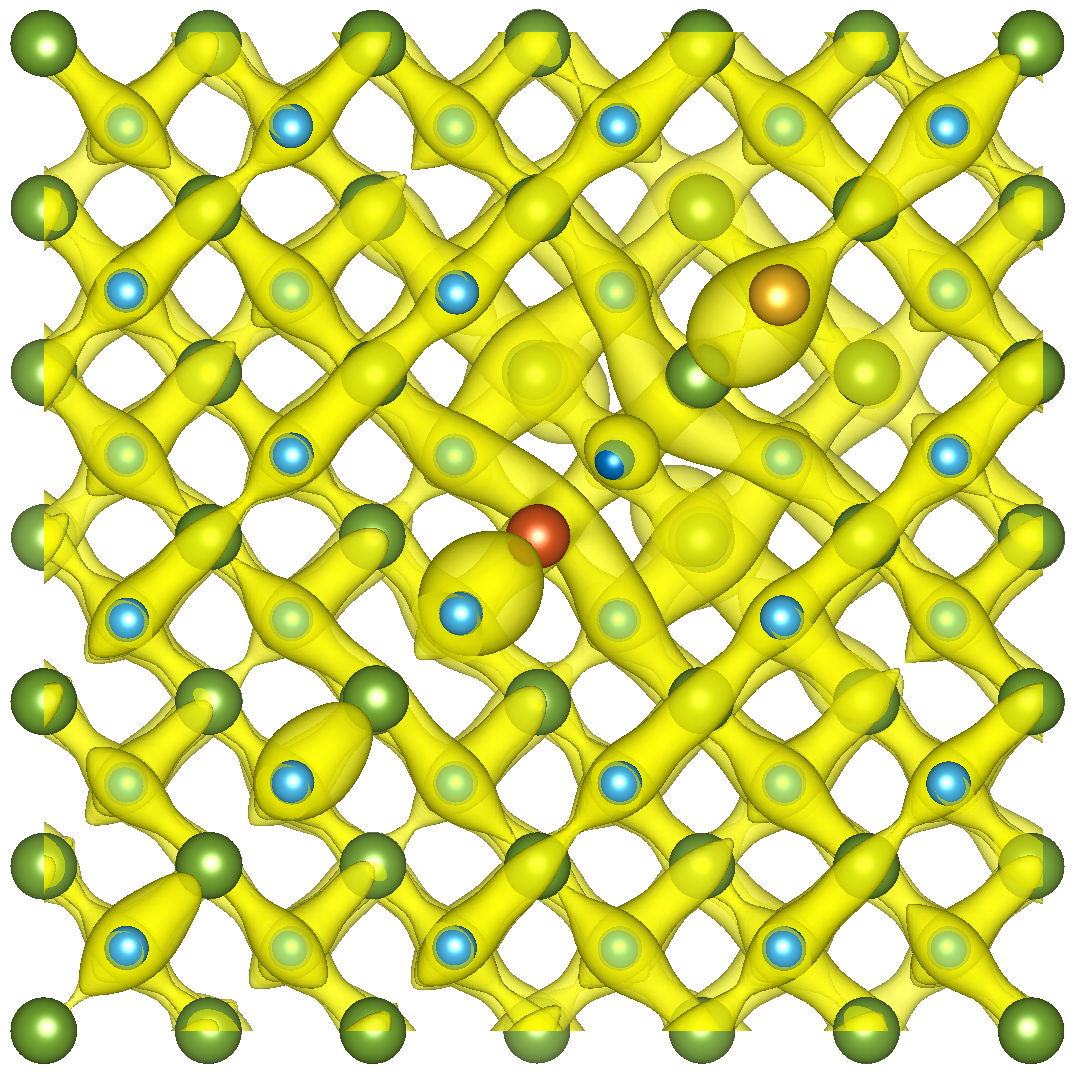}
\hspace{1 cm}
\includegraphics[width=0.23\columnwidth]
{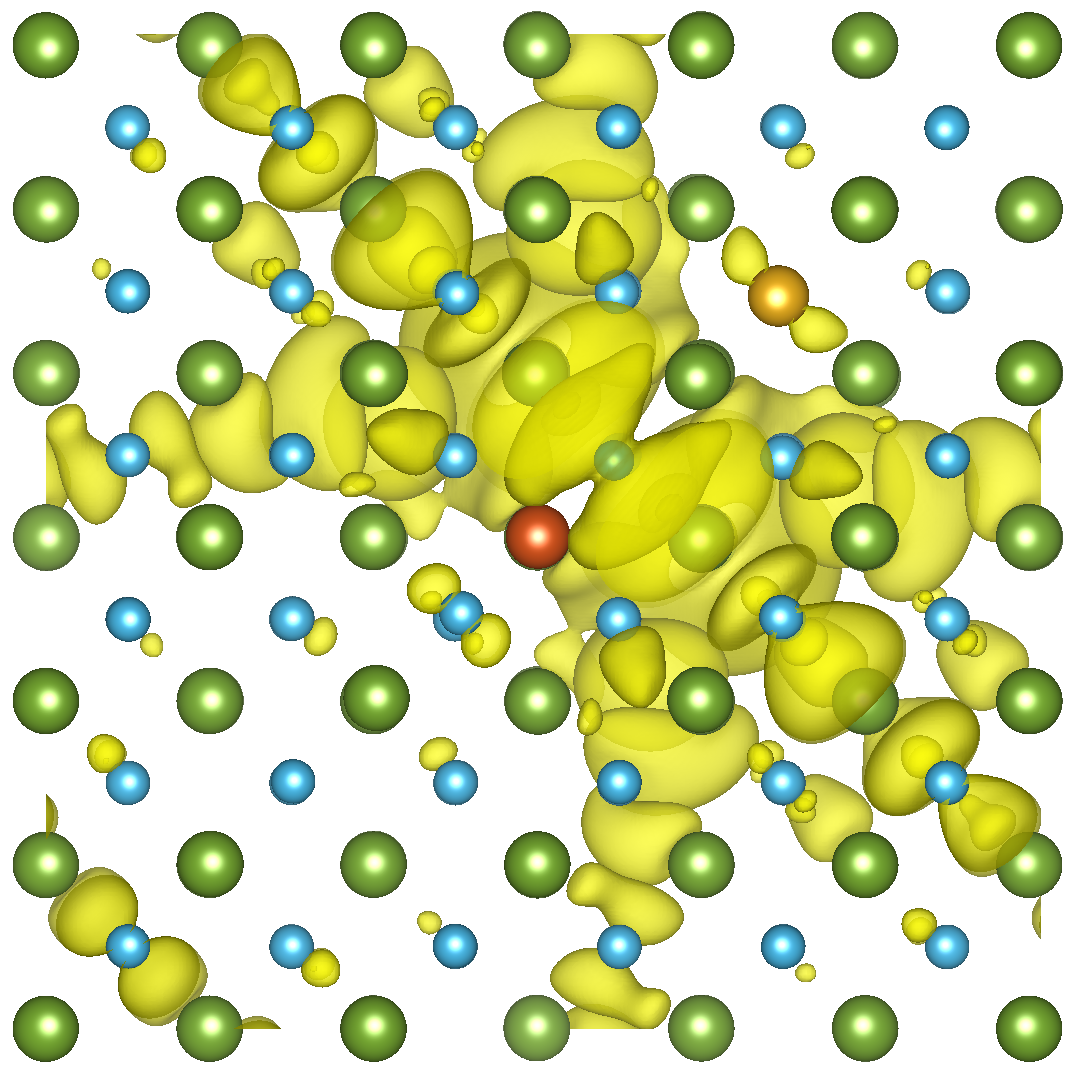}
\end{figure}

Be$_\text{B}$-B$_\text{As}$: One two-fold degenerated band at VBM, and one band in the band gap (the latter one is strongly localized and depicted at $1\times10^{-3} \AA^{-3}$)
\begin{figure}[h]
\includegraphics[width=0.23\columnwidth]{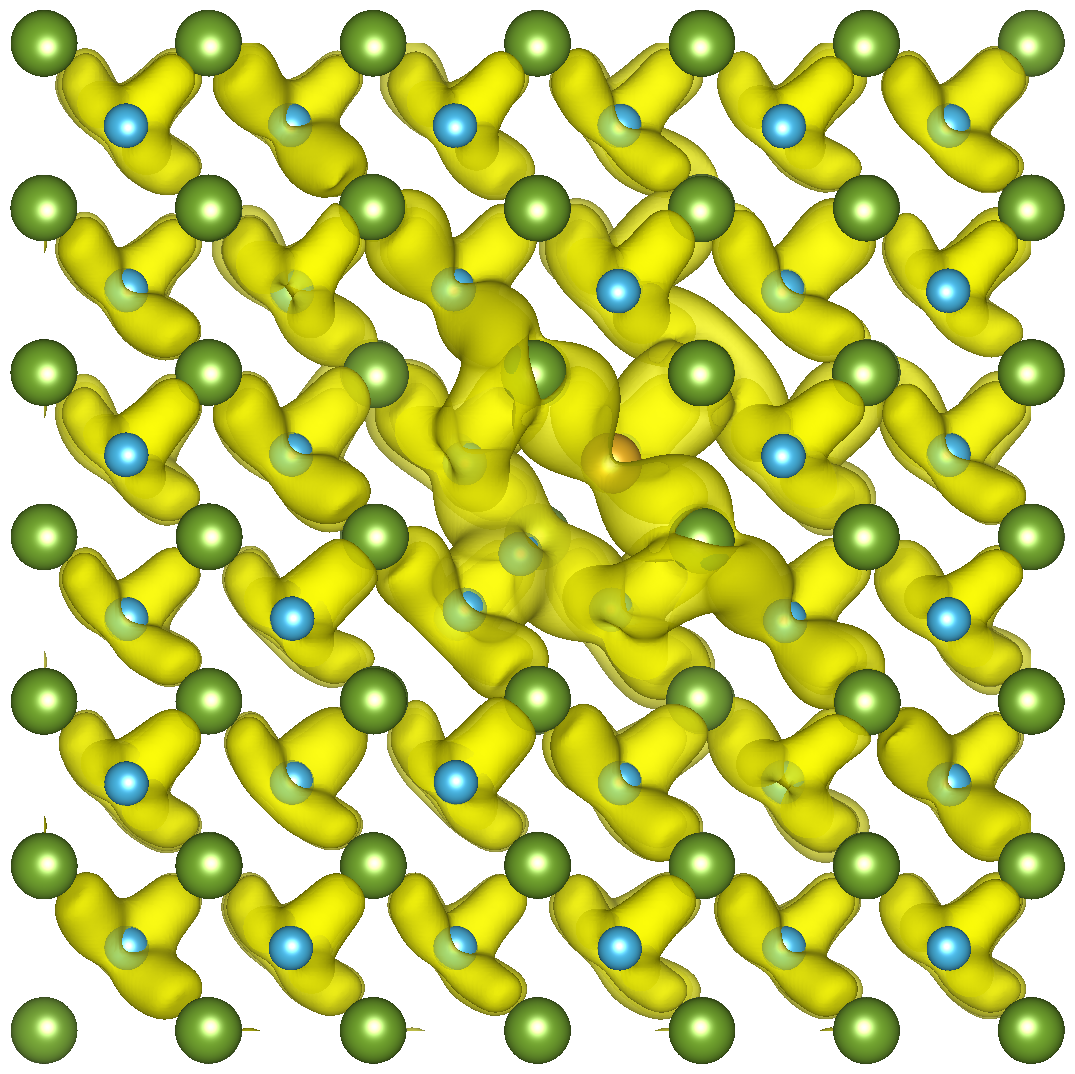}
\hspace{1 cm}
\includegraphics[width=0.23\columnwidth]
{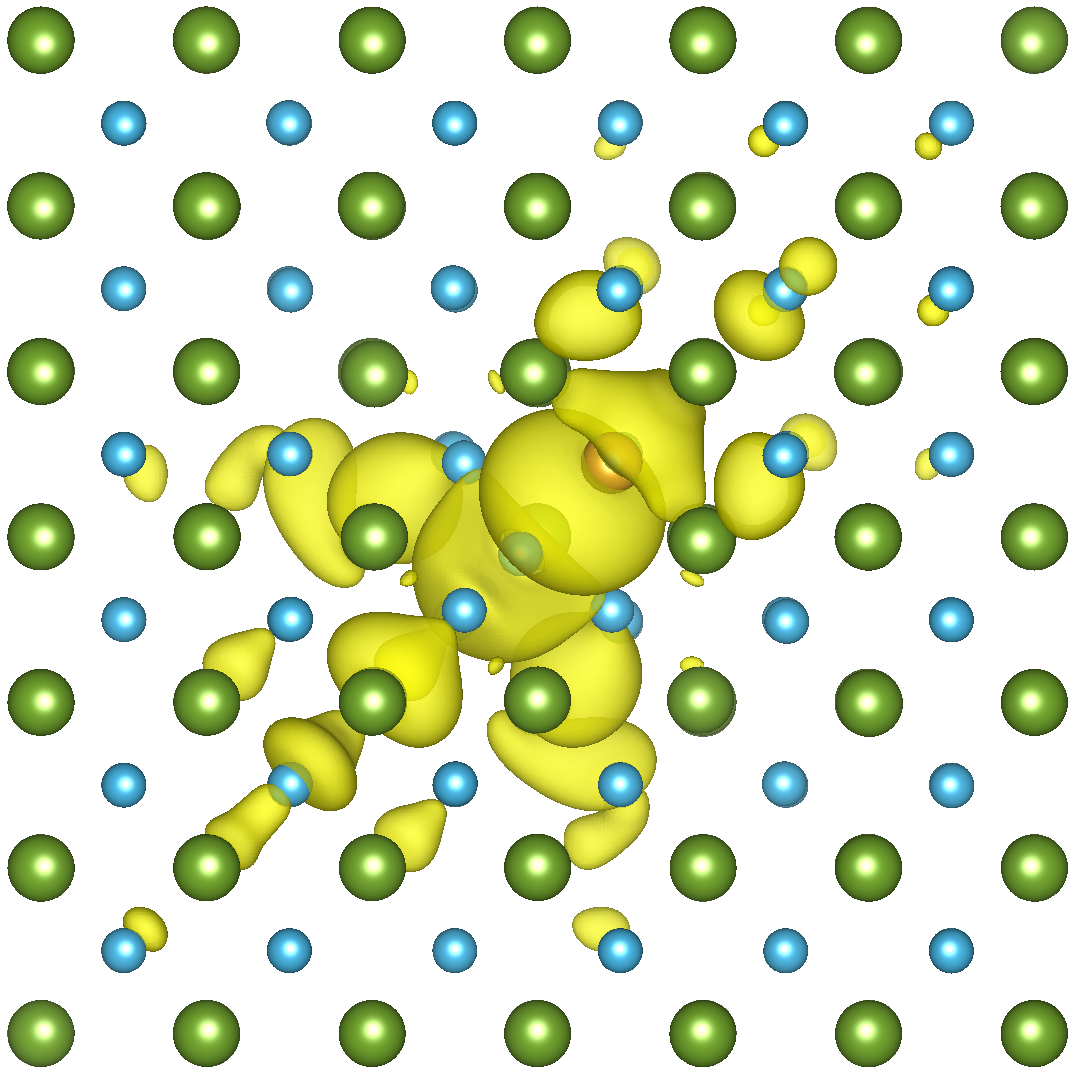}
\end{figure} 

\clearpage
Be$_\text{B}$-N$_\text{As}$: One band near VBM, and one two-fold degenerated band at VBM
\begin{figure}[h]
\includegraphics[width=0.23\columnwidth]{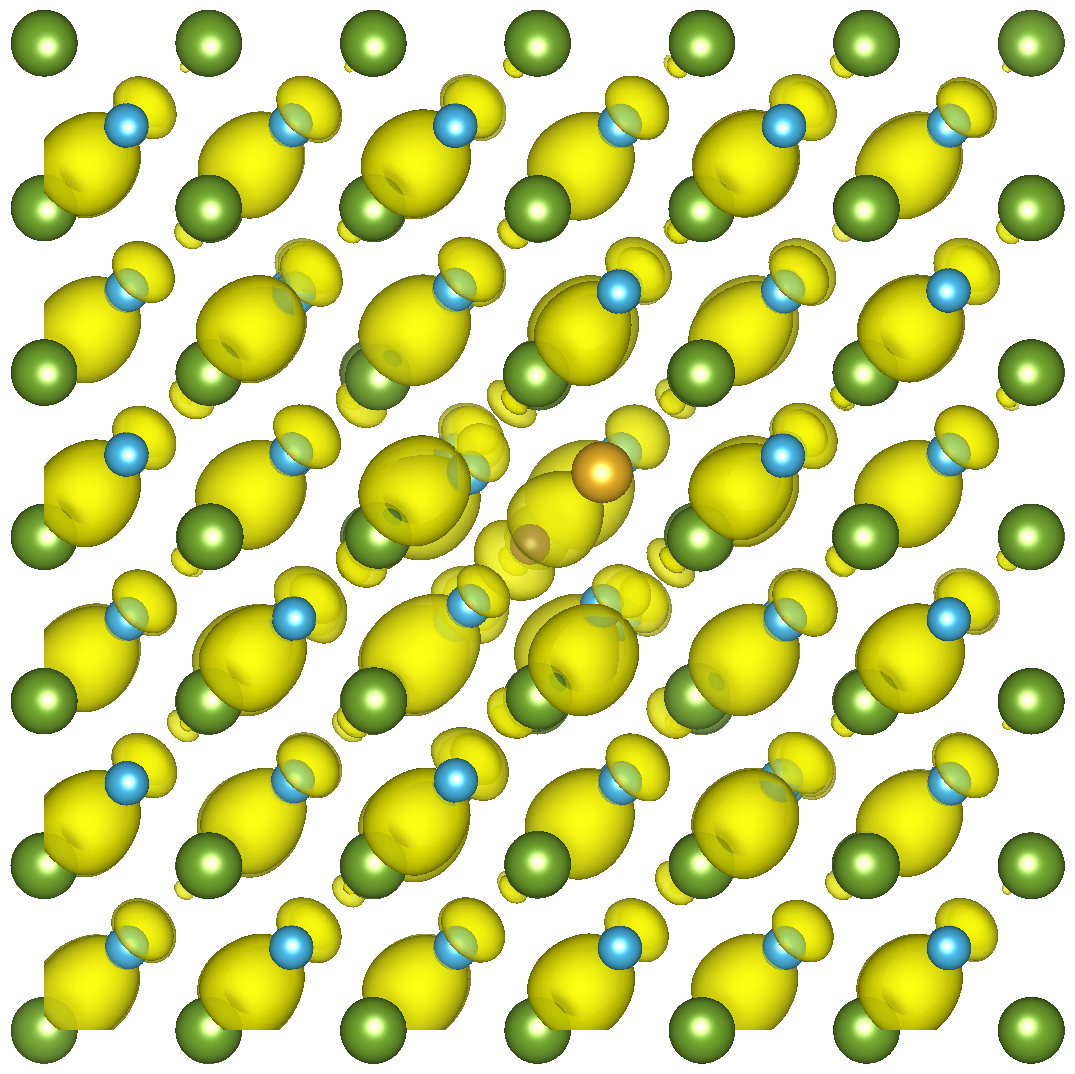}
\hspace{1 cm}
\includegraphics[width=0.23\columnwidth]
{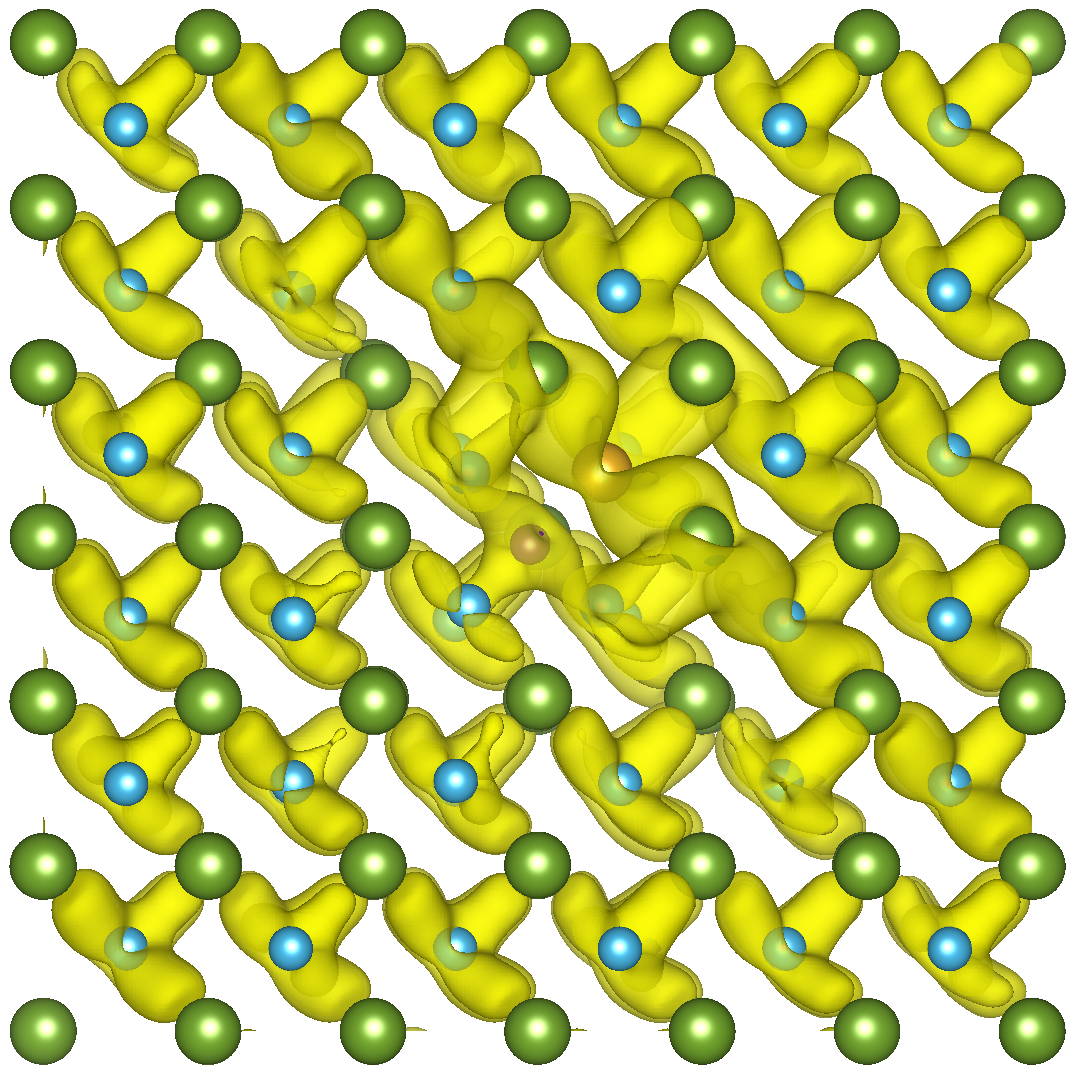}
\end{figure} 

Si$_\text{As}$-As$_\text{B}$B$_\text{As}$: Three bands near VBM
\begin{figure}[h]
\includegraphics[width=0.23\columnwidth]{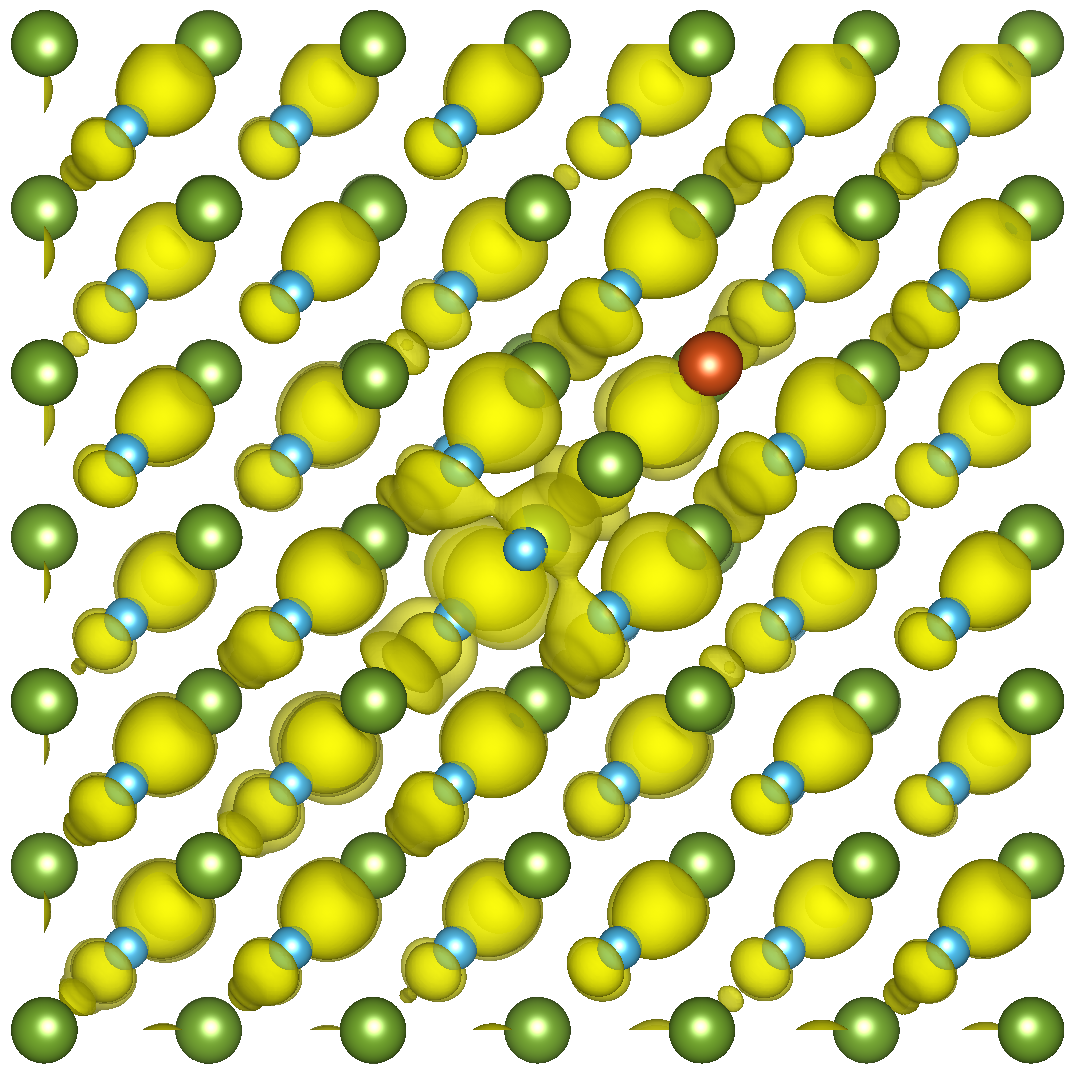}
\hspace{.5 cm}
\includegraphics[width=0.23\columnwidth]{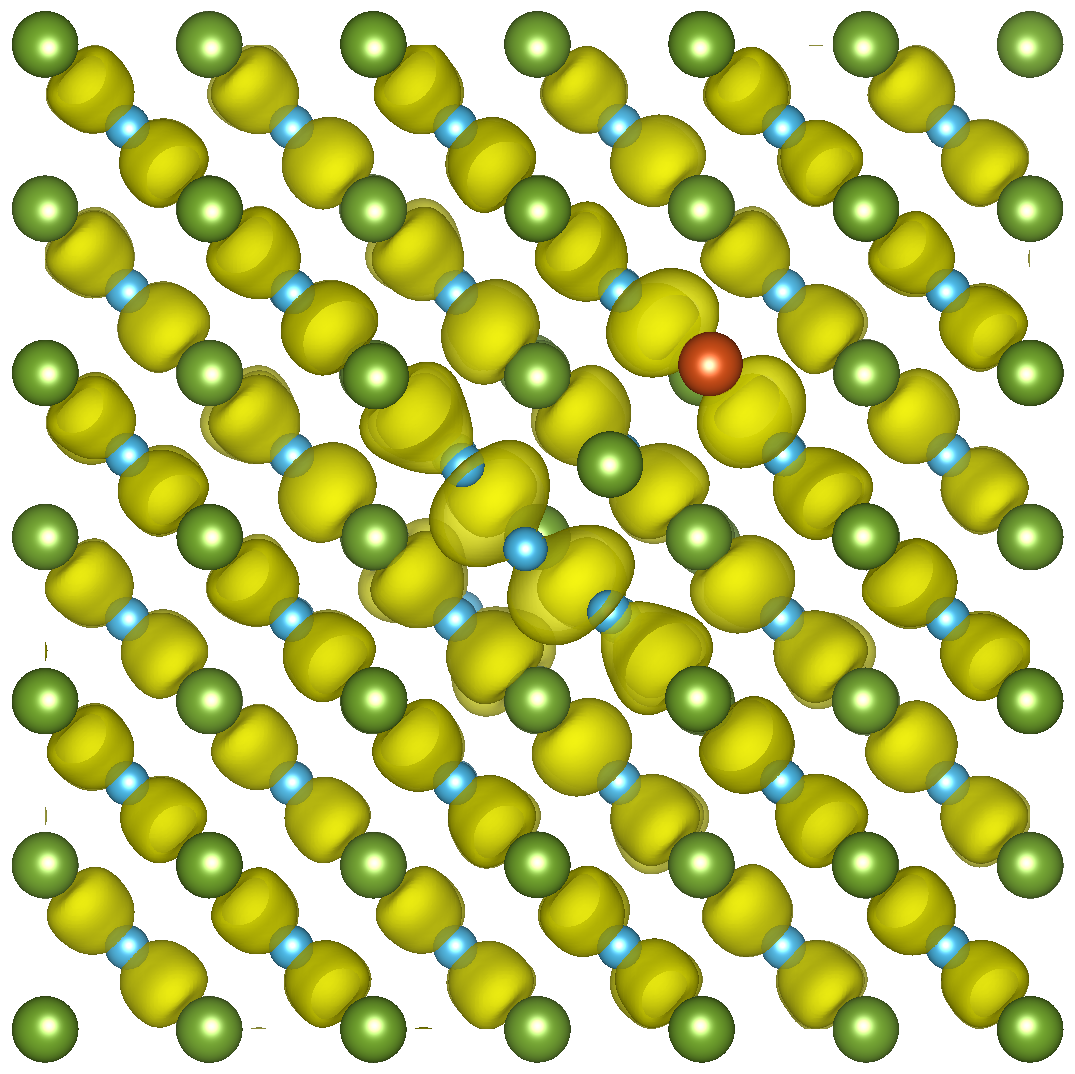}
\hspace{.5 cm}
\includegraphics[width=0.23\columnwidth]{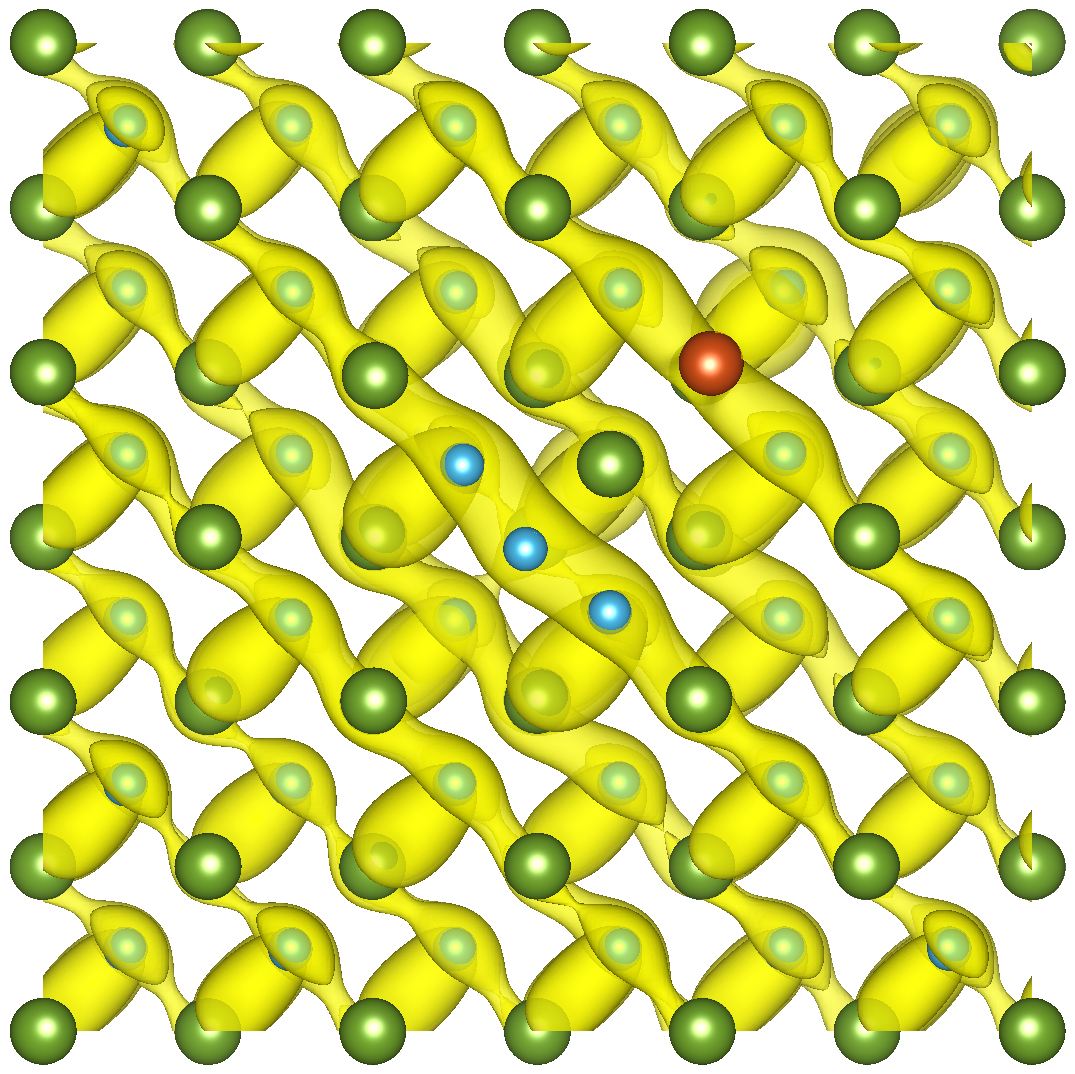}
\end{figure} 

Si$_\text{As}$-C$_\text{As}$: Three bands near VBM
\begin{figure}[h]
\includegraphics[width=0.23\columnwidth]{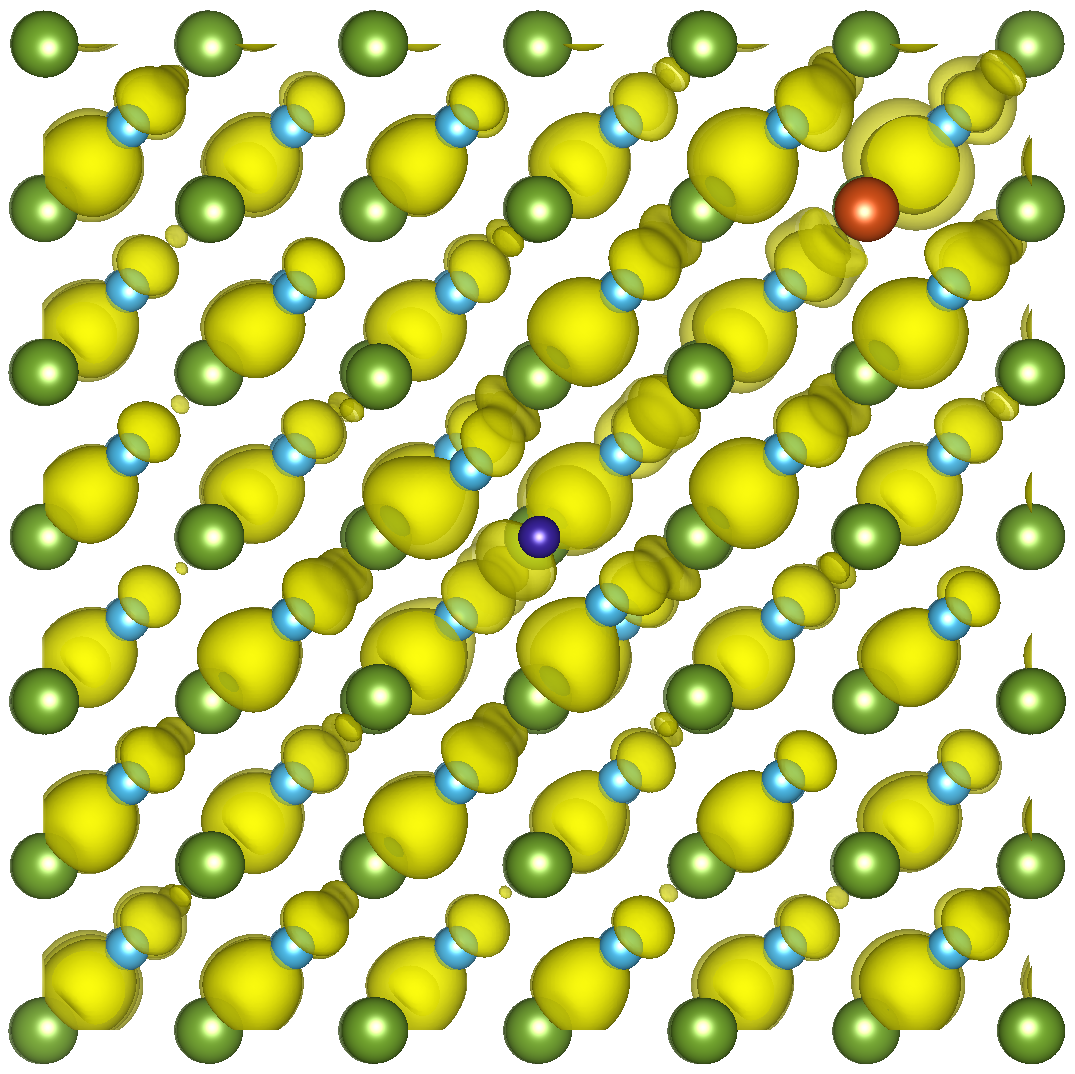}
\hspace{.5 cm}
\includegraphics[width=0.23\columnwidth]{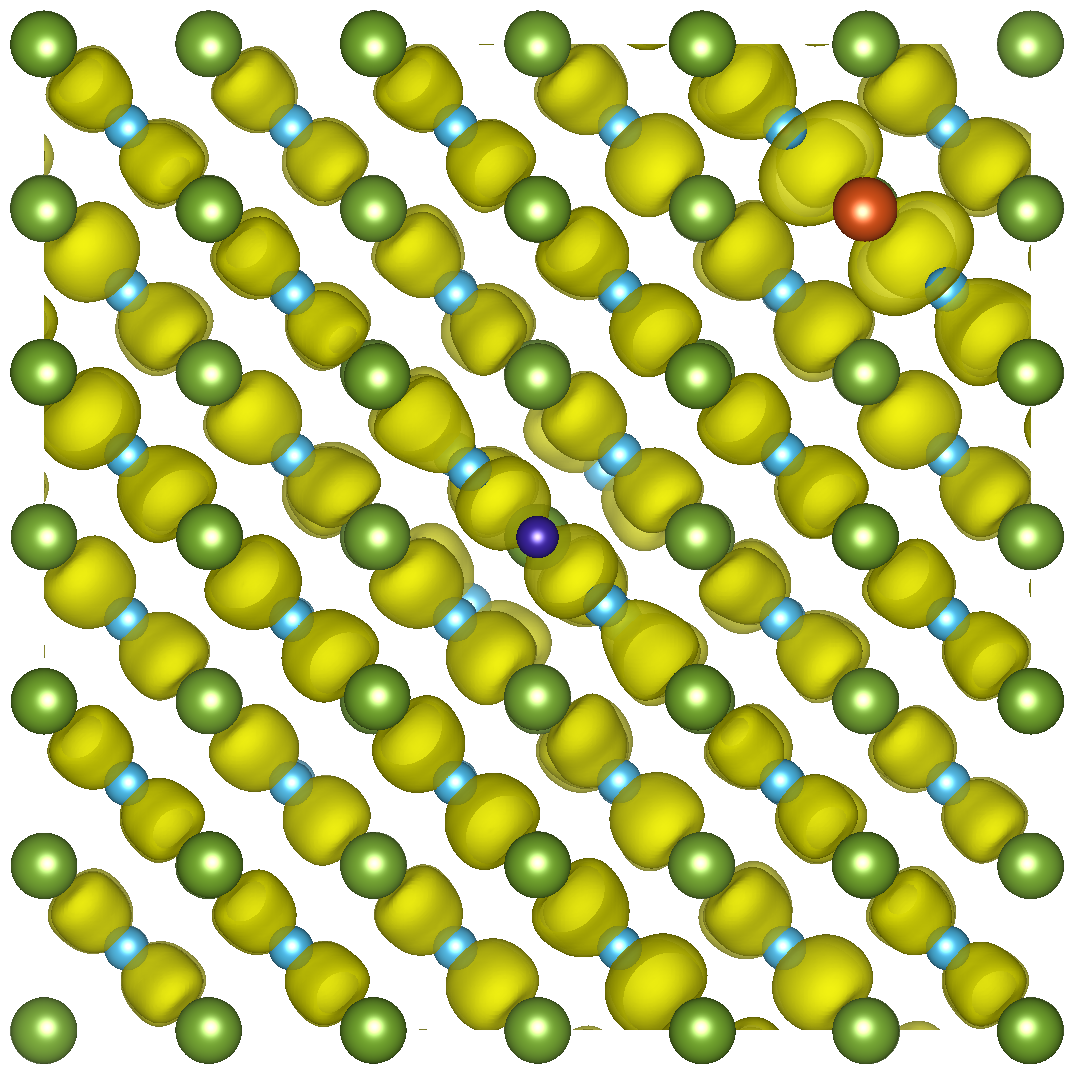}
\hspace{.5 cm}
\includegraphics[width=0.23\columnwidth]{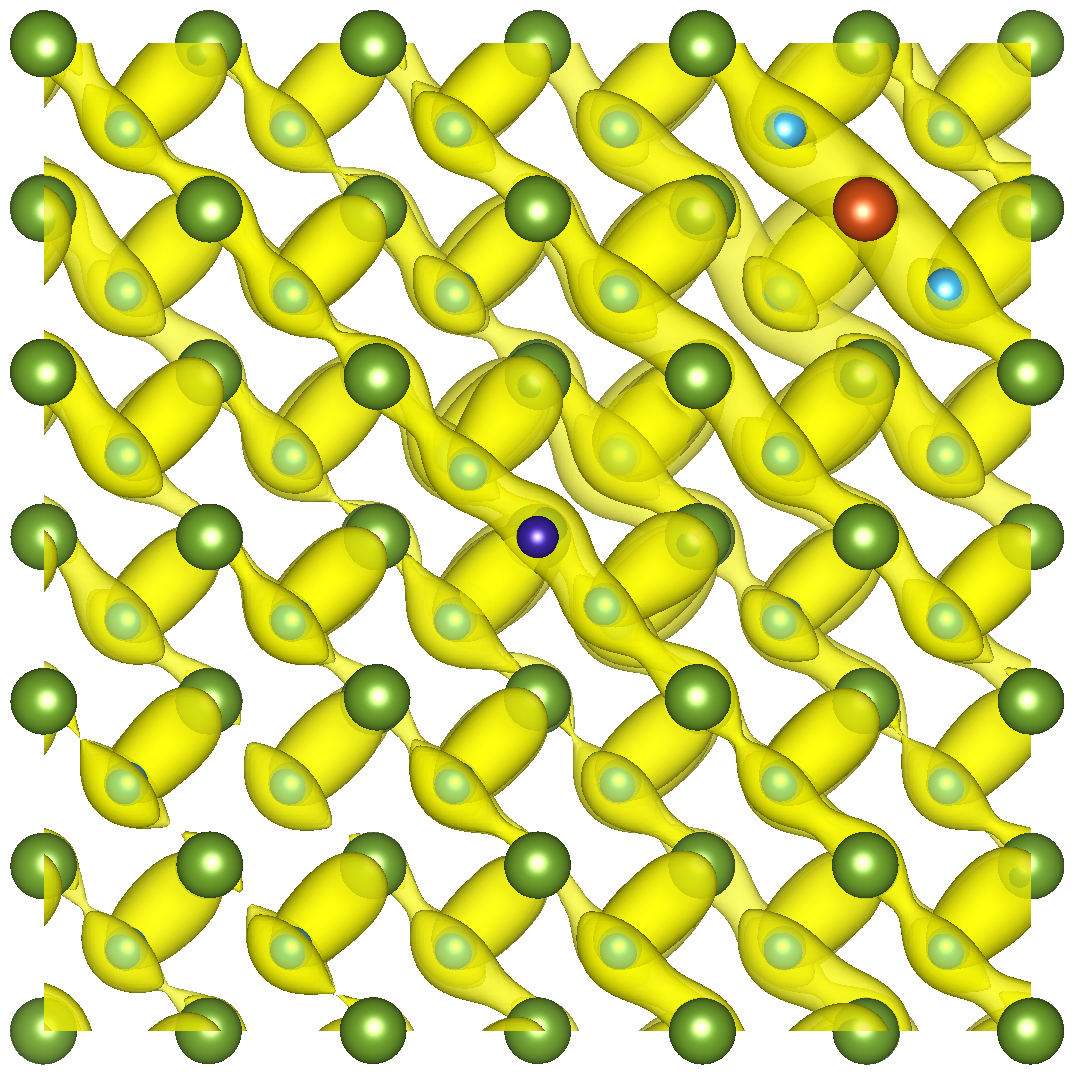}
\end{figure} 

Si$_\text{As}$-O$_\text{B}$O$_\text{As}$: One band in the band gap (it is strongly localized and depicted at $1\times10^{-3} \AA^{-3}$)
\begin{figure}[h]
\includegraphics[width=0.23\columnwidth]
{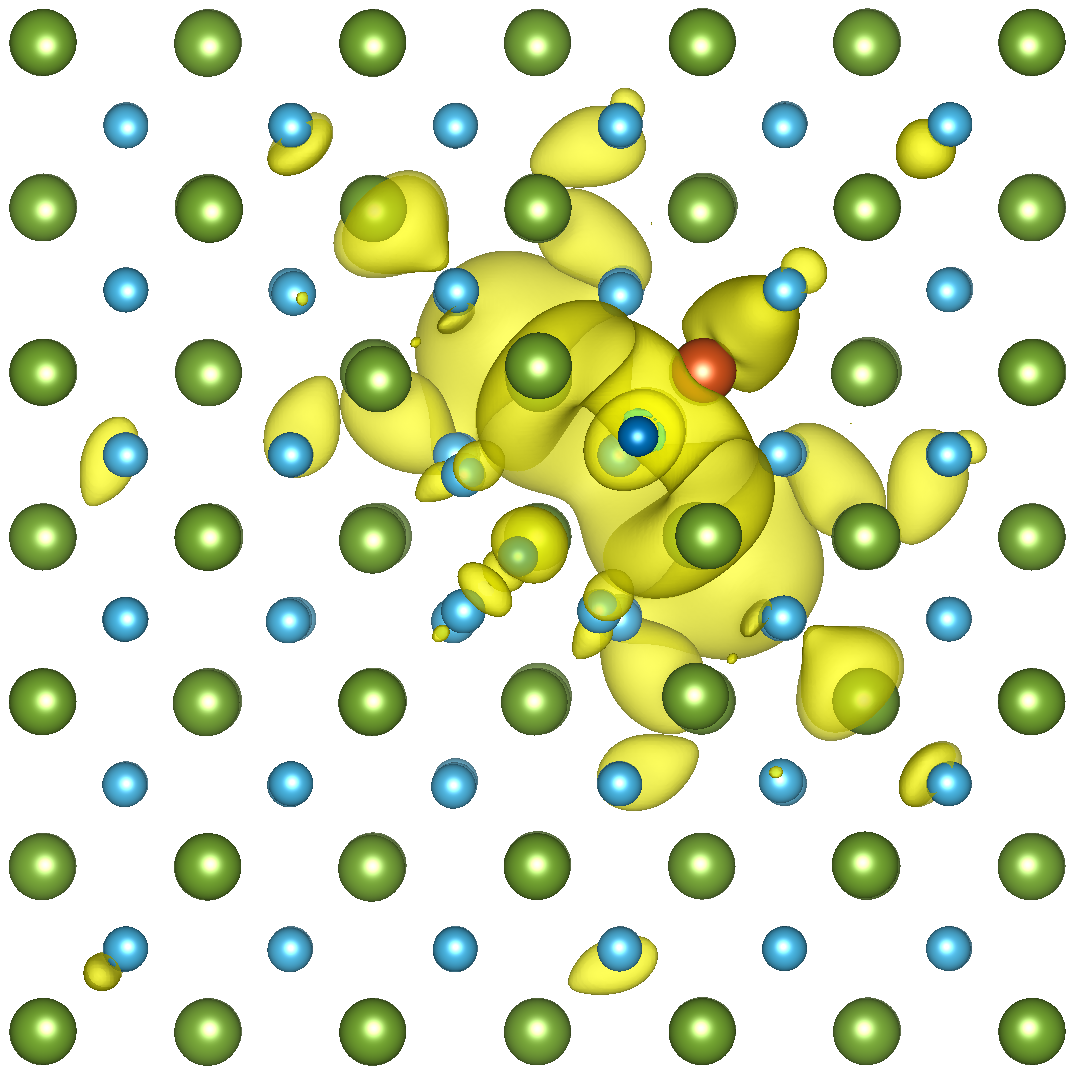}
\end{figure} 

\clearpage

Si$_\text{As}$-C$_\text{As}$Si$_\text{B}$: Three bands near VBM
\begin{figure}[h]
\includegraphics[width=0.23\columnwidth]{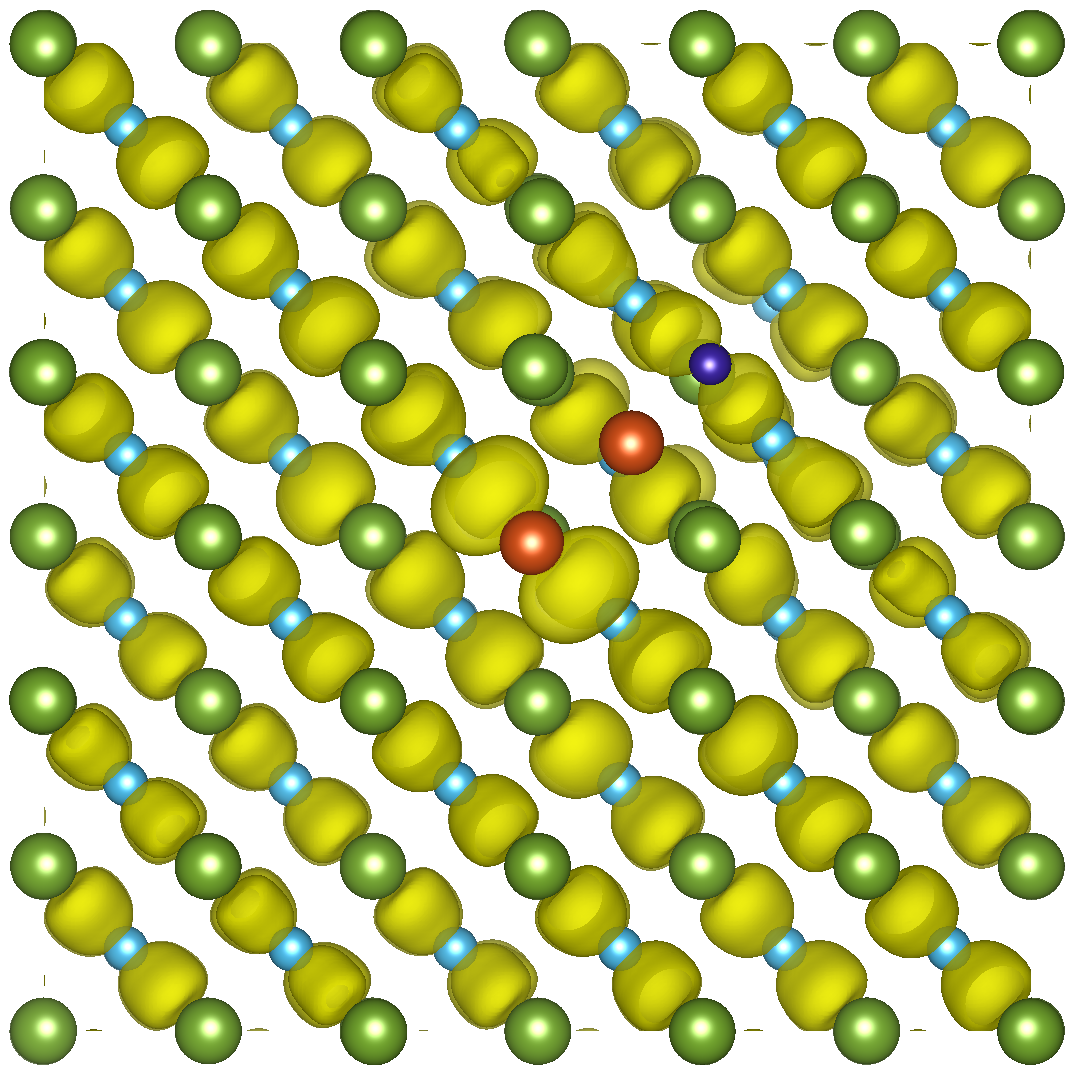}
\hspace{.5 cm}
\includegraphics[width=0.23\columnwidth]{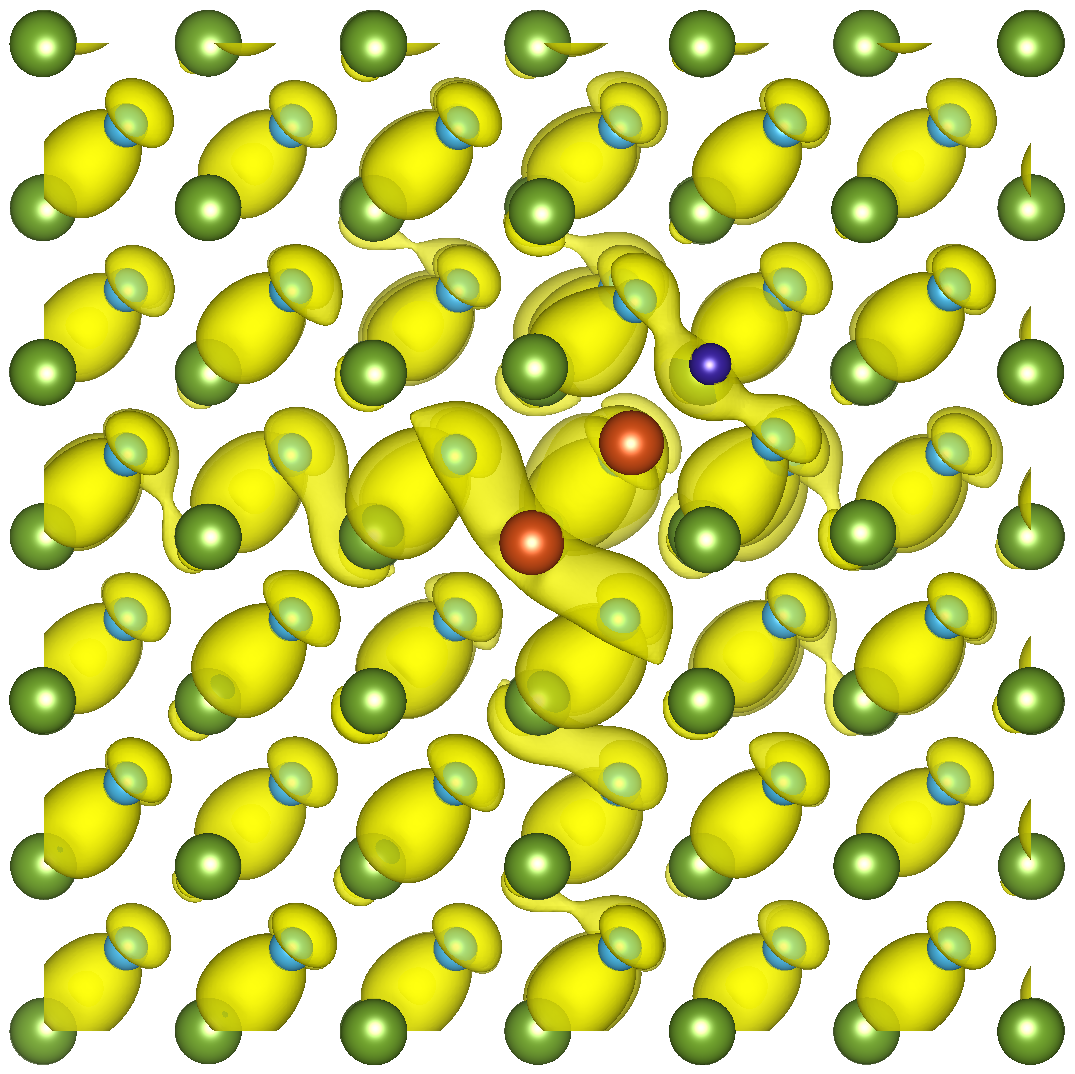}
\hspace{.5 cm}
\includegraphics[width=0.23\columnwidth]{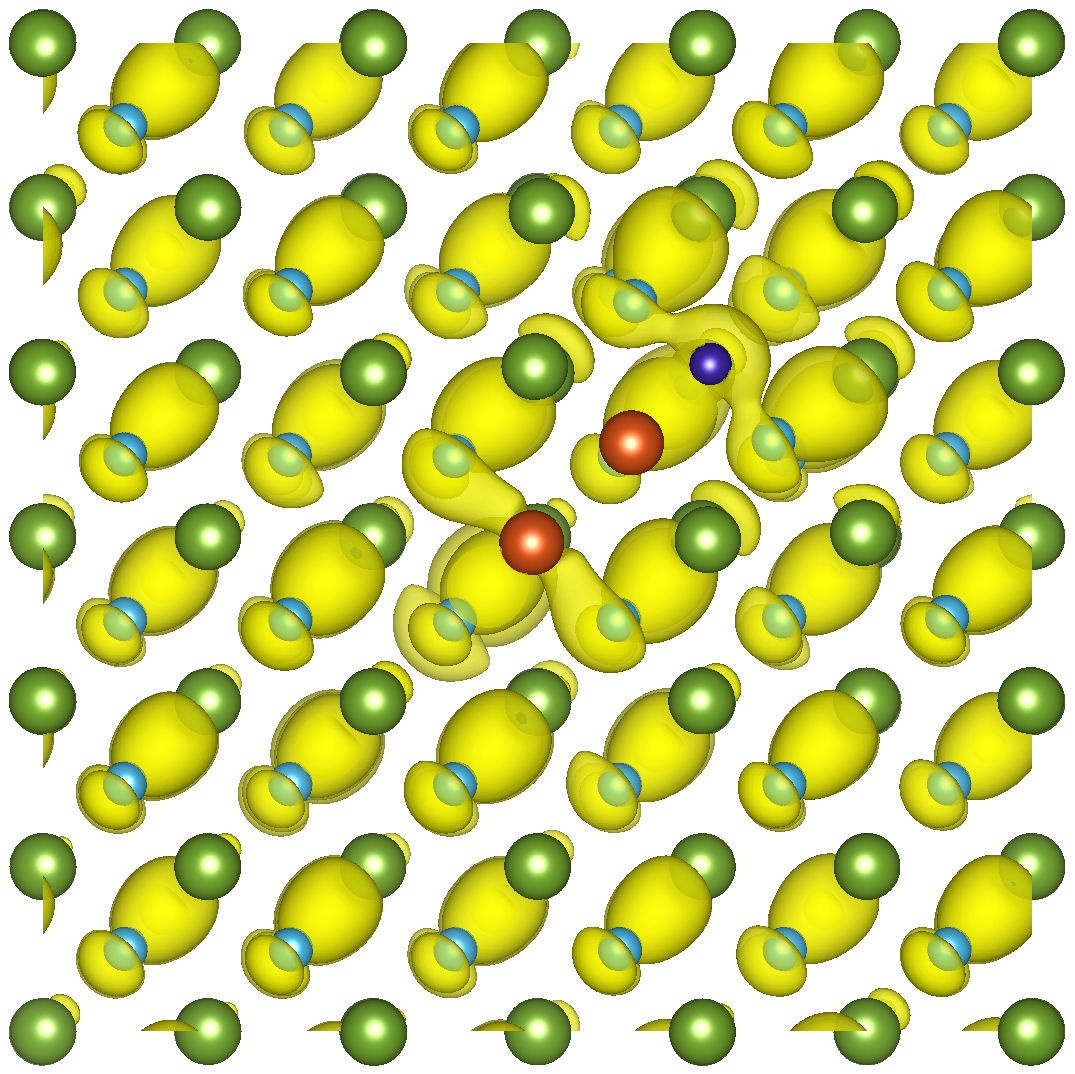}
\end{figure} 

Si$_\text{As}$-O$_\text{B}$Si$_\text{As}$: Two bands at VBM, and one band in the band gap
\begin{figure}[h]
\includegraphics[width=0.23\columnwidth]{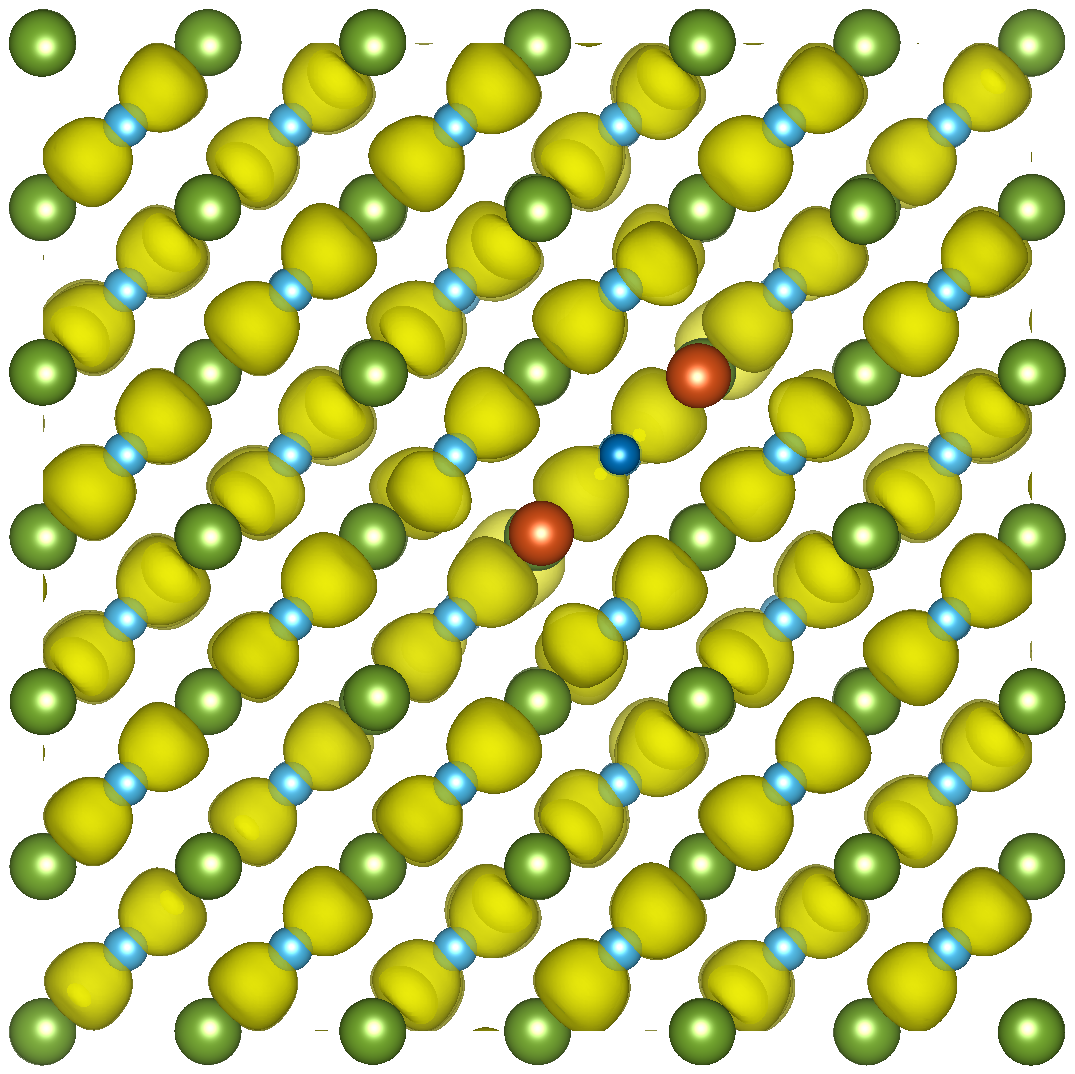}
\hspace{.5 cm}
\includegraphics[width=0.23\columnwidth]{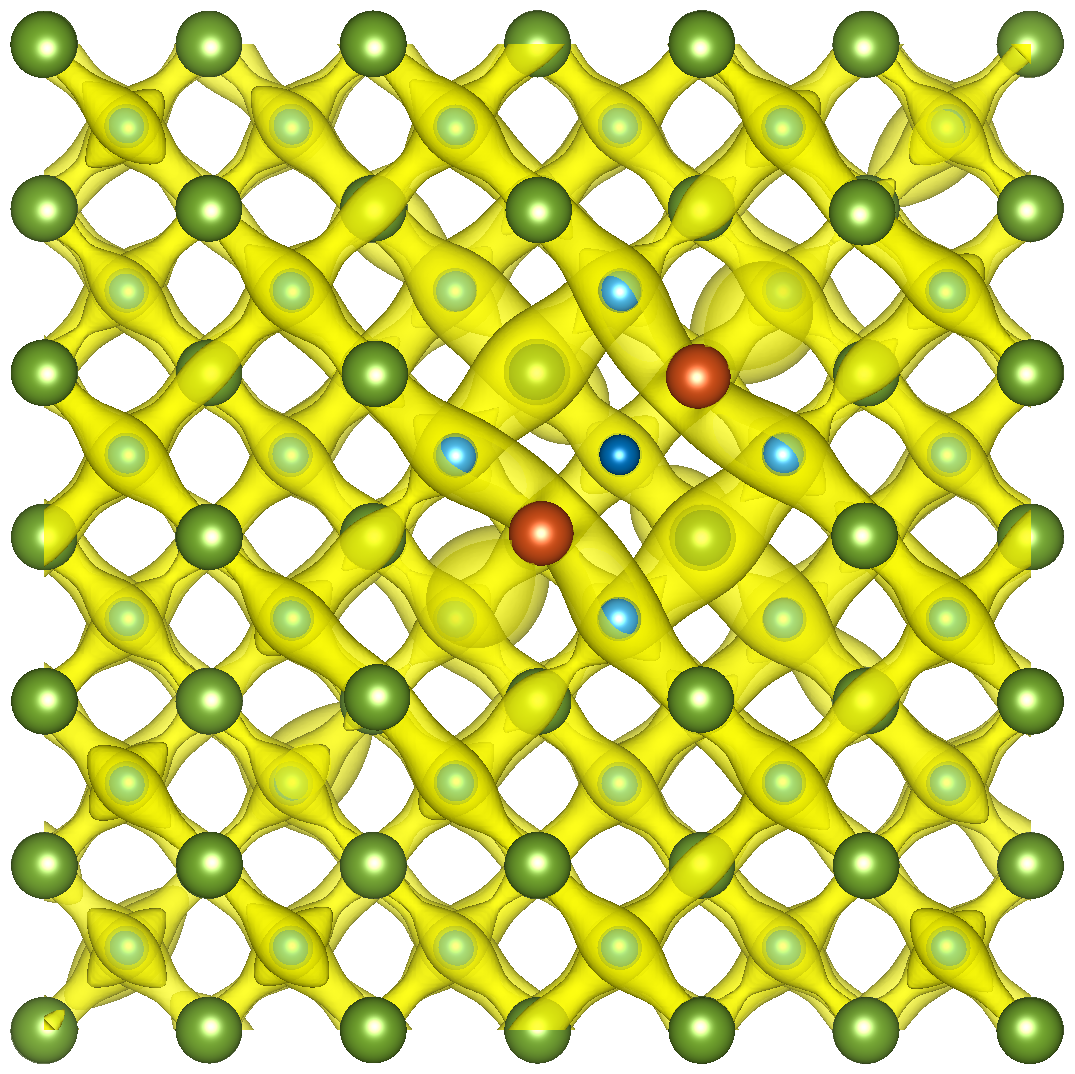}
\hspace{.5 cm}
\includegraphics[width=0.23\columnwidth]{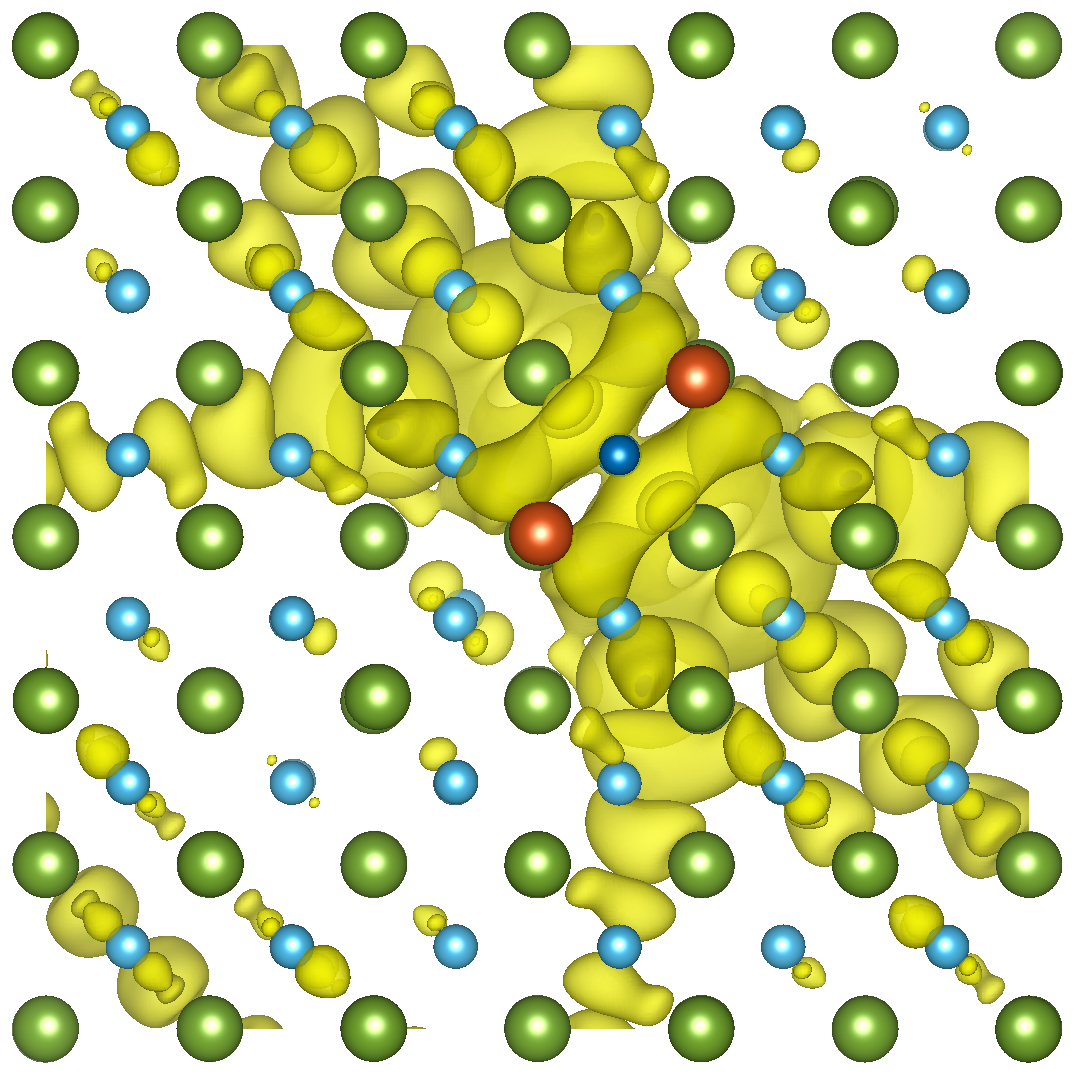}
\end{figure}

Si$_\text{As}$-B$_\text{As}$: Three bands near VBM
\begin{figure}[h]
\includegraphics[width=0.23\columnwidth]{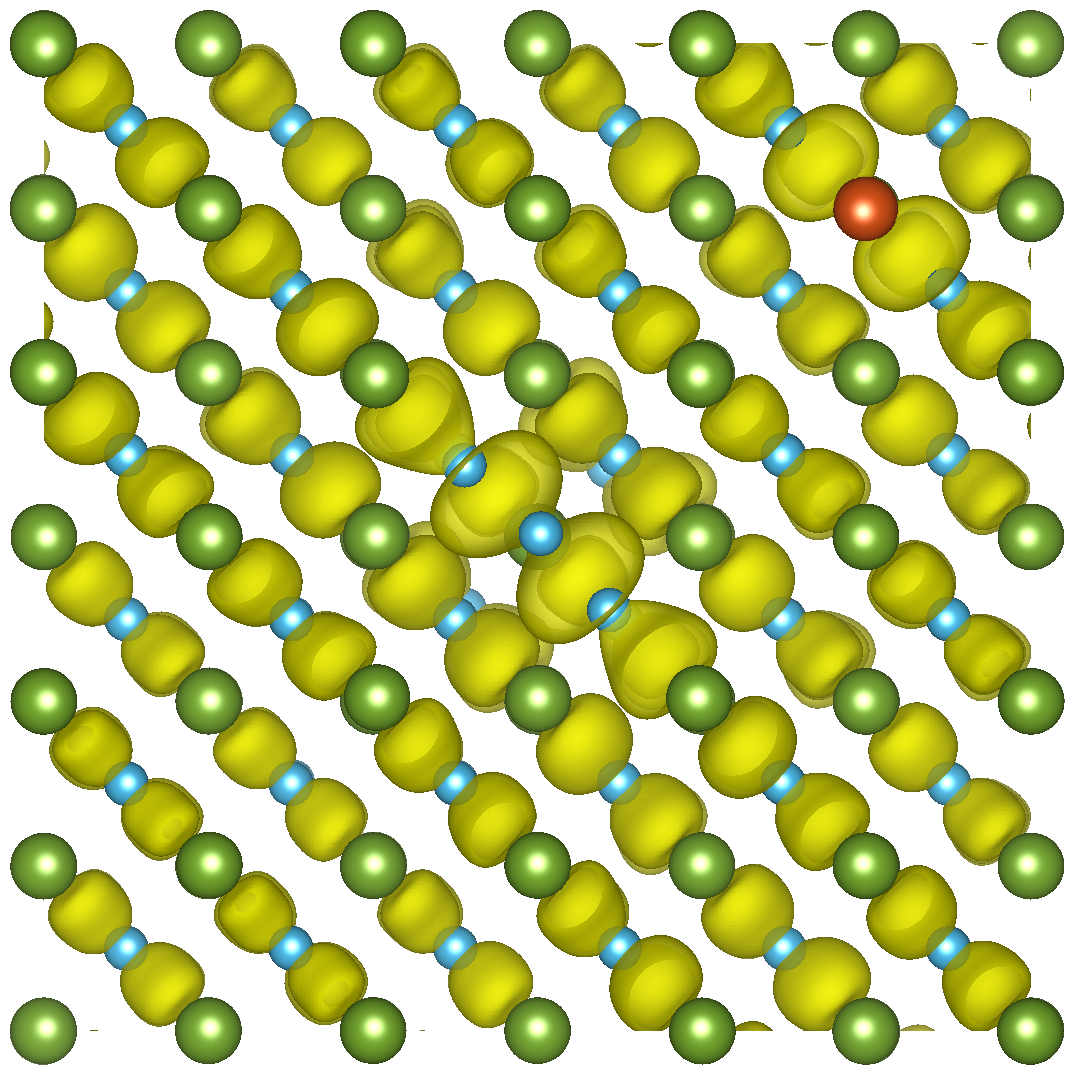}
\hspace{.5 cm}
\includegraphics[width=0.23\columnwidth]{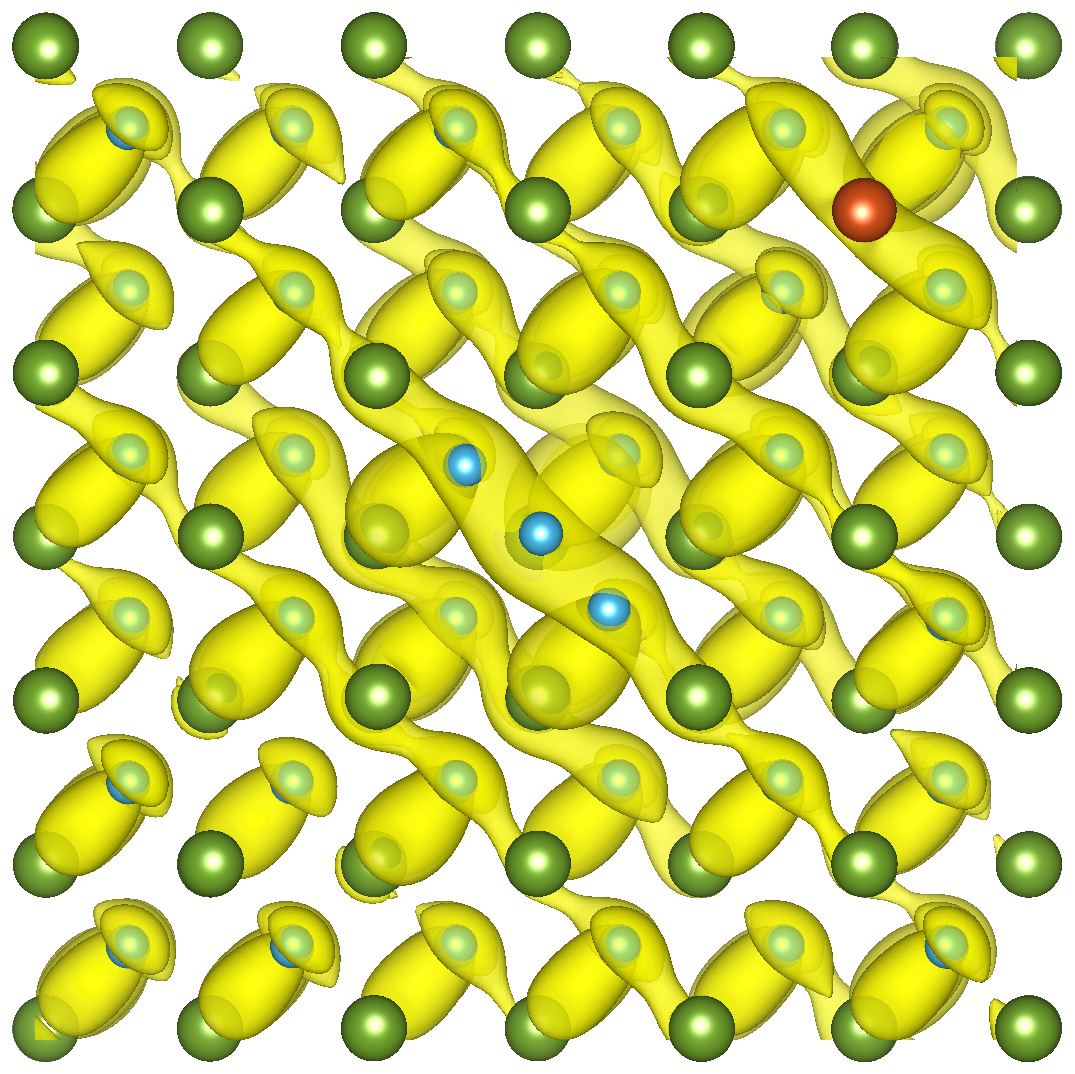}
\hspace{.5 cm}
\includegraphics[width=0.23\columnwidth]{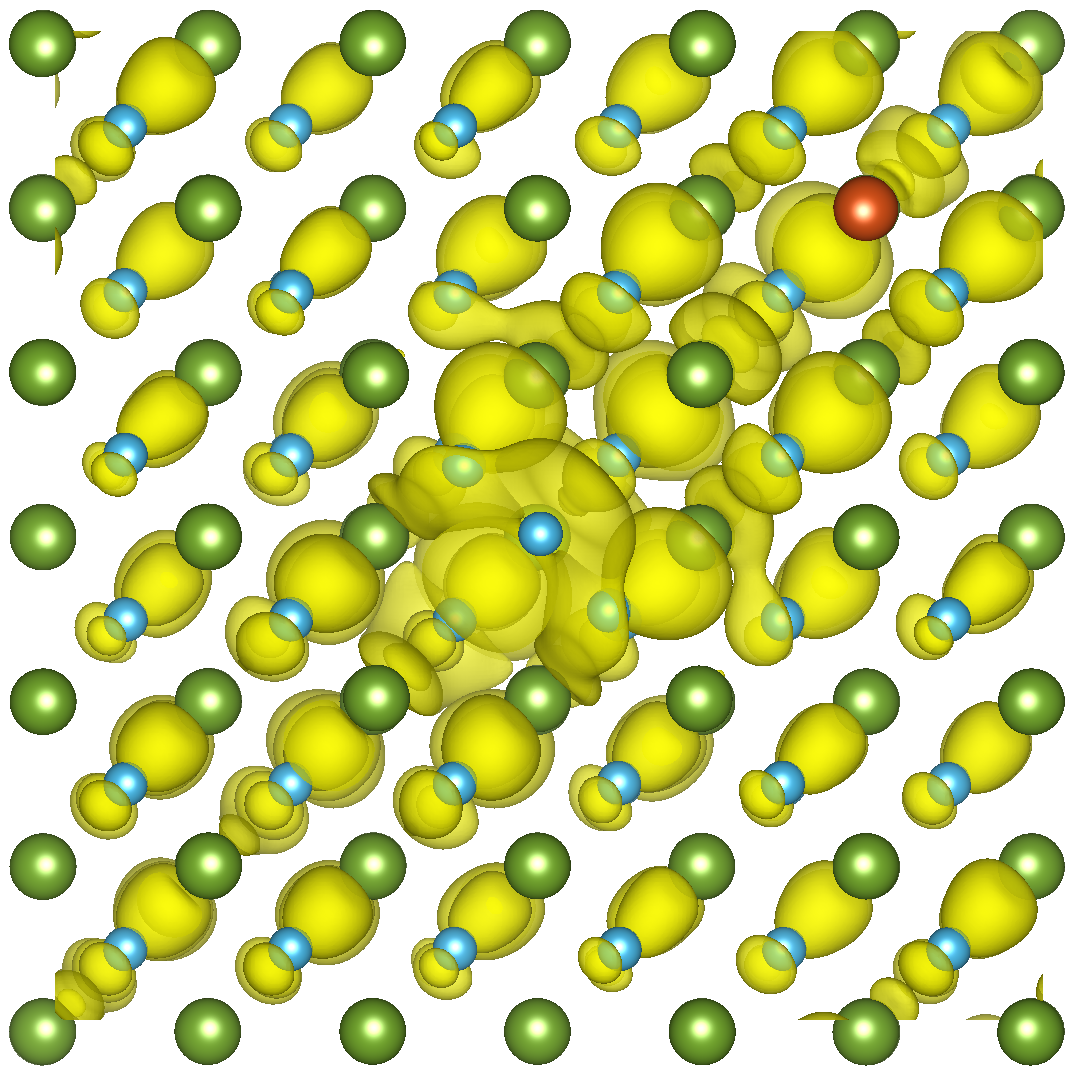}
\end{figure} 

Si$_\text{As}$-N$_\text{As}$: Three bands near VBM
\begin{figure}[h]
\includegraphics[width=0.23\columnwidth]{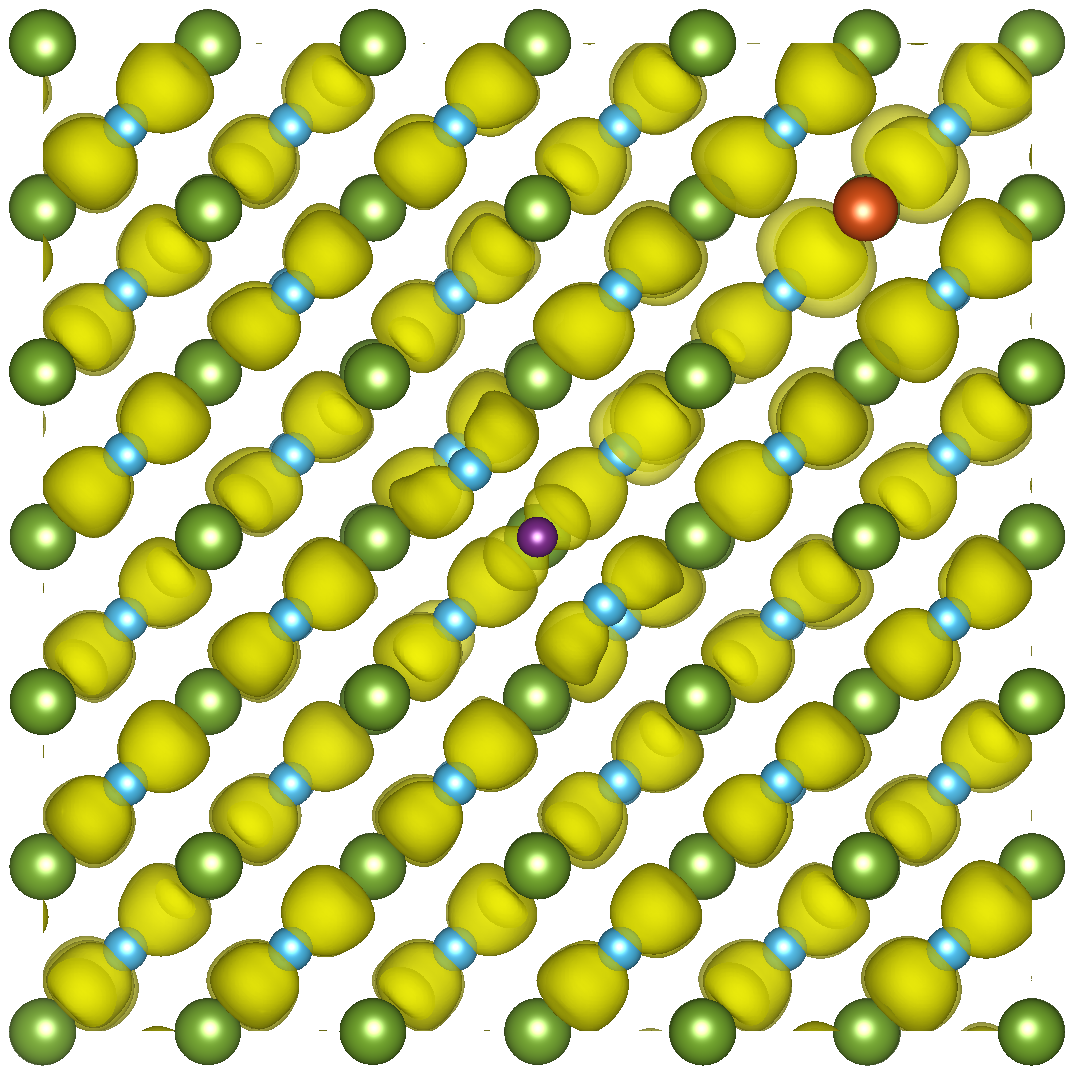}
\hspace{.5 cm}
\includegraphics[width=0.23\columnwidth]{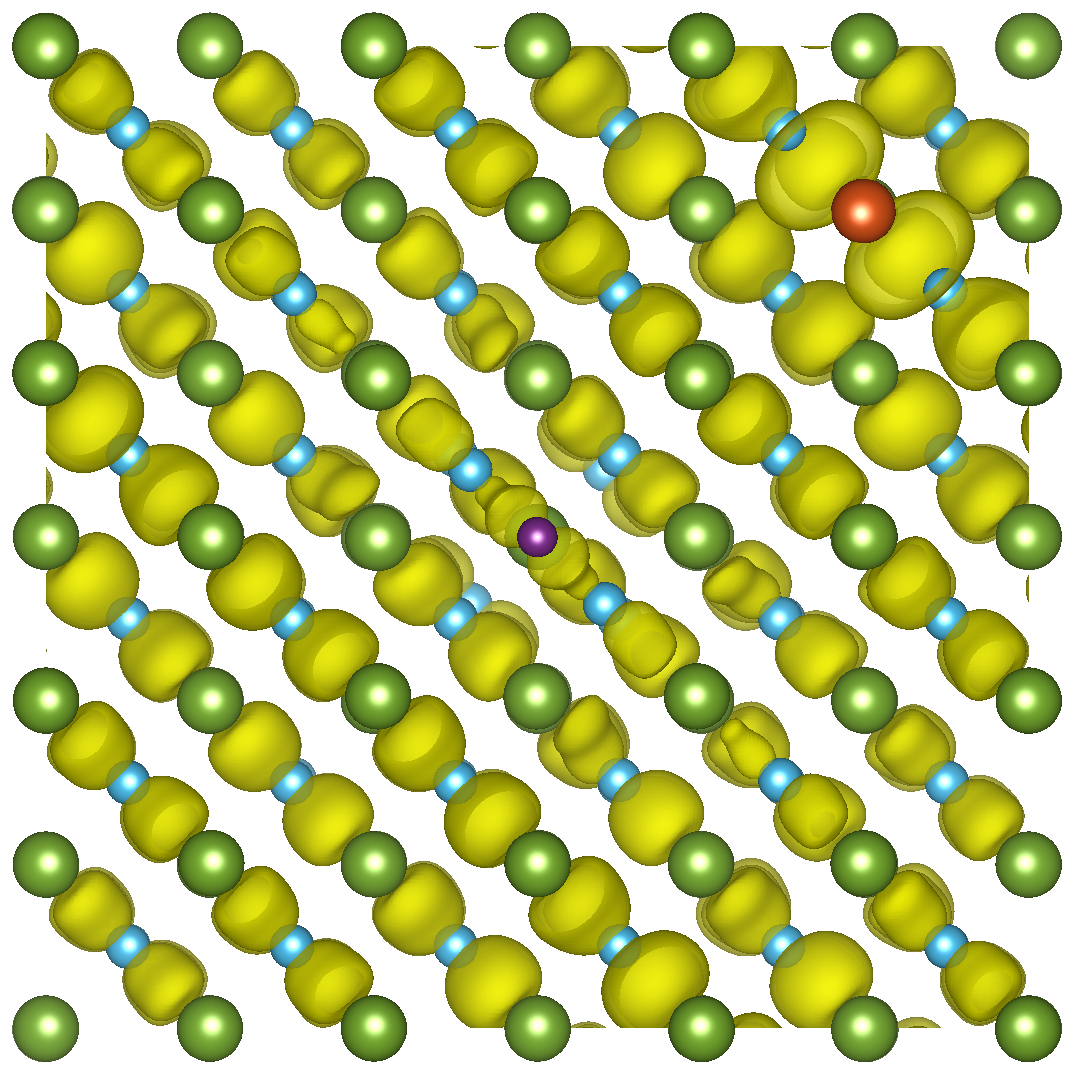}
\hspace{.5 cm}
\includegraphics[width=0.23\columnwidth]{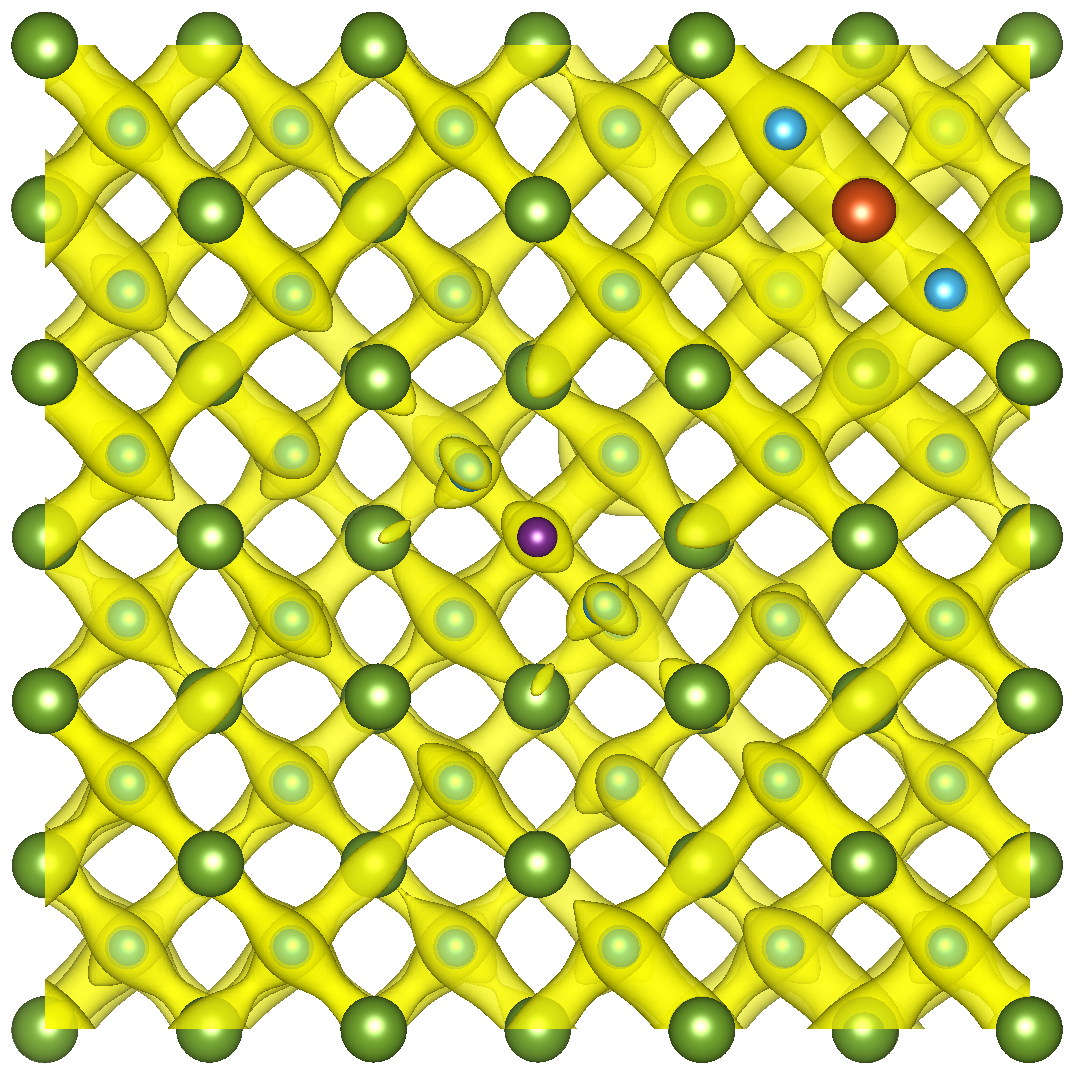}
\end{figure} 

\clearpage

Si$_\text{B}$-As$_\text{B}$B$_\text{As}$: Two bands near CBM
\begin{figure}[h]
\includegraphics[width=0.23\columnwidth]{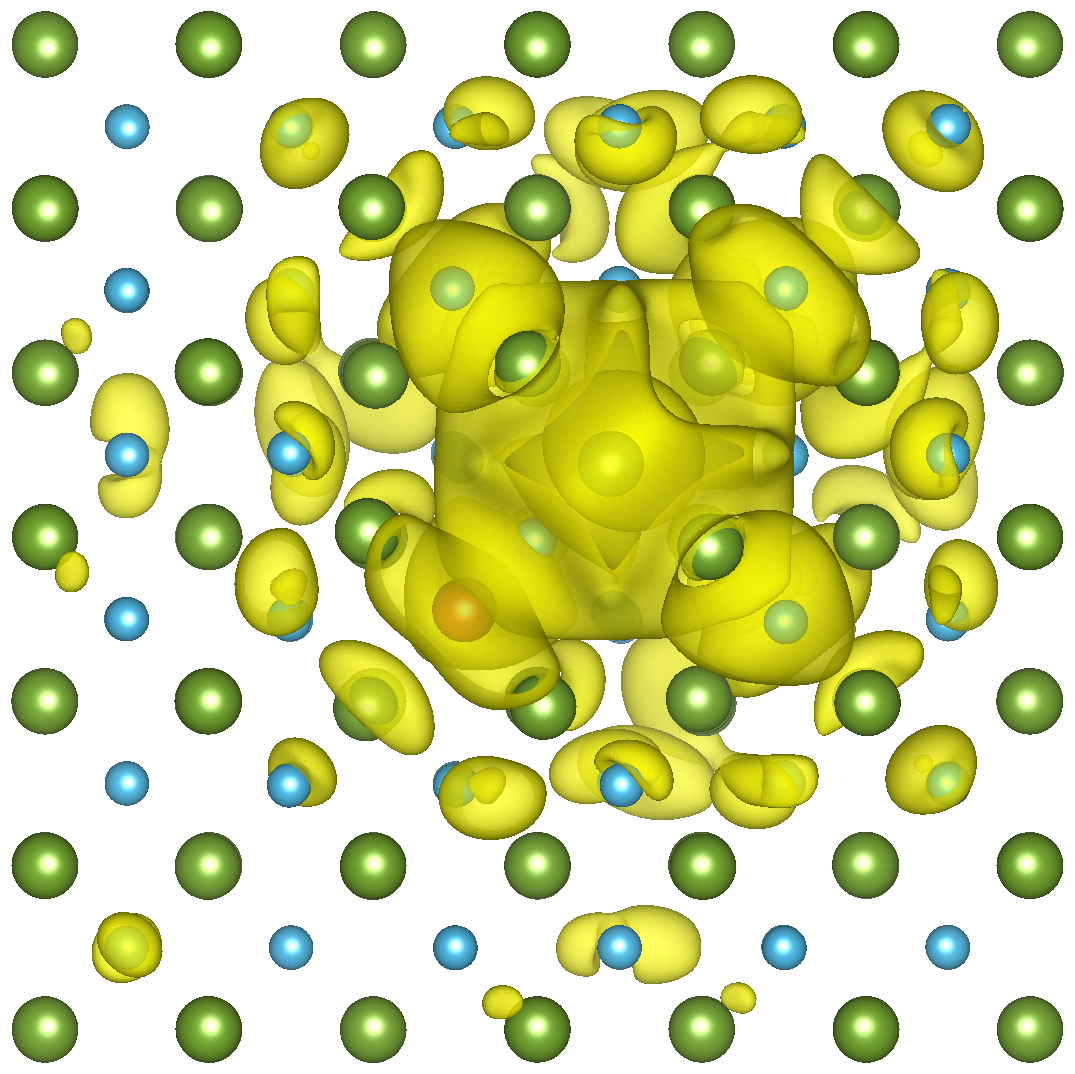}
\hspace{1 cm}
\includegraphics[width=0.23\columnwidth]{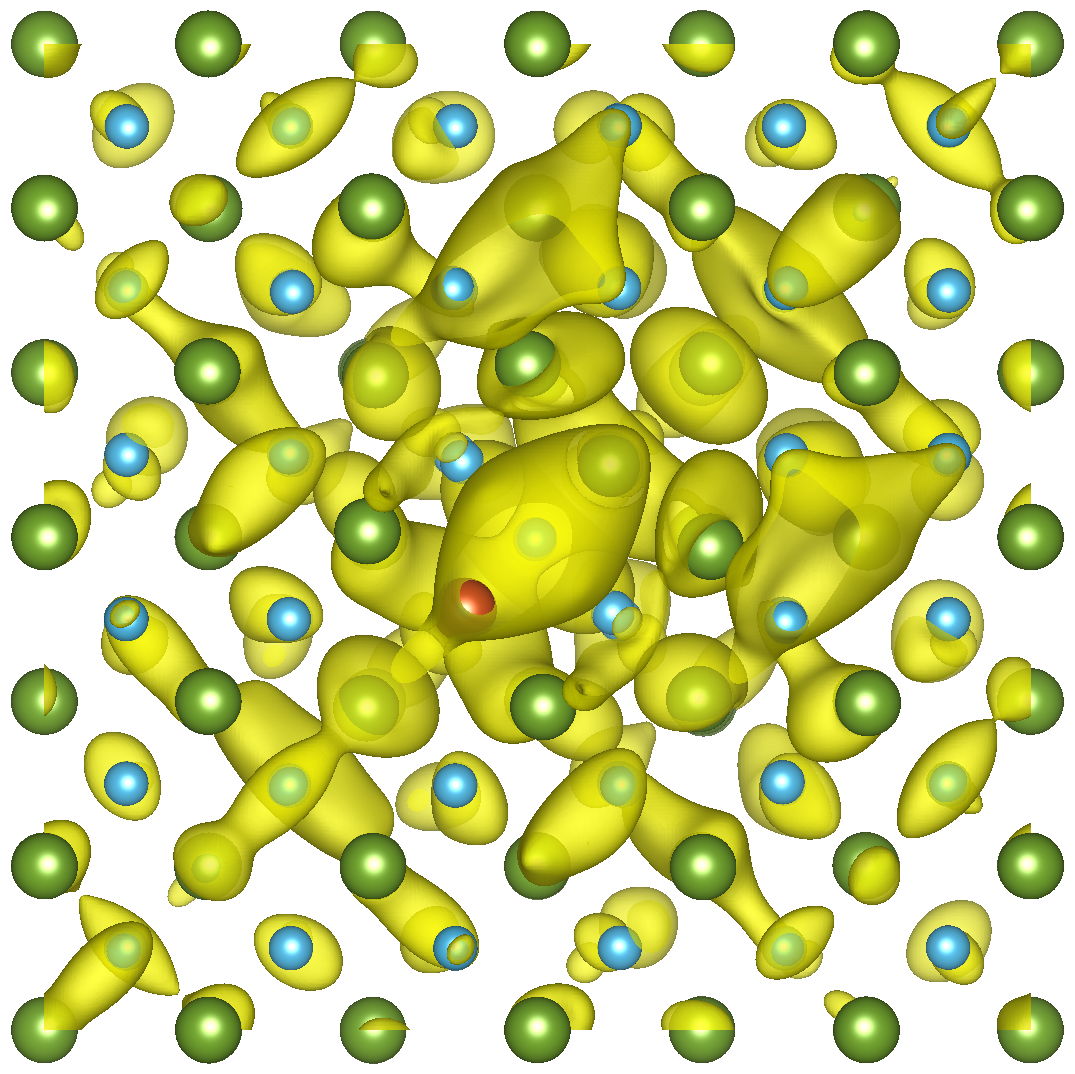}
\end{figure} 

Si$_\text{B}$-O$_\text{B}$O$_\text{As}$: One band in the band gap
\begin{figure}[h]
\includegraphics[width=0.23\columnwidth]{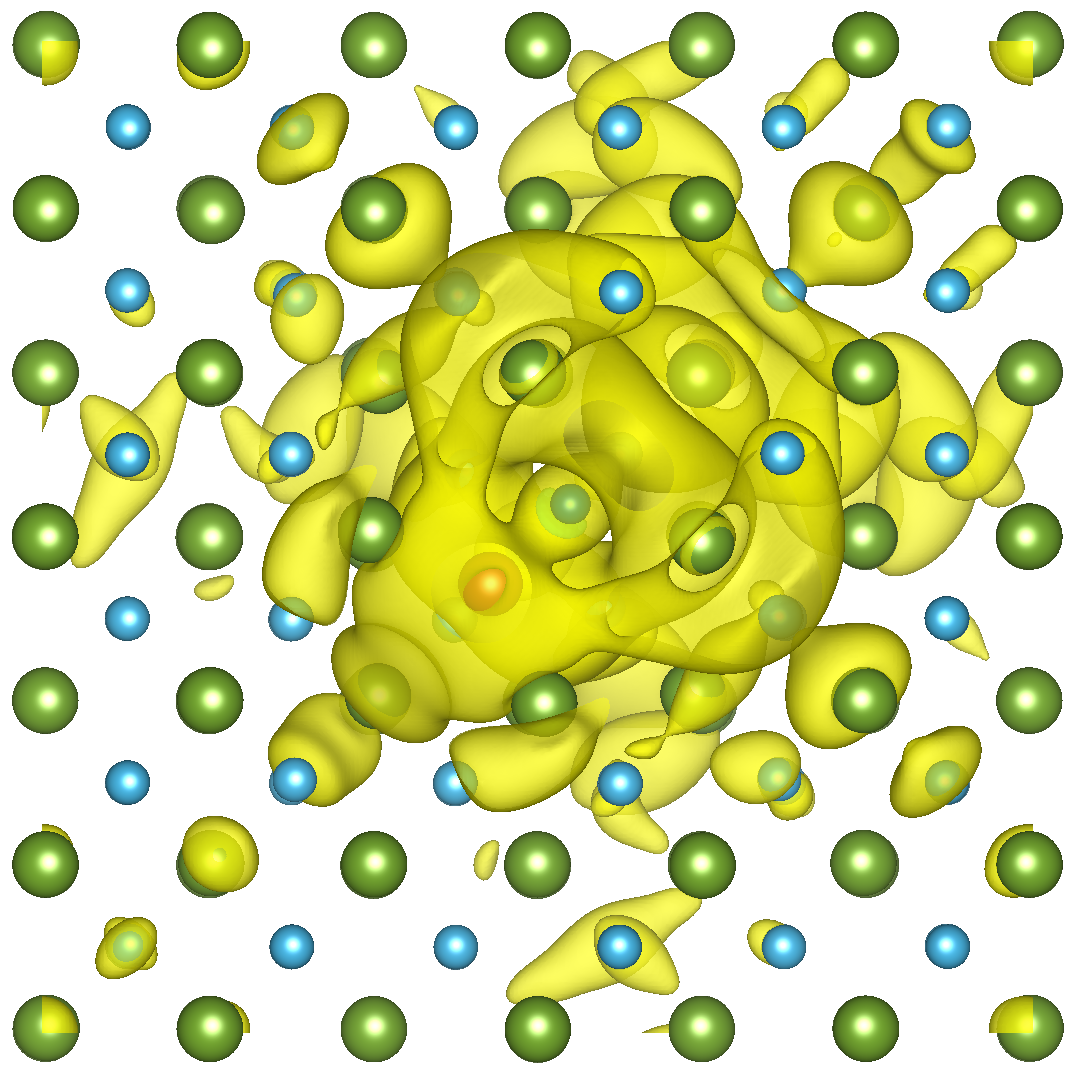}
\end{figure} 

Si$_\text{B}$-C$_\text{As}$Si$_\text{B}$: Two bands near CBM
\begin{figure}[h]
\includegraphics[width=0.23\columnwidth]{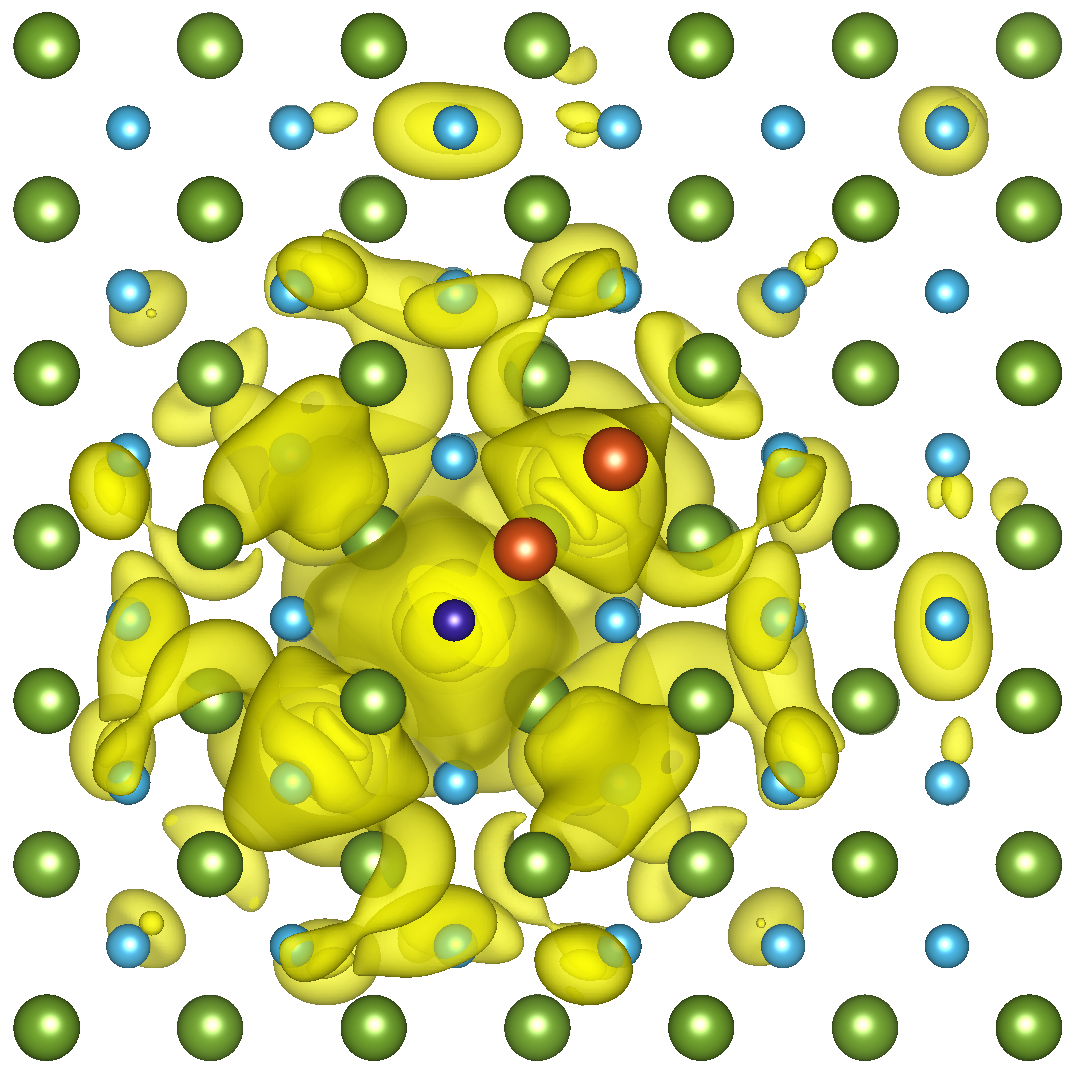}
\hspace{1 cm}
\includegraphics[width=0.23\columnwidth]{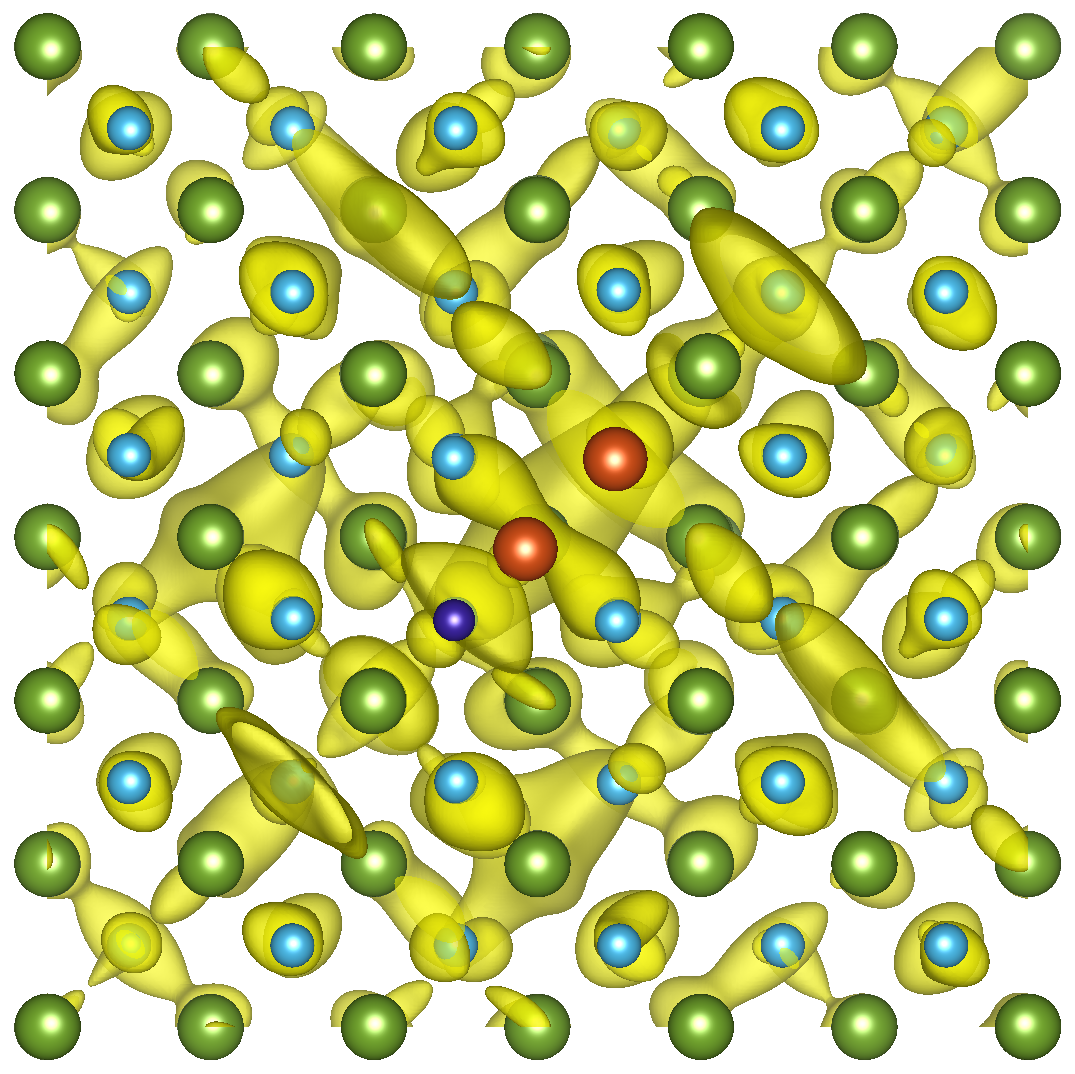}
\end{figure} 

Si$_\text{B}$-O$_\text{B}$Si$_\text{As}$: One band in the band gap
\begin{figure}[h]
\includegraphics[width=0.23\columnwidth]{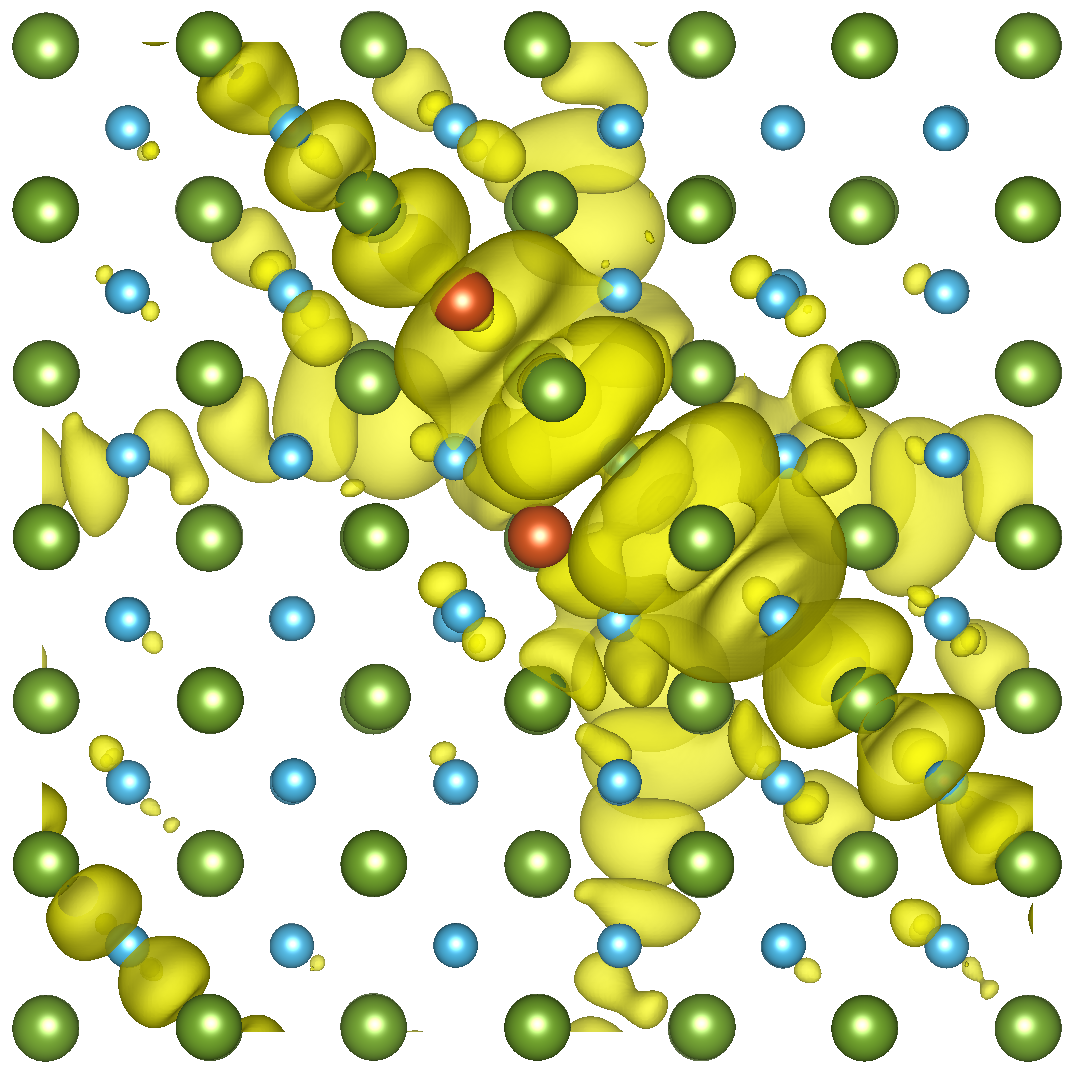}
\end{figure} 

\clearpage
Si$_\text{B}$-As$_\text{B}$: One band in the band gap (it is strongly localized and depicted at $1\times10^{-3} \AA^{-3}$), and two bands near CBM
\begin{figure}[h]
\includegraphics[width=0.23\columnwidth]{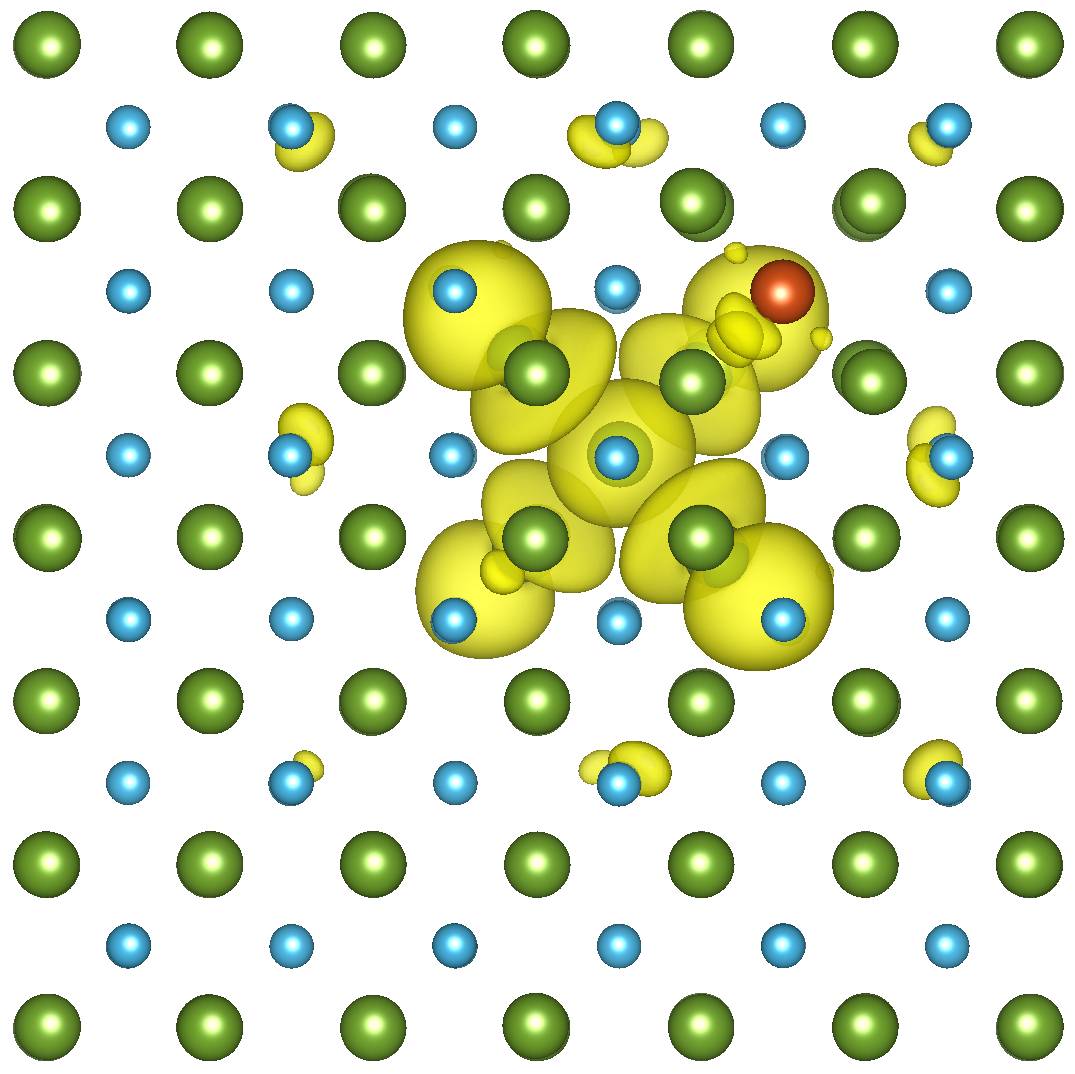}
\hspace{.5 cm}
\includegraphics[width=0.23\columnwidth]{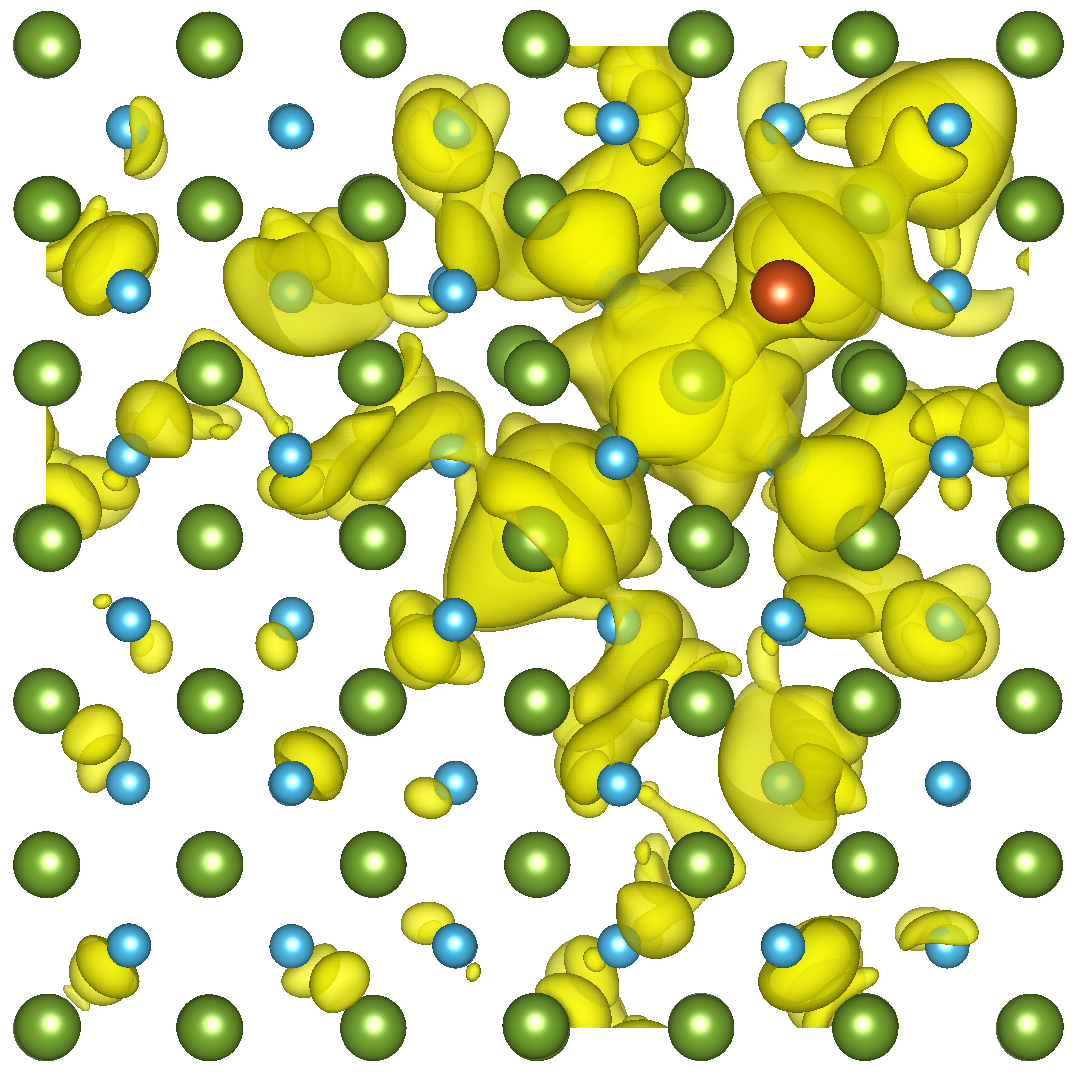}
\hspace{.5 cm}
\includegraphics[width=0.23\columnwidth]{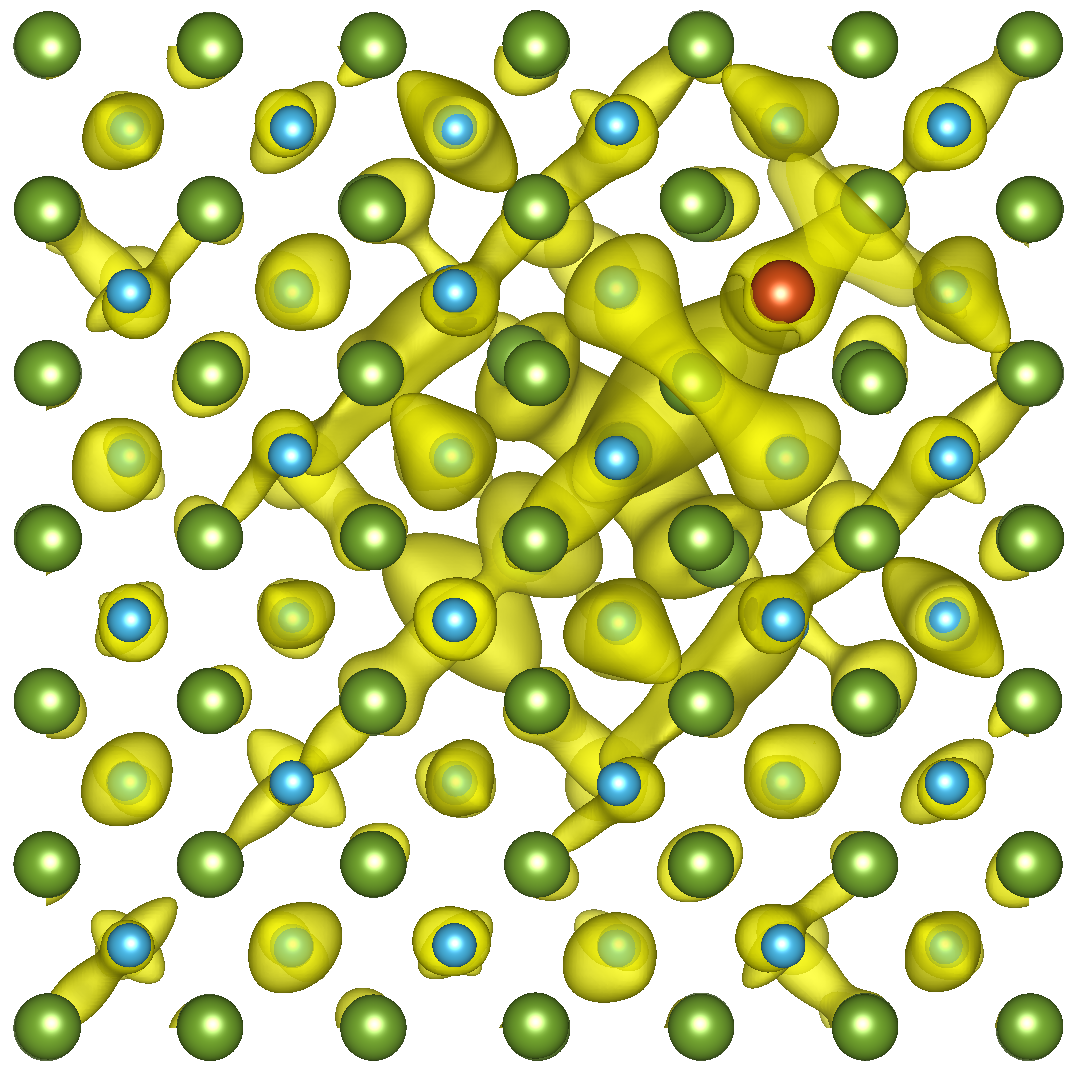}
\end{figure} 

Si$_\text{B}$-N$_\text{As}$: One band at CBM
\begin{figure}[h]
\includegraphics[width=0.23\columnwidth]{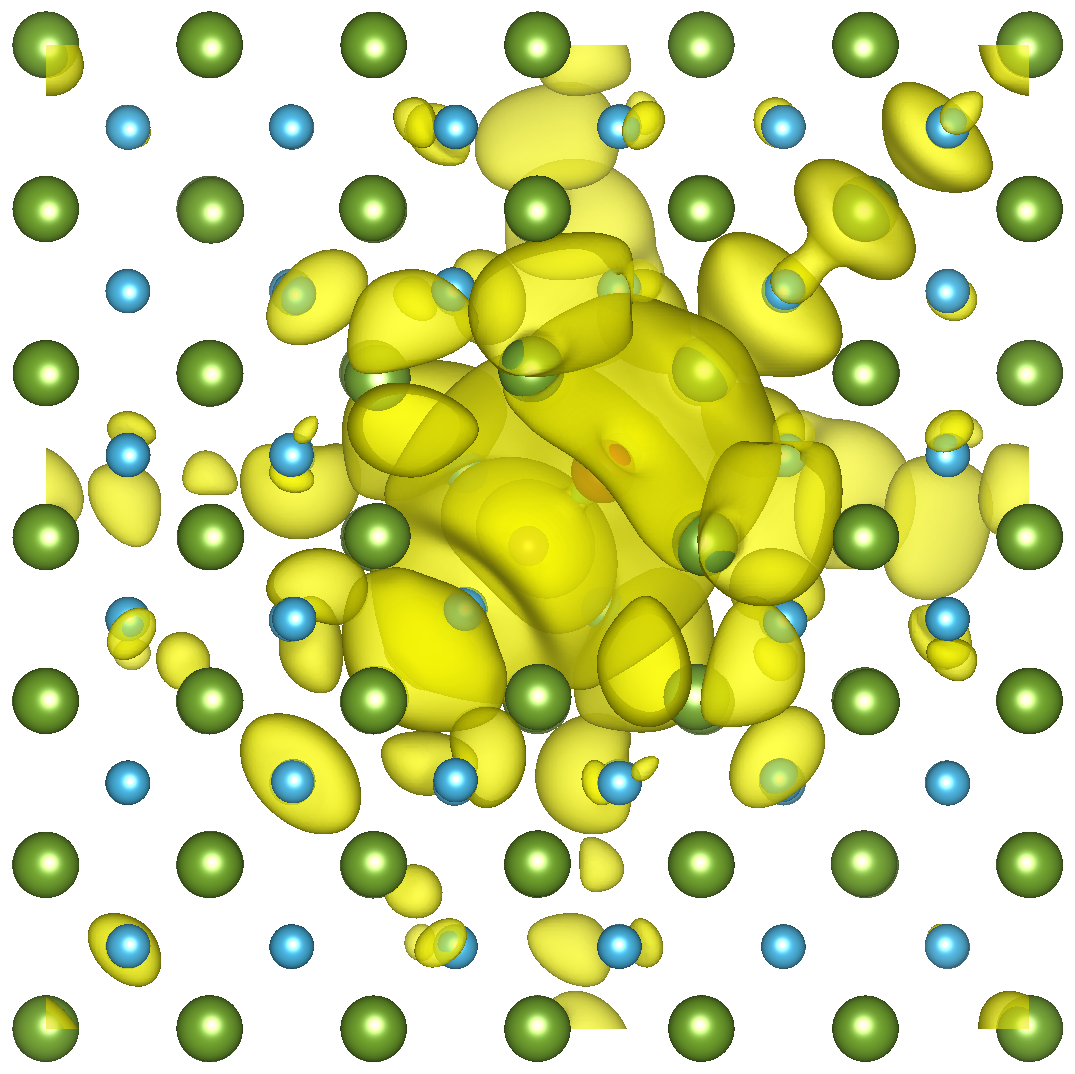}
\end{figure} 

Se$_\text{As}$-As$_\text{B}$B$_\text{As}$: Two bands near CBM
\begin{figure}[h]
\includegraphics[width=0.23\columnwidth]{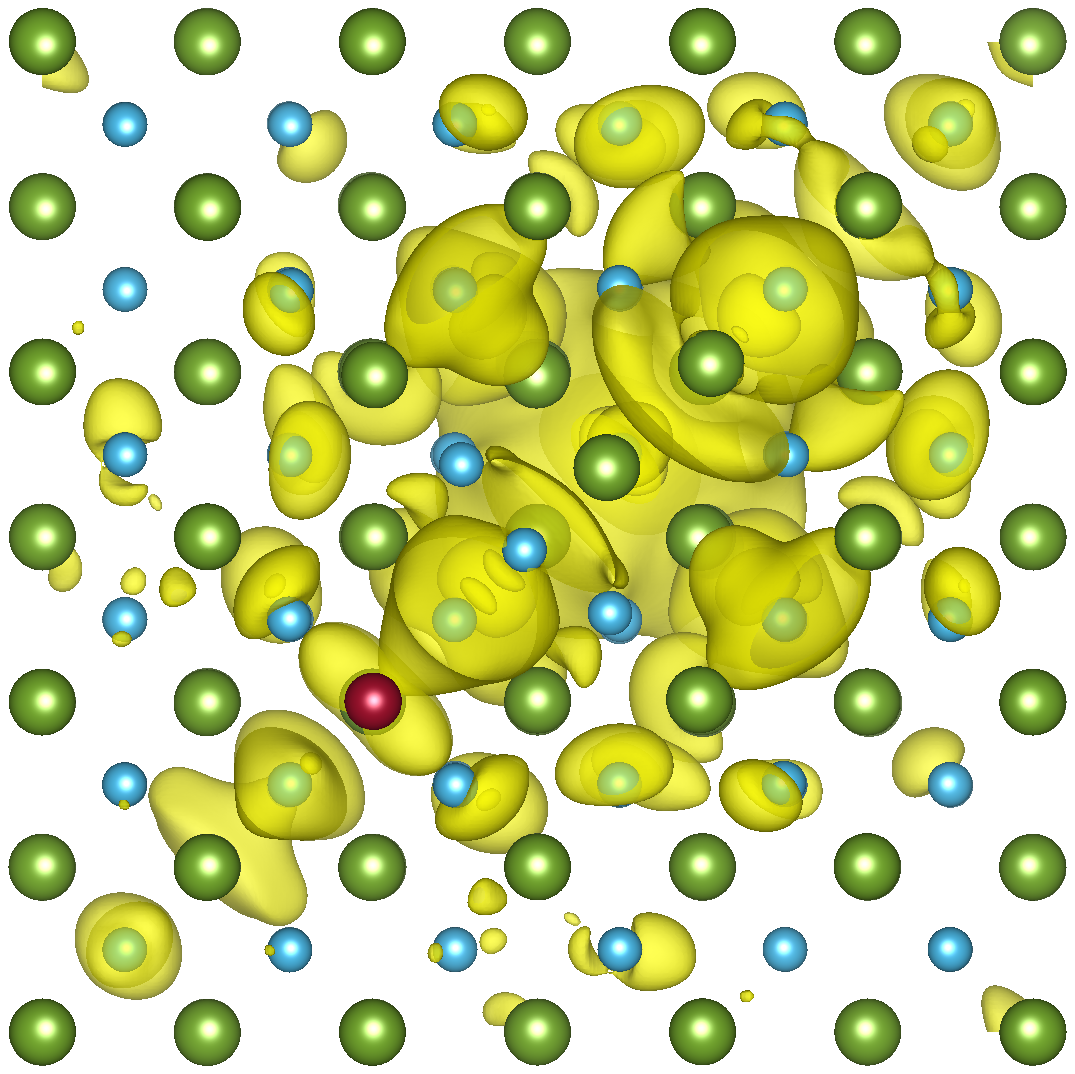}
\hspace{1 cm}
\includegraphics[width=0.23\columnwidth]{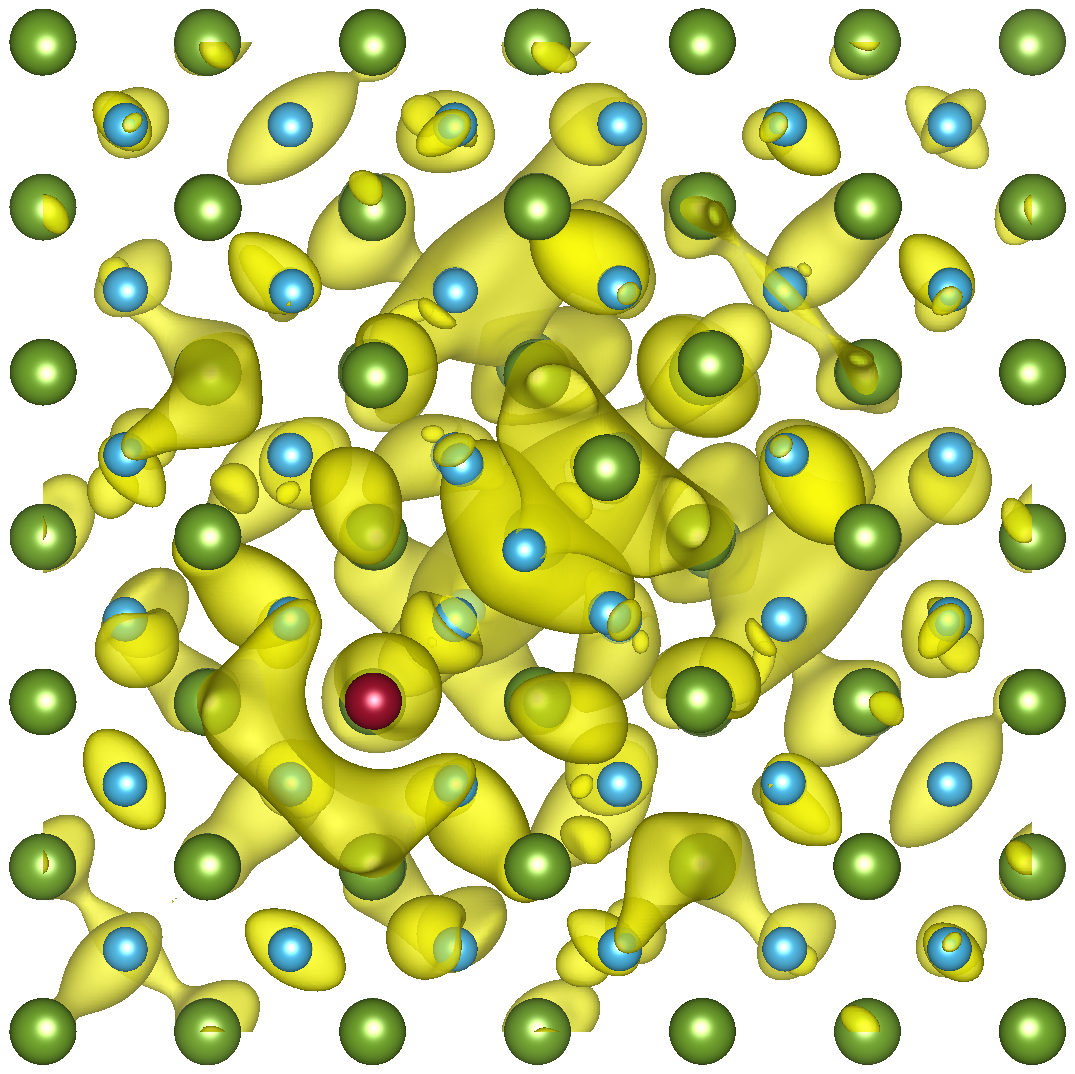}
\end{figure} 

Se$_\text{As}$-O$_\text{B}$O$_\text{As}$: One band in the band gap
\begin{figure}[h]
\includegraphics[width=0.23\columnwidth]{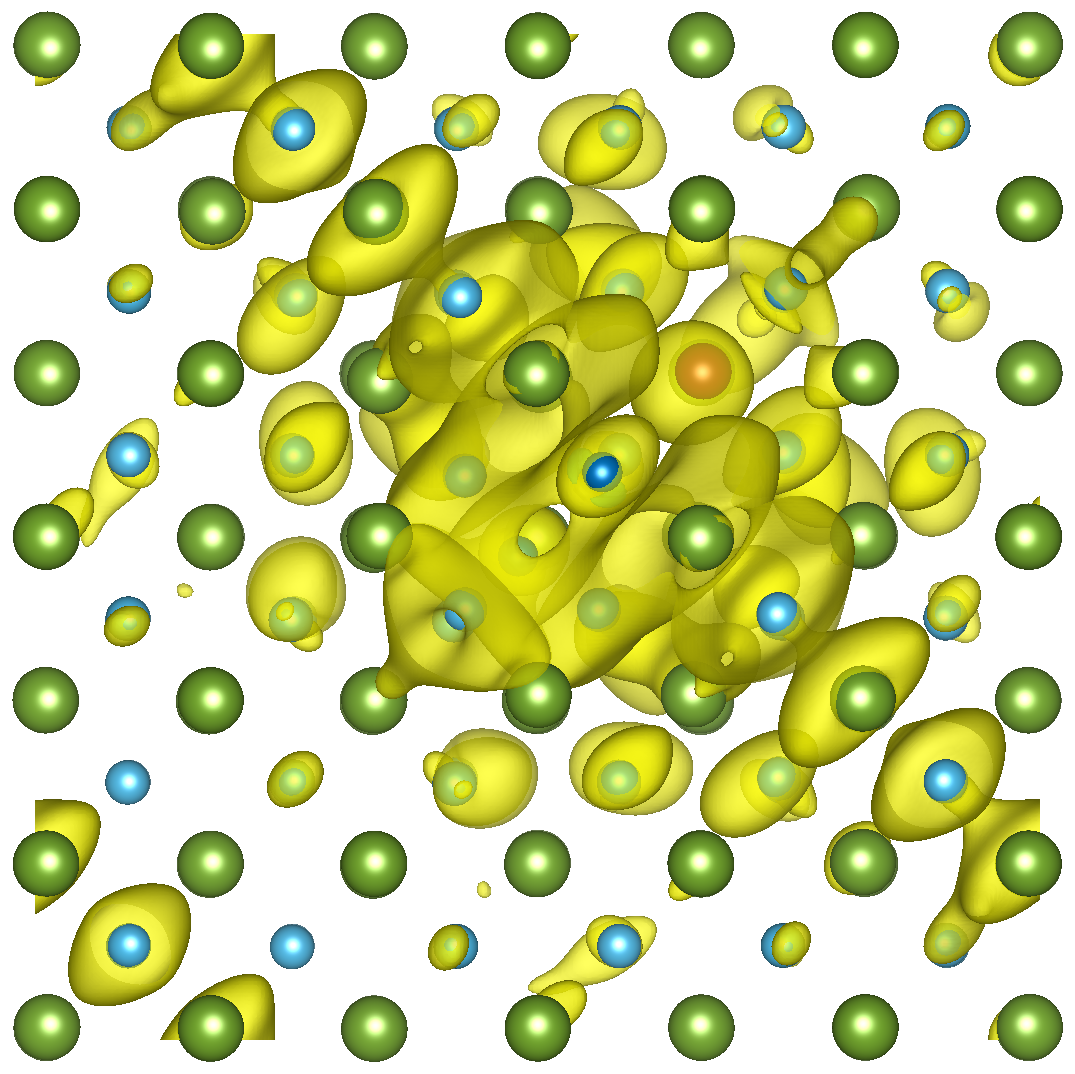}
\end{figure} 
\clearpage

Se$_\text{As}$-C$_\text{As}$Si$_\text{B}$: Two bands near CBM
\begin{figure}[h]
\includegraphics[width=0.23\columnwidth]{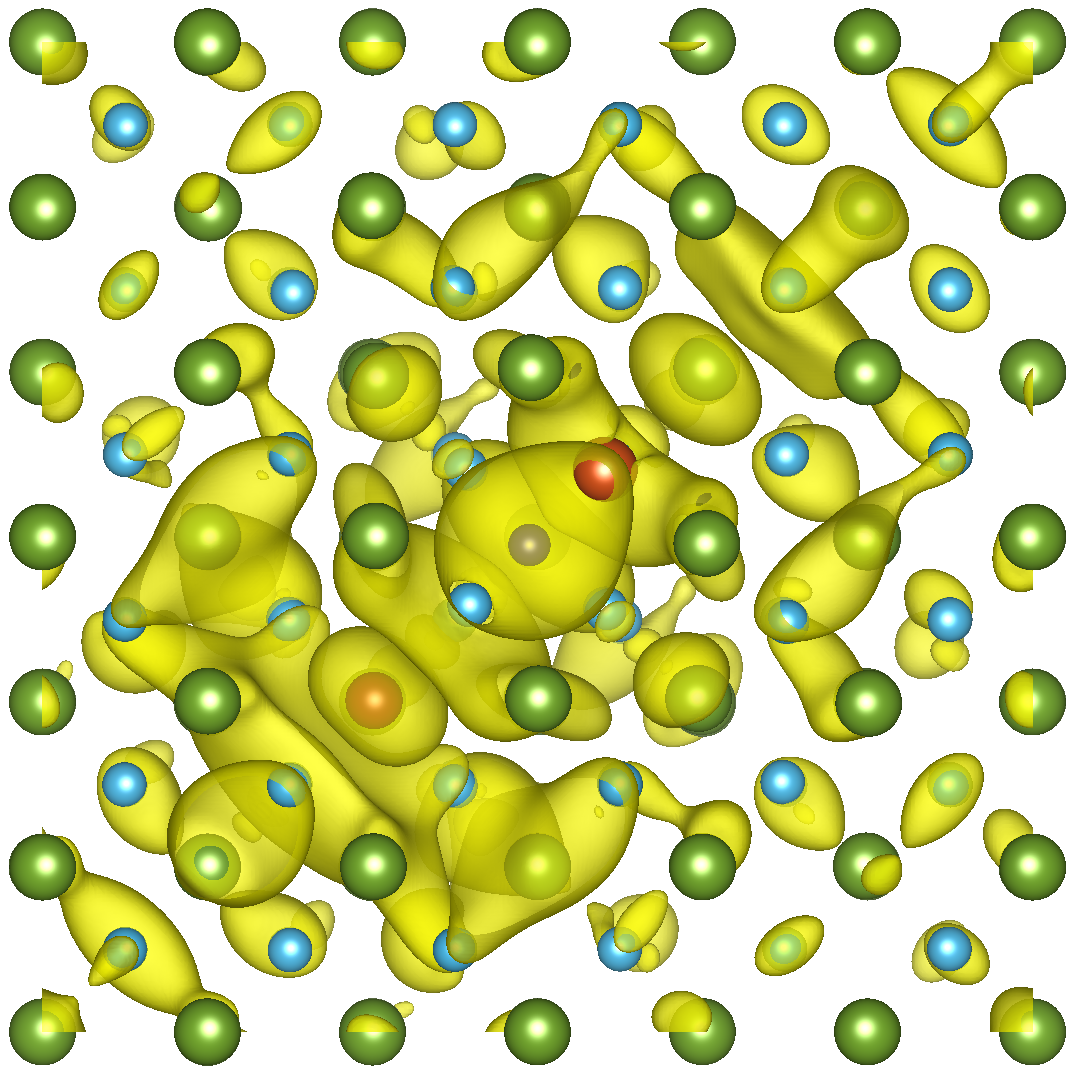}
\hspace{1 cm}
\includegraphics[width=0.23\columnwidth]{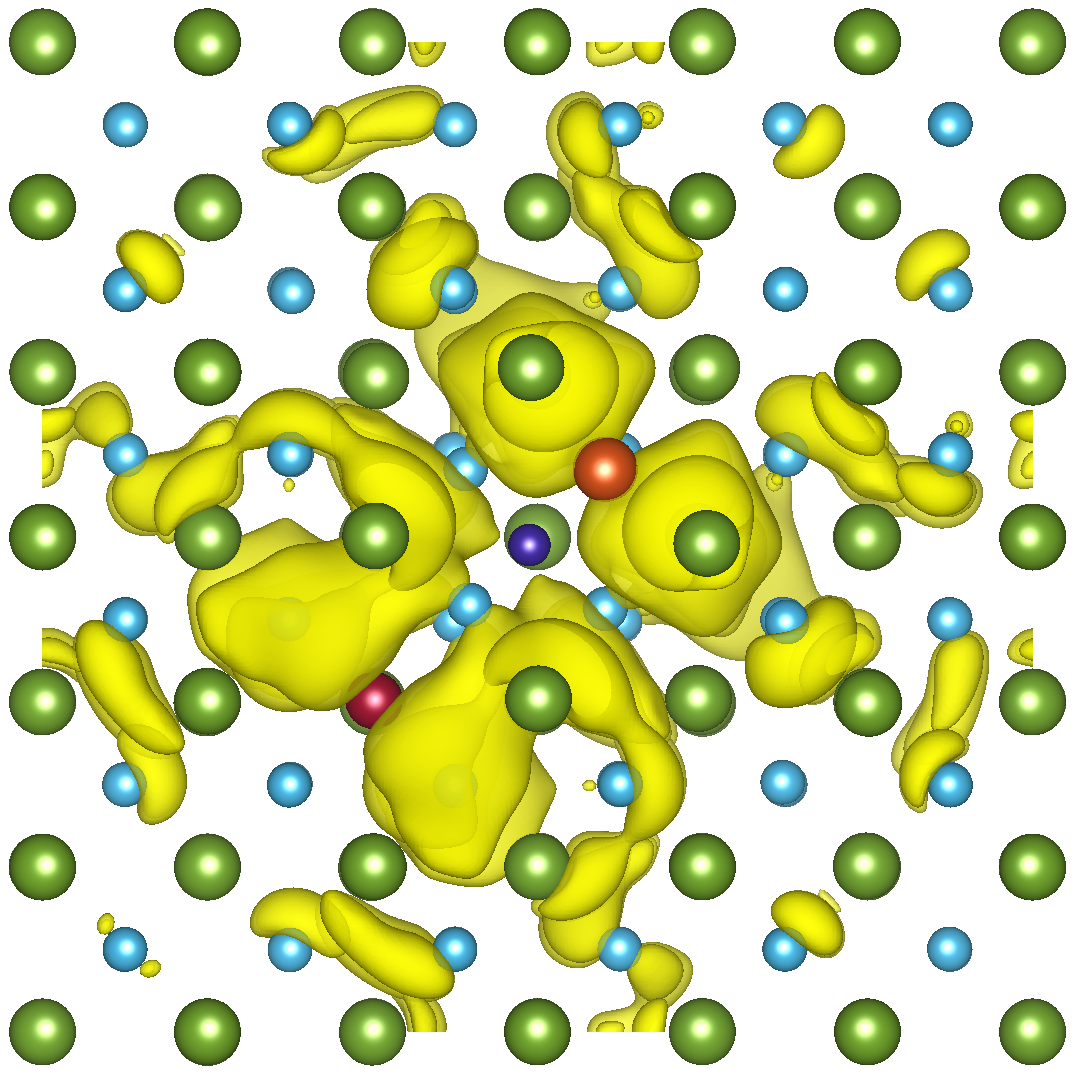}
\end{figure} 

Se$_\text{As}$-O$_\text{B}$Si$_\text{As}$: One band in the band gap
\begin{figure}[h]
\includegraphics[width=0.23\columnwidth]{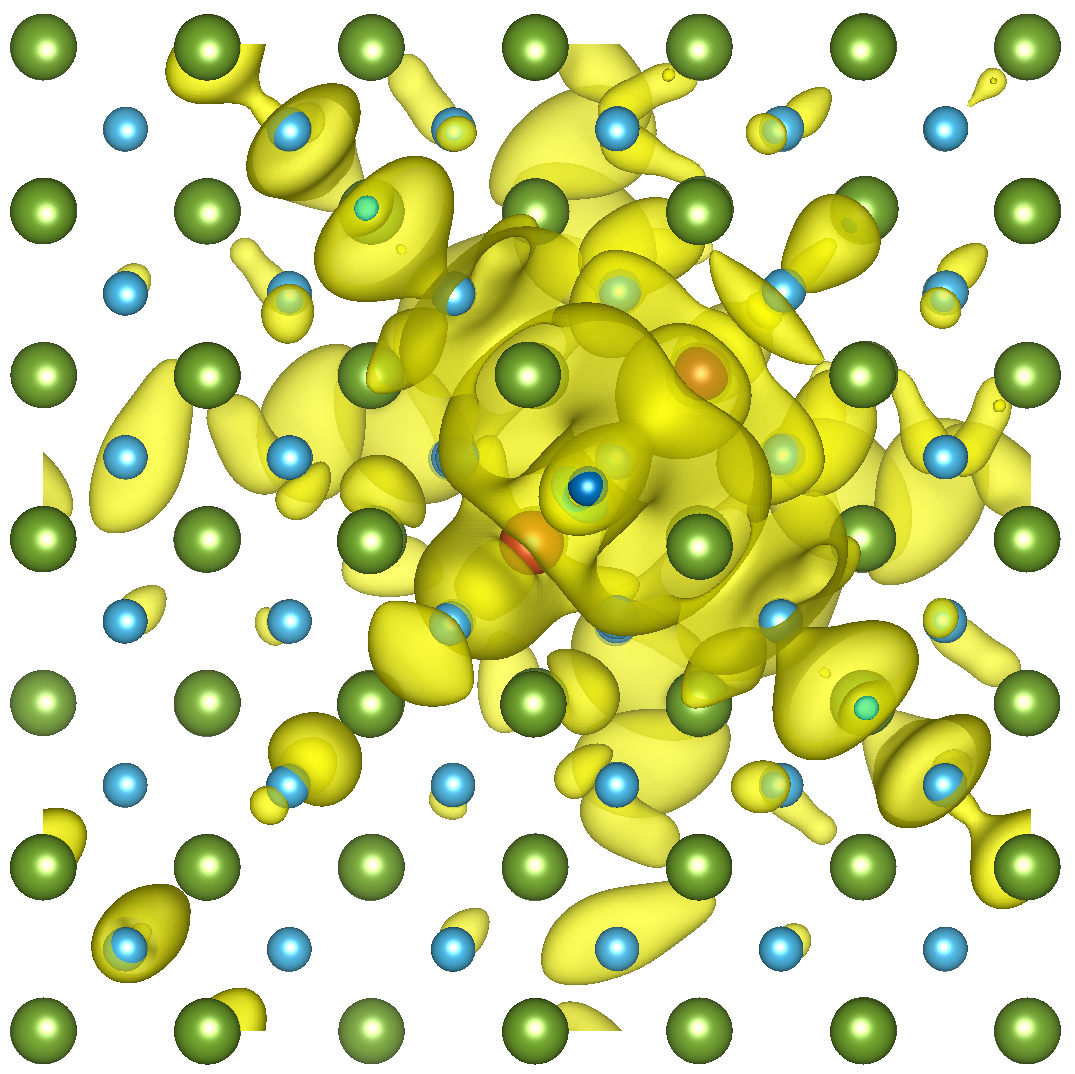}
\end{figure} 

Se$_\text{As}$-As$_\text{B}$: Two bands near CBM
\begin{figure}[h]
\includegraphics[width=0.23\columnwidth]{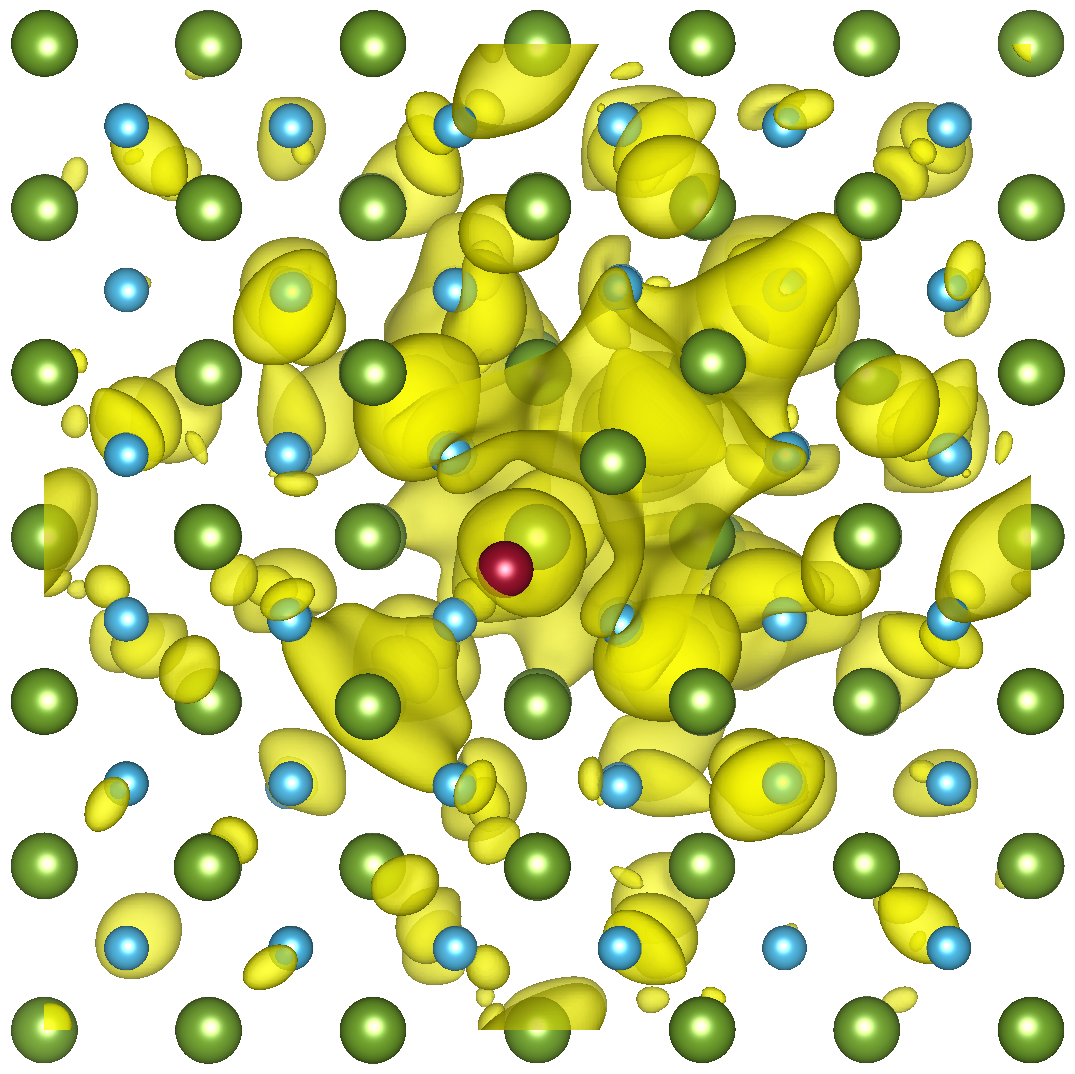}
\hspace{1 cm}
\includegraphics[width=0.23\columnwidth]{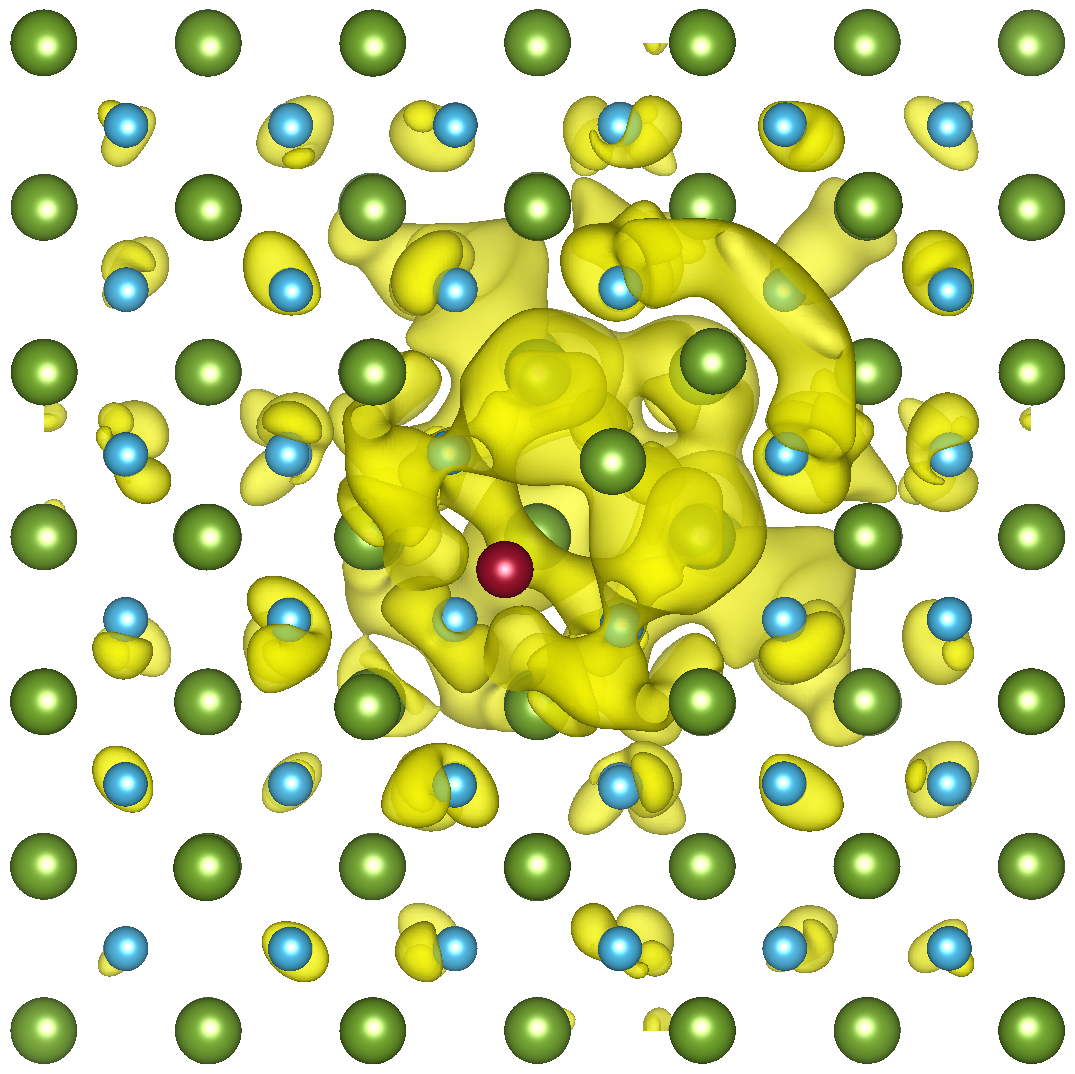}
\end{figure} 

Se$_\text{As}$-N$_\text{As}$: Two bands near CBM
\begin{figure}[h]
\includegraphics[width=0.23\columnwidth]{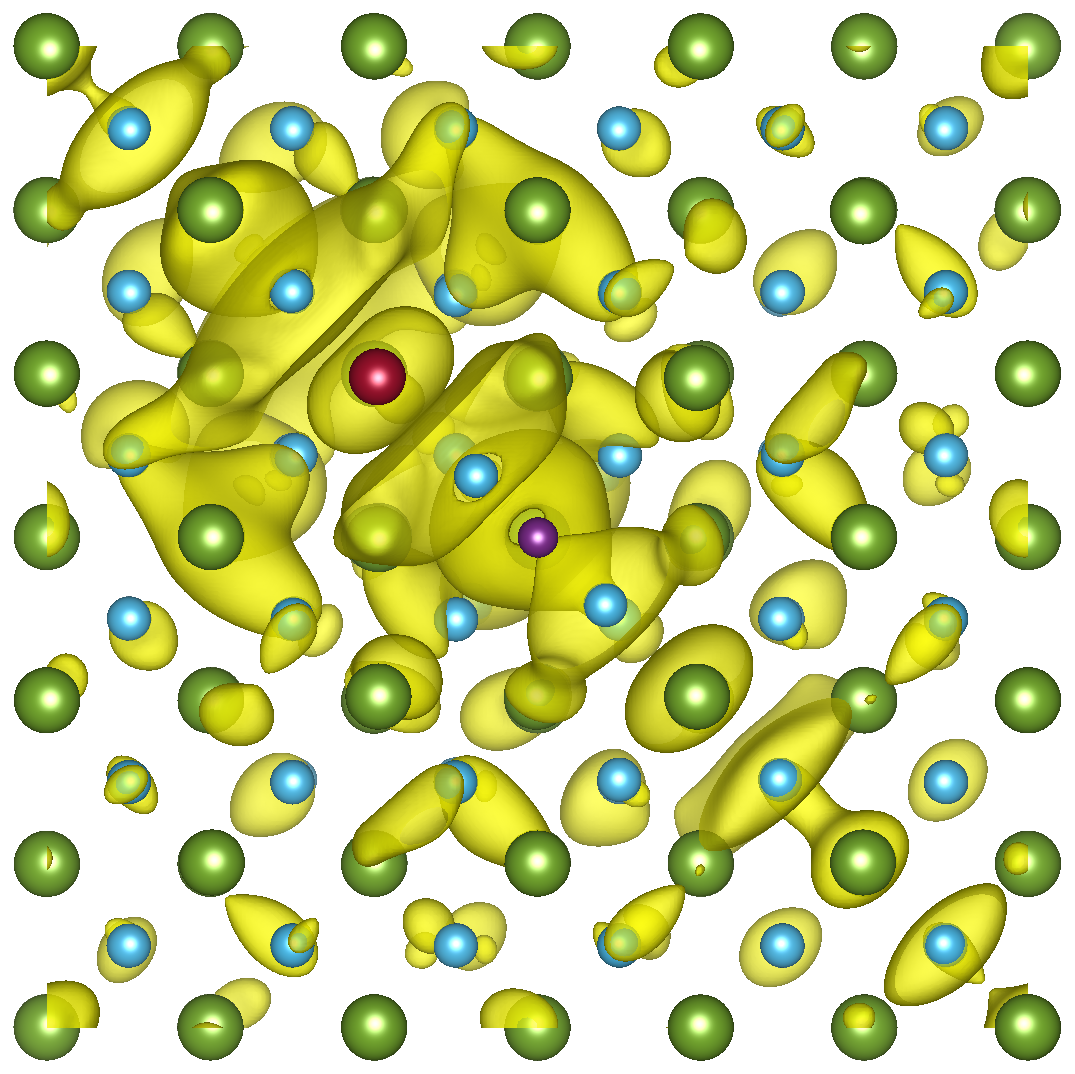}
\hspace{1 cm}
\includegraphics[width=0.23\columnwidth]{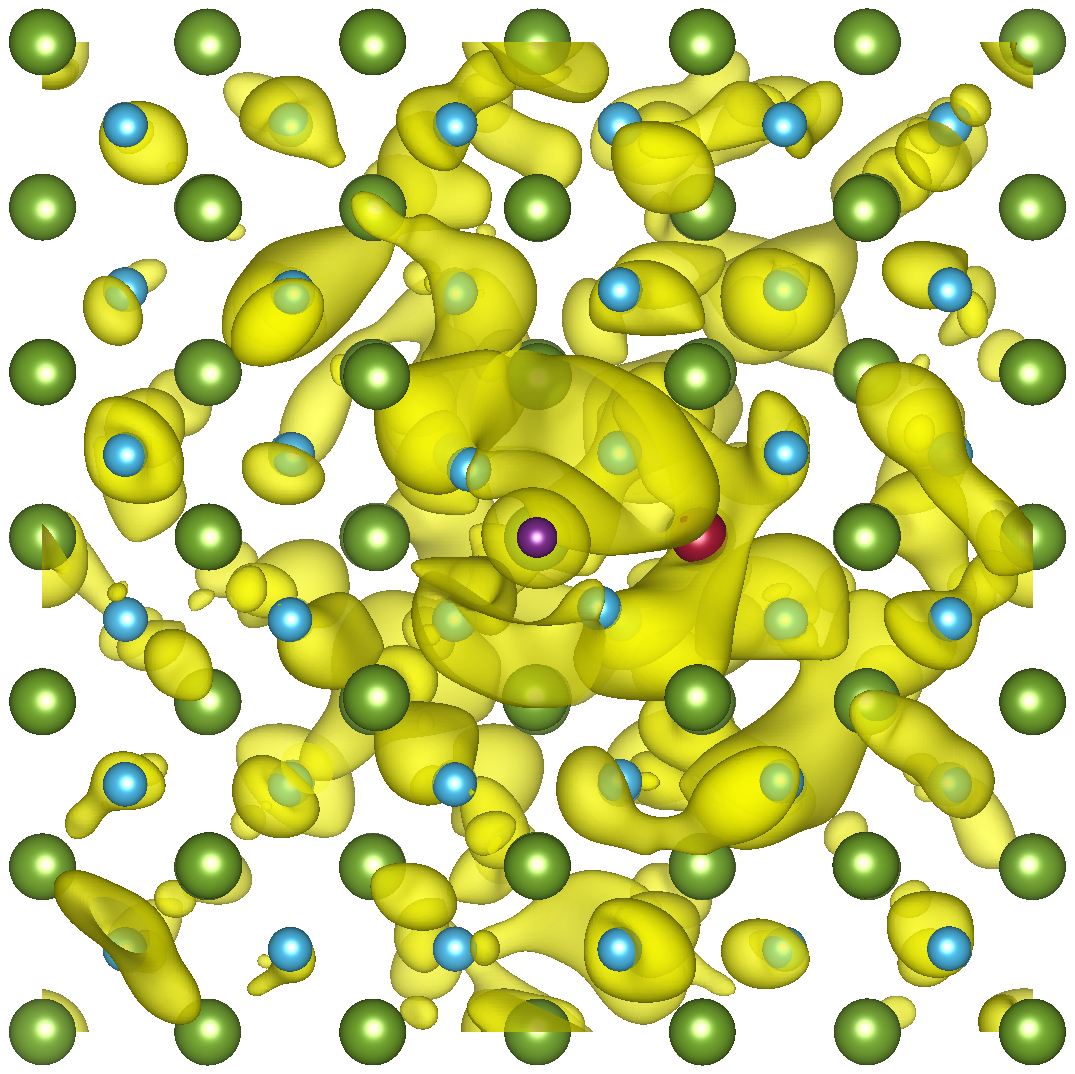}
\end{figure} 

\clearpage


\bibliographystyle{apsrev4-1} 
\bibliography{bas} 